\documentclass[a4paper,11pt]{article}
\pdfoutput=1 
\usepackage{jcappub} 
\usepackage[T1]{fontenc} 
\usepackage{amsmath}
\usepackage{amsfonts}
\usepackage{amsthm}
\usepackage{amssymb}
\usepackage{latexsym, array, multirow, verbatim, enumerate}
\usepackage{slashed}
\usepackage{booktabs}
\usepackage{tabularx}
\usepackage{cancel}
\usepackage{dcolumn}
\usepackage{bm}
\usepackage{graphicx, caption, subcaption} 
\usepackage{url}
\usepackage{pifont}
\usepackage{empheq}
\usepackage[utf8]{inputenc}
\usepackage{bm}
\usepackage{blkarray}
\usepackage[font=footnotesize,labelfont=sf]{caption}
\usepackage{cleveref}
\usepackage{float}
\usepackage{multirow}
\usepackage{physics}
\usepackage{hyperref}
\usepackage{xcolor}
\usepackage[normalem]{ulem}

\newcommand{\solarmass}{\textup{M}_\odot}

\title{\boldmath Probing Exotic Astrophysical Dark objects through Astrometric Microlensing from Gaia}

\author[a]{Lalit Singh Bhandari,}
\author[b]{Vikram Rentala,}
\author[a]{Arun M. Thalapillil,}
\author[c]{and Himanshu Verma,}

\affiliation[a]{ Department of Physics,\\ Indian Institute of Science Education and Research Pune,\\ Pune 411008, India}
\affiliation[b]{ Department of Physics,\\ Indian Institute of Technology Bombay, \\Powai, Mumbai,
Maharashtra, 400076, India}

\affiliation[c]{Department of Physics and Astronomy, Louisiana State University, Baton Rouge, LA 70803, USA}

\emailAdd{bhandari.lalitsingh@students.iiserpune.ac.in}
\emailAdd{rentala@phy.iitb.ac.in}
\emailAdd{thalapillil@iiserpune.ac.in}
\emailAdd{hverma@lsu.edu}

\abstract{We present the first comprehensive study of astrometric microlensing by  exotic astrophysical dark objects, focusing on two theoretically motivated models---Q-ball and boson star. We demonstrate that these extended objects generate distinctive signatures that depart markedly from point-mass lenses like primordial black holes. The smoking-gun signature for these exotic objects is the emergence of caustics, which form when the lens radius is below a critical threshold.  Crossing these caustics induces discontinuous jumps in the images-centroid trajectory, a distinctive feature of these extended dark objects.  We show these patterns are sensitive to the internal mass profile, with boson stars generating larger, more prominent caustic structures than Q-balls—enabling the models to be distinguished. Using the Gaia DR3 stellar catalogue, we forecast a high-yield discovery potential, up to $\sim 6000$ detectable astrometric microlensing events for a 10-year mission, peaking for $M \sim 1-10~M_\odot$ and $R \lesssim 10~\text{AU}$. In the absence of anomalous detections, Gaia can set powerful 90\%  confidence level constraints on the fractional abundance of these exotic objects, reaching $f_{\text{\tiny DM}} \le 10^{-3}$ in the peak region which covers masses from $10^{-1}-10^{7}~M_{\odot}$ and radii $R<10^{6}~\text{AU}$. Crucially, these projected astrometric microlensing constraints are significantly stronger than existing photometric microlensing limits in the $1-10~M_\odot$ mass range. This work establishes astrometric microlensing with Gaia as a powerful, complementary, and near-future probe with the potential to discover exotic astrophysical dark objects.}

\begin{document}
\maketitle

\section{Introduction}\label{sec:intro}
Existence of dark matter is a cornerstone of the standard cosmological model, yet its physical nature remains an open mystery (see, for instance\,\cite{Rubin:1970zza,Bertone:2004pz,Feng:2010gw,ParticleDataGroup:2024cfk,Cooley:2022ufh,Clowe:2006eq,Bertone:2016nfn,Planck:2015fie,Bartelmann:2001zm,Blumenthal:1984bp} and references within). Evidence for dark matter is robust and multiscale, including anomalous galactic rotation curves\,\cite{Rubin:1970zza}, collisionless dynamics within galaxy clusters\,\cite{Clowe:2006eq}, and formation of large-scale structure\,\cite{Blumenthal:1984bp}, and the cosmological microwave background (CMB) power spectrum\,\cite{Planck:2015fie}. Among these, gravitational lensing\,\cite{Bartelmann:2001zm} provides one of the most direct probes of the total mass distribution, consistently confirming that the vast majority of matter is non-luminous, and making it an indispensable tool for characterizing this dark component.

Despite extensive observational evidence, the underlying microphysics of dark matter remains elusive\,(see, for instance\,\cite{Bertone:2004pz,Feng:2010gw,ParticleDataGroup:2024cfk,Cooley:2022ufh} and references therein). The persistent lack of direct detection signals for weakly interacting massive particles (WIMPs)\,\cite{Steigman:1984ac,Bertone:2016nfn}, a historically leading candidate, has broadened the theoretical landscape to include a vast hierarchy of candidates. This landscape now encompasses both new fundamental particles—such as axions\,\cite{Preskill:1982cy,Marsh:2015xka,Adams:2022pbo}, hidden sector particles\,\cite{Fabbrichesi:2020wbt,Holdom:1985ag}, and fuzzy dark matter\,\cite{Hu:2000ke,Ferreira:2020fam}--and a rich variety of macroscopic objects\,\cite{Navarro:1995iw,Bertschinger:1985pd,Ricotti:2009bs,Delos:2018ueo,Carr:1974nx,Carr:2024nlv,Ruffini:1969qy,Colpi:1986ye,Visinelli:2021uve,Schunck:2003kk,Braaten:2015eeu,Visinelli:2017ooc,Kolb:1993zz,Coleman:1985ki,Kusenko:1997si}.

These macroscopic candidates range from dense substructures predicted within the standard $\Lambda$CDM model, like subhalos\,\cite{Navarro:1995iw,Bertschinger:1985pd} and ultra-compact minihalos (UCMHs)\,\cite{Ricotti:2009bs,Delos:2018ueo} to a diverse zoology of exotic objects. This latter category includes primordial black holes (PBHs)\,\cite{Carr:1974nx,Carr:2024nlv} , gravastars\,\cite{Mazur:2004fk},  and various non-topological solitons, such as boson stars\,\cite{Ruffini:1969qy,Colpi:1986ye,Visinelli:2021uve,Schunck:2003kk}, axion stars\,\cite{Braaten:2015eeu,Visinelli:2017ooc}, axion miniclusters\,\cite{Kolb:1993zz}, and Q-balls\,\cite{Coleman:1985ki,Kusenko:1997si,Friedberg:1986tq,Multamaki:2002wk,Tamaki:2011zza}. A key unifying characteristic of the latter class of objects is their finite size and complex internal structure, which differs fundamentally from the central singularity of a black hole or the simple cusps of standard halos.  

Among the diverse array of these exotic astrophysical dark objects (EADOs), boson stars\,\cite{Ruffini:1969qy,Colpi:1986ye,Visinelli:2021uve,Schunck:2003kk} and Q-balls\,\cite{Coleman:1985ki,Kusenko:1997si} have garnered significant theoretical interest recently. Their formation is rooted in the dynamics of scalar fields in the early universe. Q-balls are primarily expected to arise from the fragmentation of a post-inflationary scalar condensate, a process notably realized in scenarios like Affleck-Dine baryogenesis\,\cite{Kasuya:1999wu,Kasuya:2000wx,Kasuya:2001hg,Kusenko:2008zm} or through a cosmological phase transition\,\cite{Frieman:1988ut,Jiang:2024zrb}. Boson stars, on the other hand, can form through direct gravitational collapse\,\cite{Madsen:1990gg}  or by gravitational condensation\,\cite{Kolb:1993hw,Seidel:1993zk,Levkov:2018kau} of primordial scalar field fluctuations.

The complex internal structure of boson stars and Q-balls creates a rich phenomenology\,\cite{Kling:2017hjm,Kling:2020xjj,Cardoso:2022vpj,Chang:2024xjp,Saffin:2022tub,Gao:2023gof,Heeck:2020bau,Almumin:2021gax,Kusenko:1999gz,Kusenko:2004yw,CalderonBustillo:2020fyi,Ge:2024itl,Evstafyeva:2024qvp, Croon:2018ybs,Olivares:2018abq,Vincent:2015xta,Rosa:2022tfv,Rosa:2022toh,Li:2025awg,Chiba:2009zu,Hong:2024uxl,Zhou:2015yfa,Kasuya:2014bxa,Troitsky:2015mda,CrispimRomao:2024nbr,Croon:2020ouk,Croon:2020wpr,Ansari:2023cay}, with recent efforts focusing on gravitational signatures as a means of distinguishing them from point-mass candidates like primordial black holes. For boson stars, this includes studying gravitational waves from boson star mergers\,\cite{CalderonBustillo:2020fyi,Ge:2024itl,Evstafyeva:2024qvp,Croon:2018ybs}, as well as efforts to distinguish them from black holes via their accretion dynamics and shadow with the Event Horizon Telescope\,\cite{Olivares:2018abq,Vincent:2015xta,Rosa:2022tfv,Rosa:2022toh,Li:2025awg}.  Similarly, recent work on Q-balls has investigated gravitational wave signals from their formation\,\cite{Kusenko:2008zm,Chiba:2009zu,Hong:2024uxl,Zhou:2015yfa}, while also refining their connection to Affleck-Dine baryogenesis and their role as a dark matter candidate\,\cite{Kasuya:2014bxa,Troitsky:2015mda}. 

Gravitational microlensing serves as a common and powerful probe, with ongoing efforts directed towards distinguishing extended EADOs from point-mass lenses\,\cite{CrispimRomao:2024nbr,Croon:2020ouk,Croon:2020wpr,Ansari:2023cay}. Prior studies have largely focused on photometric microlensing (PML)\,\cite{CrispimRomao:2024nbr,Croon:2020ouk,Croon:2020wpr,Ansari:2023cay,DeRocco:2023hij,Fairbairn:2017dmf,Fairbairn:2017sil,Fujikura:2021omw,Sugiyama:2021xqg,Mroz:2025xbl,Pappas:2025zag}, where the extended nature of these objects would manifest as a characteristic deviation from the standard point-mass light curve. However, the unique signatures imprinted by these extended lenses have not yet been fully exploited, particularly in the realm of high-precision astrometry\,\cite{Boden:1998me,Erickcek:2010fc,Li_2012,Jankovi_2025,OGLE:2022gdj,Mondino:2023pnc}, which provides a novel and complementary avenue for discovery.

		\begin{figure}[h!]
			\begin{center}
				\includegraphics[scale=0.4]{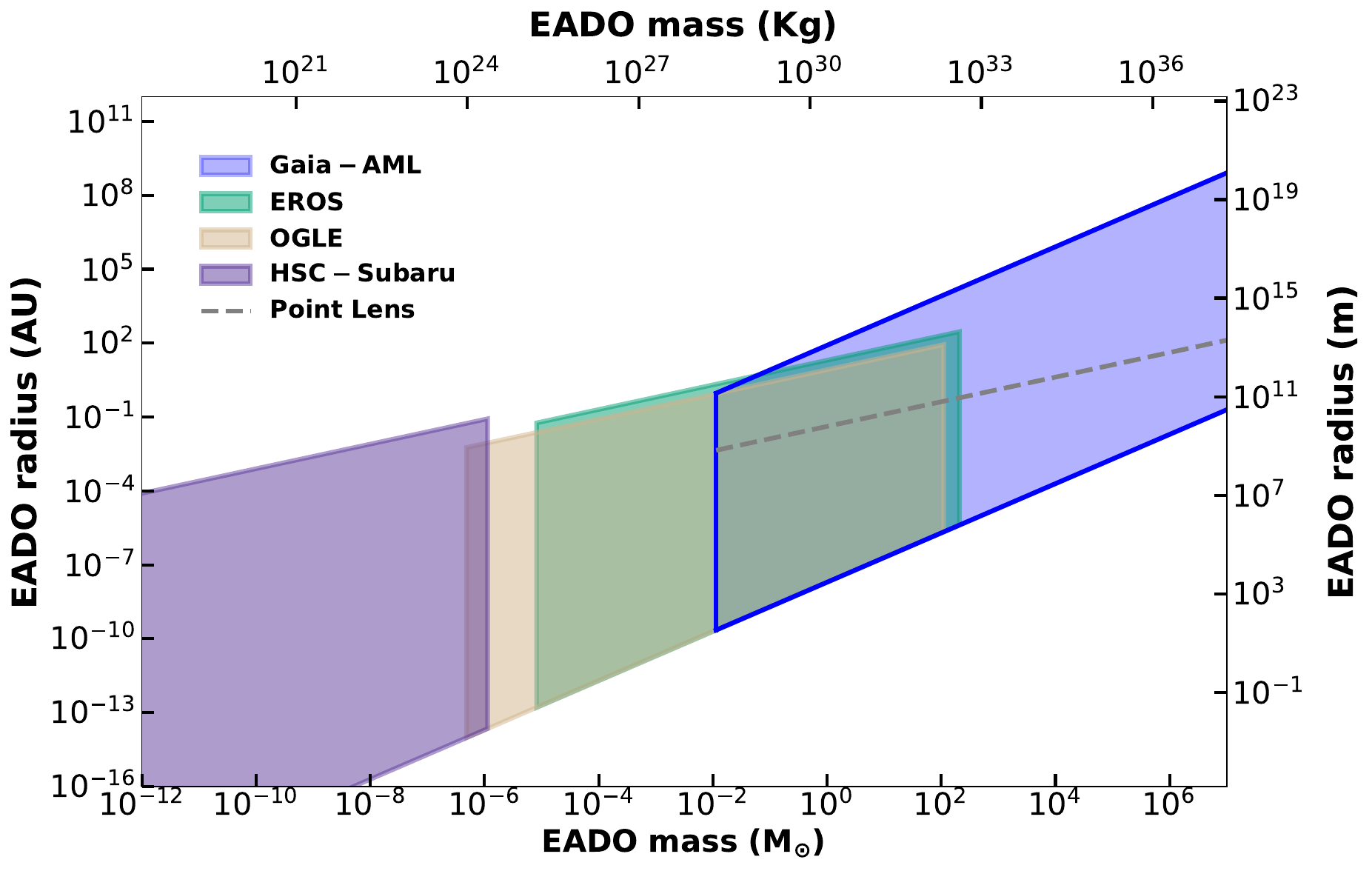}
			\end{center}
			\caption{The plot displays a heuristic estimate of the mass and size of EADOs that can be probed by Gaia\,\cite{Gaia:2016zol}. The blue-shaded region indicates Gaia’s sensitivity to such lenses for AML. For comparison, we also show the PML sensitivity of other surveys: EROS\,\cite{Croon:2018ybs}, OGLE\,\cite{Croon:2018ybs}, and HSC-Subaru\,\cite{Croon:2020ouk}, highlighting the complementarity of these observational probes in constraining EDS populations. For Gaia, the maximum lens radius relevant for AML is approximately $\sim R_{\text{\tiny E}}^2/(\sigma_{\text{\tiny a,min}}D_{\text{\tiny L}})$. For the estimation of the maximum probe-able radius, $R_{\text{\tiny L,max}}(M_{\text{\tiny L}})$, we assume a source distance of $D_{\text{\tiny S}}=8.5$ kpc, a lens distance of $D_{\text{\tiny L}}=D_{\text{\tiny S}}/2$, and a minimum astrometric uncertainty of $\sigma_{\text{\tiny a,min}}=0.05~\text{mas}$. Furthermore, the lower bound on the lens radius for each survey is given by the Schwarzschild radius, $R_{\text{\tiny Sch}}$ corresponding to the EADO, while the upper bound for surveys probing PML signatures is of the order of the maximum Einstein radius, $R_{\text{\tiny E}}$, accessible in that survey. Below the grey dashed line (drawn at $\sim 0.01 R_{\text{\tiny E}}$ ), the AML signal of EADOs effectively behave as point-like objects. The minimum and maximum EDS masses accessible to a given survey are determined by comparing the survey cadence and total observation time with the characteristic microlensing event timescale. Refer to Sec.\,\ref{sec:AML_class} for further details.}
			\label{fig:Gaia_sens}
		\end{figure}
Traditional photometric microlensing, which relies on detecting the temporary brightening of a background star, has been used to discover numerous invisible objects like free-floating planets\,\cite{Sumi_2011,Oz_2017,OGLE:2022gdj,Lam_2022}.    These studies\,\cite{Croon:2020ouk,Croon:2020wpr,Ansari:2023cay} have examined PML signatures of EADOs and derived corresponding abundance constraints (see Fig.\,\ref{fig:Gaia_sens}). For boson stars, PML analyses\,\cite{Croon:2020ouk,Croon:2020wpr} exclude masses in the range $10^{-11} - 10~\solarmass$ with radii less than $\sim 100~R_{\odot}$, limiting their contribution to the dark matter fraction  $f_{\text{\tiny DM}}\lesssim 0.1\%$. In addition, for Q-balls\,\cite{Ansari:2023cay}, constraints have been placed around $f_{\text{\tiny DM}}\lesssim 0.01\%$  within the mass range $10^{-7} - 10~\solarmass$ for radii around $1~R_{\odot}$. It is noteworthy that the PML signal is strongly enhanced only in low-impact parameter events. Consequently, distinguishing an EADO, like a boson star, from a point-mass singularity requires a rare, fine-tuned event with a very low impact parameter\,\cite{Croon:2020ouk,Croon:2020wpr,CrispimRomao:2024nbr,Ansari:2023cay}.

Astrometric microlensing (AML)\,\cite{Hog:1994xt,Walker_1995,Miyamoto_1995,Gould:1996nb,Dominik_2000,Lee:2010pj} provides a more powerful probe, as it relies on detecting the apparent positional shift of a background star. This is different from a PML event, which relies on a transient apparent brightening of the star due to magnification of the source star.   In the AML events due to point-lens, the signal's sensitivity falls off as $1/\theta_{\text{\tiny S}}$\,\cite{Dominik_2000}, whereas for PML the corresponding signal scales as $1/\theta_{\text{\tiny S}}^4$\,\cite{Dominik_2000}, where $\theta_{\text{\tiny S}}$ is the angle between the true position of the star and the lens.   As a result, improvements in astrometric precision translate into sensitivity to significantly higher impact parameter events, whereas comparable gains in photometric precision do not extend the reach nearly as effectively.

The capability to perform precise astrometric measurements is not mere speculation; the Hubble Space Telescope (HST)\,\cite{Kains_2017} has already been used to monitor microlensing events with milliarcsecond precision and\,\cite{OGLE:2022gdj} has even measured the mass of an isolated stellar-mass black hole via astrometric deflection. Looking ahead, the Nancy Grace Roman Space Telescope\,\cite{Spergel:2015sza} is a key science driver in this arena: its sub-milliarcsecond astrometric capabilities are projected to routinely detect astrometric microlensing, due to primordial black holes\,\cite{Fardeen:2023euf}. At present, it is the European Space Agency's (ESA) Gaia mission\,\cite{Gaia:2016zol,Lindegren_2012,Gaia:2023fqm} that has already revolutionized astrometry, delivering extremely precise positional and proper-motion data for over a billion stars and enabling the identification of candidate astrometric microlensing events across the sky. 

The Gaia mission\,\cite{Gaia:2016zol,Lindegren_2012,Gaia:2023fqm} is conducting an all-sky survey, repeatedly measuring the positions, parallaxes, and proper motions of nearly $1.5$ billion stars with unprecedented accuracy. Reaching sub-milliarcsecond precision, Gaia is constructing the most detailed three-dimensional map of our galaxy to date.  The mission's vast and exquisitely precise dataset is particularly well-suited for the detection of AML events, as it enables a systematic search for small deviations caused by extended dark objects like boson stars and Q-balls. With its ongoing data releases\,\cite{Gaia:2023fqm}, Gaia serves as the definitive instrument for probing the nature of dark compact objects through AML in this new era of precision astronomy. 

In this work, building on Gaia’s exquisite capabilities, we undertake a systematic investigation of AML signatures due to  EADOs, focusing on two astrophysical models---thin-wall Q-balls\,\cite{Coleman:1985ki}, modeled as uniform density spheres (UDSs), and boson stars\,\cite{Ruffini:1969qy} (BSs) without any self-interactions of the constituent bosons. In Fig.\,\ref{fig:Gaia_sens}, we provide an approximate estimate of the mass and size ranges accessible to Gaia and have compared them with the sensitivities of other photometric surveys. Previous work has examined AML due to dark matter subhaloes such as Navarro–Frenk–White (NFW) profile haloes\,\cite{Erickcek:2010fc}, isothermal subhaloes\,\cite{Erickcek:2010fc}, and ultra-compact minihalos\,\cite{Li_2012}, but detailed studies of  EADOs has so far remained unexplored. We aim to characterize the unique AML signatures of   EADOs and to evaluate the limits that Gaia observations can place on their fractional contribution to dark matter. 

A central focus of our analysis is the detailed characterization of AML signals from these exotic objects and their distinction from conventional point-lens signatures (see Sec.\,\ref{sec:aml_eds}). We demonstrate that the most prominent feature of these lenses is the occurrence of discontinuous jumps in the apparent trajectory of the source. These jumps are a direct consequence of the source star crossing the caustics of the lens (see Sec.\,\ref{sec:AML_UDS_BS}). We also estimate how these AML signals depend upon the parameters of these  EADOs, i.e., their density profile, mass, and radius. 

Building upon this, in Sec.\,\ref{sec:aml_lens_prob_no_lens_prob} we estimate the probability for a single star in the Gaia catalogue to undergo a detectable AML event caused by these  EADOs. By convolving this single-star probability with the full Gaia DR3 stellar catalogue\,\cite{Gaia:2023fqm} and a standard NFW dark matter halo model\,\cite{Navarro:1995iw}, we compute the expected total number of events Gaia should observe over its mission lifetime in Sec.\,\ref{sec:aml_lens_prob_no_no}.

Finally, in Sec.\,\ref{sec:gaia_detectibility}, we leverage our number of event predictions to forecast the exclusion limits Gaia can place on the population of thin-wall Q-balls and boson stars BS with different masses and radii.  Our comprehensive analysis, detailed in the subsequent sections, yields several key predictions:
\begin{itemize}
    \item Extended EADOs can produce distinct astrometric signatures, most notably discontinuous jumps in a star's apparent trajectory due to caustic crossings. These features arise when their radii fall below a critical threshold, specifically $R_{\text{\tiny Sch.}}<R_{\text{\tiny u}} < \sqrt{3/2} ~R_{\text{\tiny E}}$ for UDS, and $R_{\text{\tiny Sch.}}<R_{\text{\tiny BS}} \lesssim2.73 ~R_{\text{\tiny E}}$ for BS (see Sec.\,\ref{sec:aml_eds} for details)\footnote{Here, $R_{\text{\tiny Sch.}}$ and $R_{\text{\tiny E}}$ denote the Schwarzschild radius and the Einstein radius of the corresponding  EADOs, respectively.}.
    \item A key physical distinction we identify is that BS generates caustics at a larger source position than a UDS of comparable mass and radius. This directly increases the observational impact parameter for caustic crossing events for BS, providing a robust method to distinguish between these two astrophysical objects. (see Sec.\,\ref{sec:AML_UDS_BS} for details).
    \item Gaia is expected to detect AML signals caused by  EADOs with masses $10^{-1} - 10^{7}\,\solarmass$ and radii $R_{\text{\tiny L}}\lesssim 10^6~\text{AU}$. For both UDS and BS, the number of AML events peaks for radii $R_{\text{\tiny L}} \lesssim 10~\text{AU}$ and mass $M_{\text{\tiny L}} \simeq 1-10,\solarmass$, yielding up to $\sim 2500$ events for an observation time of $5$ years and $\sim 6000$ events for an observation time of  $10$ years (see Sec.\ref{sec:aml_lens_prob_no} for details). 
    \item Assuming no detection of anomalous events\footnote{Anomalous events are those not produced by standard astrophysical objects such as stars, white dwarfs, neutron stars,  astrophysical black holes, or statistical noise.}, Gaia can place stringent constraints on the abundance of UDS and BS, limiting their contribution to the dark matter fraction up to $f_{\text{\tiny L}}\lesssim 10^{-3}$ for masses $10^{-1} - 10^{7}\,\solarmass$ and radii $R_{\text{\tiny L}} \lesssim  10^6~\text{AU}$ (see Sec.\ref{sec:gaia_detectibility} for details)
\end{itemize}

The paper is organized as follows. In Sec.\,\ref{sec:formalism} we review EADOs of interest—-thin-wall Q-balls and boson stars--along with the formalism of astrometric microlensing. In Sec.\,\ref{sec:aml_eds}, we then analyze the characteristic astrometric features induced by UDS and BS lenses. Sec.\,\ref{sec:AML_sig_Gaia_det} discusses the corresponding AML signals and their detectability prospects with Gaia. In Sec.\,\ref{sec:aml_lens_prob_no}, we compute the lensing probabilities and estimate the total number of events expected to be observable by Gaia. Constraints on the abundance of  EADOs are presented in Sec.\,\ref{sec:gaia_detectibility}. We summarize our results in Sec.\,\ref{sec:conclusion}, and present a discussion of key assumptions and possible improvements to our rate estimates in Sec.\,\ref{sec:discussion}.

\section{Theoretical Framework}\label{sec:formalism}
To investigate the astrometric microlensing signatures of exotic astrophysical dark objects, it is necessary to first establish the foundational framework upon which our analysis is built. This framework rests on two essential components---the model of the dark matter structures and a mathematical description of the gravitational lensing effects that they produce.

This section discusses some of these foundational components. In Sec.\,\ref{sec:eds_rev}, we review the theoretical models for  EADOs under study. Sec.\,\ref{sec:aml_rev} outlines the formalism of astrometric microlensing, focusing on the lens equation and defining the key observable—the astrometric centroid shift ($\Delta \theta_{\text{\tiny cent}} $).  Together, these elements provide the framework necessary for forecasting the AML signatures analyzed in the subsequent sections.

\subsection{Exotic astrophysical dark objects  }\label{sec:eds_rev}
This section reviews the EADOs of interest  considered in our study, namely thin-wall Q-balls and boson stars.
The fundamental description of these objects arises from a complex scalar field $\Phi$ minimally coupled to gravity, formulated through the Einstein–Klein–Gordon (EKG) action (see\,\cite{Visinelli:2021uve} for details), given by\footnote{For this subsection, we use natural units i.e. $\hbar=c=1$ for convenience.}
\begin{equation}\label{eq:eds_action}
    \mathcal{S}=\int d^4x \sqrt{-g}\left(\frac{\mathcal{R}}{16 \pi G_{\text{\tiny N}}}-g^{\mu\nu}\partial_\mu \Phi^* \partial_\nu \Phi - V(\abs{\Phi}^2)\right)\;.
\end{equation}
Here, $\mathcal{R}$ is the Ricci scalar, $g$ is the determinant of the spacetime metric $g_{\mu\nu}$, $G_{\text{\tiny N}}$ is Newton's gravitational constant, and $V(\abs{\Phi}^2)$ represents the potential of the scalar field. 

Boson stars are stabilized against gravitational collapse by a pressure sourced by the bosonic field itself. However, the origin of this pressure and the nature of these objects is fundamentally dictated by the scalar potential, $V(\abs{\Phi}^2)$.  For instance,  in the case of a mini–boson star, where the potential contains only a mass term, $V(\abs{\Phi}^2)=m^2\abs{\Phi}^2$, stability is a purely quantum mechanical effect arising from the Heisenberg uncertainty principle\,\cite{Ruffini:1969qy}. This inherent quantum pressure resists gravitational compression, serving a role analogous to the electron degeneracy pressure that stabilizes white dwarfs. This balance between quantum pressure and gravity imposes a limit to the maximum stable mass, known as the Kaup limit\,\cite{Kaup:1968zz,Lee:1988av}, given by
\begin{equation}\label{eq:kaup_limit}
    M_{\text{max}} \approx 0.633 \frac{M_{\text{PL}}^2}{m}\;,
\end{equation}
where $M_{\text{PL}} = 1/\sqrt{G_{\text{\tiny N}}}$ is the Planck mass. For ultra-light bosons mass of $m \sim 10^{-10}\,\text{eV}$, this corresponds to stellar-mass boson star ($M_{\text{\tiny BS}} \sim M_\odot$). By contrast, the inclusion of self-interaction terms, like a quartic potential $\lambda\abs{\Phi}^4/4$,\,\cite{Colpi:1986ye} fundamentally alters the stability conditions by introducing additional pressure from the self-interaction, thereby modifying the mass–radius relation.

Q-balls possess additional features and existence criteria beyond those of generic boson stars. A potential where the function $V(\abs{\Phi}^2)/\abs{\Phi}^2$ has a global minimum at a non-zero field value is the essential condition for Q-balls \,\cite{Coleman:1985ki,Kusenko:1997si,Friedberg:1986tq,Multamaki:2002wk,Tamaki:2011zza}. As a consequence, the stability, density profile, and macroscopic scaling relations of these configurations emerge directly from the specific form of the potential\,\cite{Coleman:1985ki,Kusenko:1997si,Friedberg:1986tq,Multamaki:2002wk,Heeck:2020bau}. This establishes a profound link---by probing an object’s mass distribution—for example, through gravitational lensing observations—one can, in principle, infer properties of the underlying scalar potential.

In our study, we focus on two primary models of  EADOs---mini-boson stars\,\cite{Ruffini:1969qy}, which correspond to self-gravitating scalar configurations with no self-interaction terms, and thin-wall Q-balls\,\cite{Coleman:1985ki}, which exist as a flat spacetime solution to the EKG system with a characteristic scalar potential\footnote{The thin-wall Q-balls considered in this study arise in the weak-field limit of gravitating Q-balls when the attractive self-interaction dominates the gravitational interaction\,\cite{Ansari:2023cay}.}.

		\begin{figure}[h!]
			\begin{center}
                \includegraphics[scale=0.4]{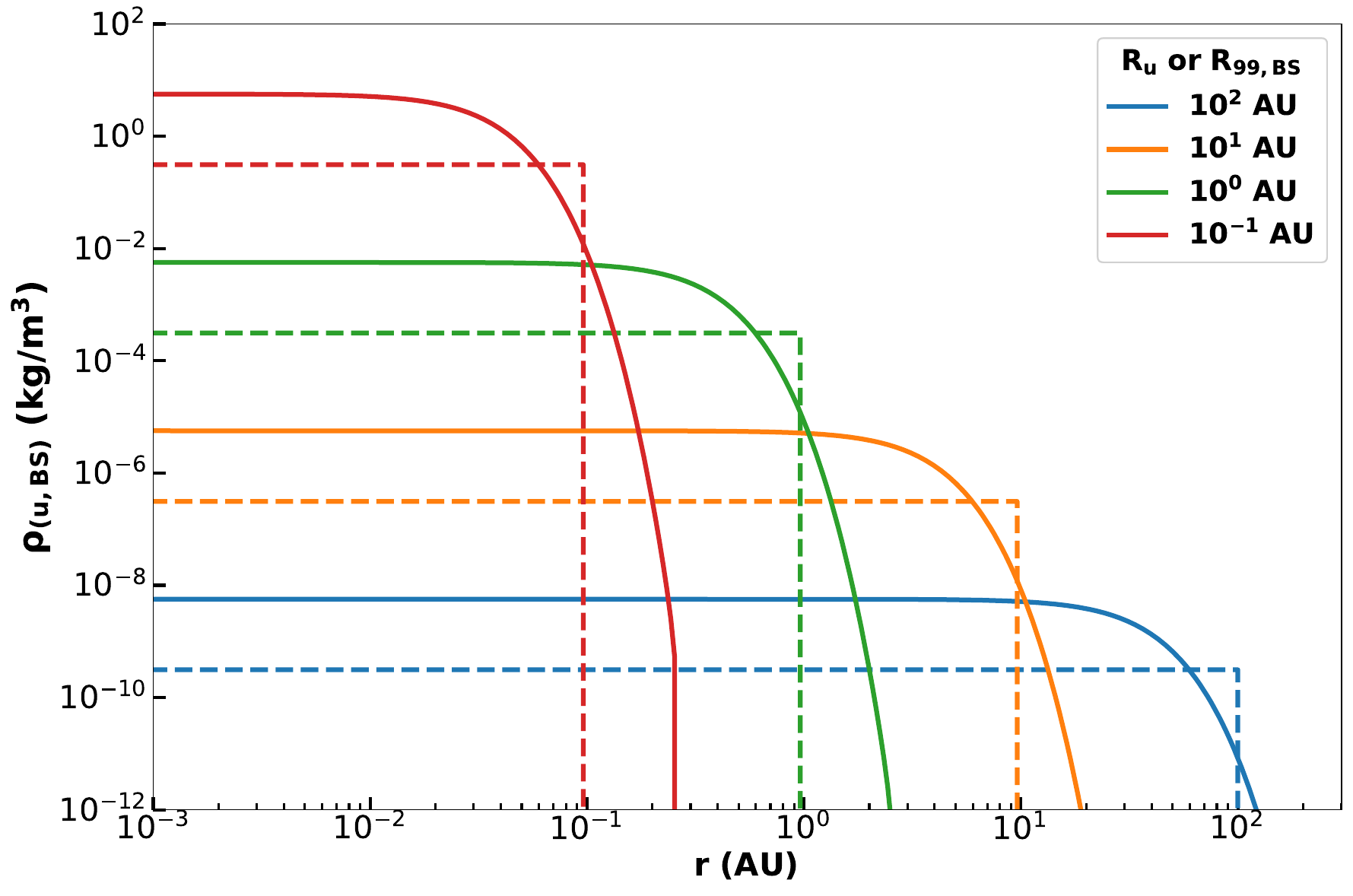}
			\end{center}
			\caption{We display the density profiles of mini-boson stars (BS, solid lines) and thin-wall Q-balls (u, dashed lines) for EADO mass $M_{\text{\tiny L}} \sim 2.2\,\solarmass$ (for which we get the most stringent constraint) with different radii. The mini-boson star profiles are obtained by solving the Schrödinger–Poisson system of equations, Eqs.\,\eqref{eq:app_bs_einstein_tt_final}, \eqref{eq:app_bs_kg_weak_eq_final_sch}, and \eqref{eq:app_bsmini_bos_star_density}, with boson mass $m \simeq (0.28,1.59,2.8,8.9)\times 10^{-14}\,\text{eV}$ corresponding to radii $R_{\text{\tiny 99,BS}} \simeq (10^2,10,1,0.1)\,\text{AU}$ respectively. As expected, boson stars have a higher mass distribution closer to the center compared to thin-wall Q-balls (modeled as uniform-density spheres), which instead display an almost flat core profile followed by a sharp drop at the boundary. }
			\label{fig:UDS_BS_dens_prof}
		\end{figure}
Mini-boson star\,\cite{Ruffini:1969qy} corresponds to self-gravitating scalar field configurations described by a purely quadratic potential,  $V(\abs{\Phi}^2)=m^2\abs{\Phi}^2$. The ground-state configuration of the mini-boson star is spherically symmetric and is assumed to take the stationary form, i.e.	
\begin{equation}\label{eq:bs_kg_weak_ground}
			\Phi_{\text{\tiny BS}}(t,r) = \left(\frac{N}{2\omega}\right)^{1/2}\phi_{\text{\tiny BS}}(r)e^{i \omega t}\;,
		\end{equation}
where $\omega \equiv  m+e$ is the energy of the ground state, with $e~(\ll m)$  the binding energy, $N$ the number of bosonic particles in the configuration, and $\phi_{\text{\tiny BS}}(r)$ is the ground state radial wavefunction. The above stationary solution possesses a time-dependent phase, which allows it to evade Derrick’s theorem \cite{Derrick:1964ww}. The ground-state configuration for mini-boson star can be derived from the action in Eq.\,\eqref{eq:eds_action}. In particular, the variation with respect to the metric $g_{\mu\nu}$ yields the Einstein field equations (see Eq.\,\eqref{eq:app_bs_einstein_eq}), while variation with respect to the complex conjugate of the scalar field $\Phi^*$ yields the Klein–Gordon equation (see Eq.\,\eqref{eq:app_bs_kg_eq}).
 In the non-relativistic and Newtonian limit, the Einstein equation Eq.\,\eqref{eq:app_bs_einstein_eq} and the Klein-Gordon equation \eqref{eq:app_bs_kg_eq} can be simplified to obtain
\begin{equation}\label{eq:bs_einstein_tt_final}
			\nabla^2 \Psi_{\text{\tiny N}} = 4\pi G \rho_{\text{\tiny BS}}\;.	
\end{equation}
\begin{equation}\label{eq:bs_kg_weak_eq_final_sch}
			 -\frac{1}{2m}\nabla^2 \phi_{\text{\tiny BS}} + m\Psi_{\text{\tiny N}}\phi_{\text{\tiny BS}} =  e\phi_{\text{\tiny BS}}\;.
\end{equation}
where $M_{\text{\tiny BS}}\approx N m$ is the total mass of the mini-boson star, and $\rho_{\text{\tiny BS}}({r})=M_{\text{\tiny BS}}\phi_{\text{\tiny BS}}^2(r)$ is the density profile of the boson star, and $\Psi_{\text{\tiny N}}$ is the Newtonian potential.
Eqs.\,\eqref{eq:bs_einstein_tt_final}, and \eqref{eq:bs_kg_weak_eq_final_sch} together constitute the Schrödinger–Poisson system for a mini-boson star, which governs its ground-state configuration. (For details, see Appendix\,\ref{app:BS})

In Fig.\,\ref{fig:UDS_BS_dens_prof}, we show the density profiles for boson stars obtained by solving Schrodinger-Poisson equations (i.e. Eqs.\,\eqref{eq:bs_einstein_tt_final}, and \eqref{eq:bs_kg_weak_eq_final_sch}) for total masses $M_{\text{\tiny BS}} \sim 2.2\,\solarmass$, for radii $R_{\text{\tiny 99,BS}} \sim 0.1-100$ AU, where, $R _{\text{\tiny 90,BS}}$ is the radius enclosing the $99\%$ of total mass  $M_{\text{\tiny BS}}$ of BS. The plot indicates that boson stars exhibit centrally concentrated densities that fall off smoothly with radius, in contrast to the nearly flat core structure of thin-wall Q-balls.

Additionally, the mini-boson star has a well-defined relationship between its mass $M_{\text{\tiny BS}}$, radius $R_{\text{\tiny BS}}$, and the mass of the constituent scalar particle $m$. This relation\,\cite{Ruffini:1969qy} is approximately given by,
\begin{eqnarray}
    \label{eq:mini_bos_star_mass_function_90_99_SI}
    R _{\text{\tiny 99,BS}}&\approx&  17.5 \left(\frac{10^{-14} eV/c^2}{m}\right)^2\left(\frac{1 \solarmass}{M_{\text{\tiny BS}}}\right) \text{AU}\;.
\end{eqnarray}
For our lensing calculations, we will treat the mass $M_{\text{\tiny BS}}$, radius $R _{\text{\tiny BS}}\equiv R_{\text{\tiny 99,BS}}$ as the fundamental parameters, keeping in mind this underlying physical relation. 

Turning to Q-balls, these are non-topological solitonic solutions that arise in flat spacetime as localized, coherent configurations of a complex scalar field carrying a conserved global $U(1)$ charge. For the flat-spacetime configuration ($g_{\mu\nu}=\eta_{\mu\nu}$), the action in Eq.\,\eqref{eq:eds_action} reduces to
\begin{equation}\label{eq:qball_action}
 \mathcal{S}=\int d^4x \left(-\partial_\mu \Phi^* \partial^\mu \Phi - V(\abs{\Phi}^2)\right)\;.
\end{equation}
Unlike the dynamical equilibrium that stabilizes a boson star, a Q-ball relies on specific features of the scalar potential, $V(\abs{\Phi}^2)$. A stable Q-ball can form if the function $V(\abs{\Phi}^2)/\abs{\Phi}^2$ has its global minimum at a non-zero field value, $\abs{\Phi}_\text{\tiny c}> 0$\,\cite{Coleman:1985ki} satisfying
\begin{equation}\label{eq:qballexist}
		0\leqslant\frac{U(\Phi_\text{\tiny 0})}{\Phi^2_\text{\tiny 0}}\equiv \omega^2_\text{\tiny Q} <m^2\;.
		\end{equation}
where $m$ is the mass of the bosonic field. 
 This condition ensures that the Q-ball configuration represents the state of lowest energy for a fixed, non-zero charge. It is therefore energetically forbidden for a Q-ball to dissipate by breaking apart into a gas of free scalar quanta, as this would require an input of energy\,\cite{Coleman:1985ki}. Similar to the boson star case, the minimum energy configuration is spherically symmetric and can be written as 
		\begin{equation}\label{eq:phitdepen}
		\Phi(r,t)=\phi_{\text{\tiny Q}}(r) e^{i\omega t}\;.
		\end{equation}
The equation of motion is obtained by varying the action (see Eq.\,\ref{eq:qball_action}) with respect to $\Phi_{\text{\tiny Q}}^\dagger$. Using the spherically symmetric solution in Eq.\,\eqref{eq:phitdepen},   we get the equation of motion of the field as 
		\begin{equation}\label{eq:eom}
	\phi''_{\text{\tiny Q}}+\frac{2}{r}\phi_{\text{\tiny Q}}'=-\frac{1}{2}\frac{\partial}{\partial\phi_{\text{\tiny Q}}}\left[\omega^2\phi_{\text{\tiny Q}}^2-V(\phi_{\text{\tiny Q}})\right]\;.
		\end{equation} 
An analysis of the equation of motion (see Eq.\,\ref{eq:eom}) reveals that the effective potential $\omega^2\phi_{\text{\tiny Q}}^2$ $-V(\phi_{\text{\tiny Q}})$ permits suitable Q-ball solutions only within the frequency range, $\omega>\omega_\text{\tiny Q}$ and $\omega<m$\,\cite{Coleman:1985ki}. This, in conjunction with the condition from Eq.\,(\ref{eq:qballexist}), then implies $\omega_\text{\tiny Q}<\omega<m$ for viable Q-ball solutions.  

 The thin-wall Q-ball solution\,\cite{Coleman:1985ki} proposed by Coleman, corresponds to the lower-frequency limit of the viable parameter space of $\omega$. Specifically, it arises in the limit where $\omega \to \omega_{\text{\tiny Q}}$ and is characterized by a nearly uniform density core that drops to zero around a well-defined radius over a very thin crust.  This makes the uniform density sphere an excellent and analytically tractable phenomenological model for this class of configurations. 

The constant-density profile provides a stark contrast to the centrally peaked, extended profile of a boson star, creating a clear distinction for observational tests. For our analysis, the uniform density sphere is defined simply by its total mass, $M_{\text{\tiny  u}}$, and its physical radius, $R_{\text{\tiny  u}}$. These two parameters are considered independent, and the density profile is given by
\begin{equation}\label{eq:aml_eds_uds_dens_prof}
    \rho_{\text{\tiny  u}}(r)=\begin{cases}
        \frac{3M_{\text{\tiny  u}}}{4\pi R_{\text{\tiny  u}}^3}\,,& r\le R_{\text{\tiny  u}}\,, \\
        0\,,& r> R_{\text{\tiny  u}}\,.
    \end{cases}
\end{equation}

This phenomenological approach allows us to explore a wide parameter space of mass and compactness of boson stars and Q-balls without being tied to a specific underlying field theory.

While our analysis focuses on mini-boson stars and thin-wall Q-balls, the theoretical landscape of  EADOs is considerably broader, encompassing axion stars\,\cite{Braaten:2015eeu,Visinelli:2017ooc}, axion miniclusters\,\cite{Kolb:1993zz}, and gravastars\,\cite{Mazur:2004fk}. A key implication of this theoretical landscape is that different EADOs have distinct internal density profiles. For instance, boson stars exhibit a cored density distribution, NFW subhaloes follow a power-law cusp, and thin-wall Q-balls are well described by a uniform-density configuration. This structural diversity is of direct observational relevance, as microlensing is uniquely sensitive to the lens density profile. The characteristic distortions imprinted on a source star’s apparent trajectory provide a powerful probe, allowing the search for  EADOs to be recast as a direct test of its underlying microphysical properties. In the current study, we demonstrate this density profile dependence by analyzing and comparing the microlensing signatures for boson stars and Q-balls.

\subsection{Astrometric microlensing}\label{sec:aml_rev}
In this section, to fix notations and relevant concepts, we will first review basic principles of gravitational microlensing by extended lenses with a specific focus on the idea of AML (see Appendix\,\ref{app:mlrev} for details). We will then derive in the subsequent sections the microlensing signals specific to boson stars and Q-balls, which are of interest in the present study.

When a massive object, referred to as the lens, passes sufficiently close to the line of sight of a background source star, the gravitational field of the lens bends the light from the source. This phenomenon—gravitational lensing\,\cite{Schneider_1992}—typically results in a transient brightening of the source star, known as photometric microlensing. Additionally, it causes a shift in the apparent angular position of the source star in the sky, which is referred to as astrometric microlensing\,\cite{Hog:1994xt,Walker_1995,Miyamoto_1995,Gould:1996nb,Dominik_2000,Lee:2010pj}.

\begin{figure}[h!]
			\begin{center}
				\includegraphics[scale=0.625]{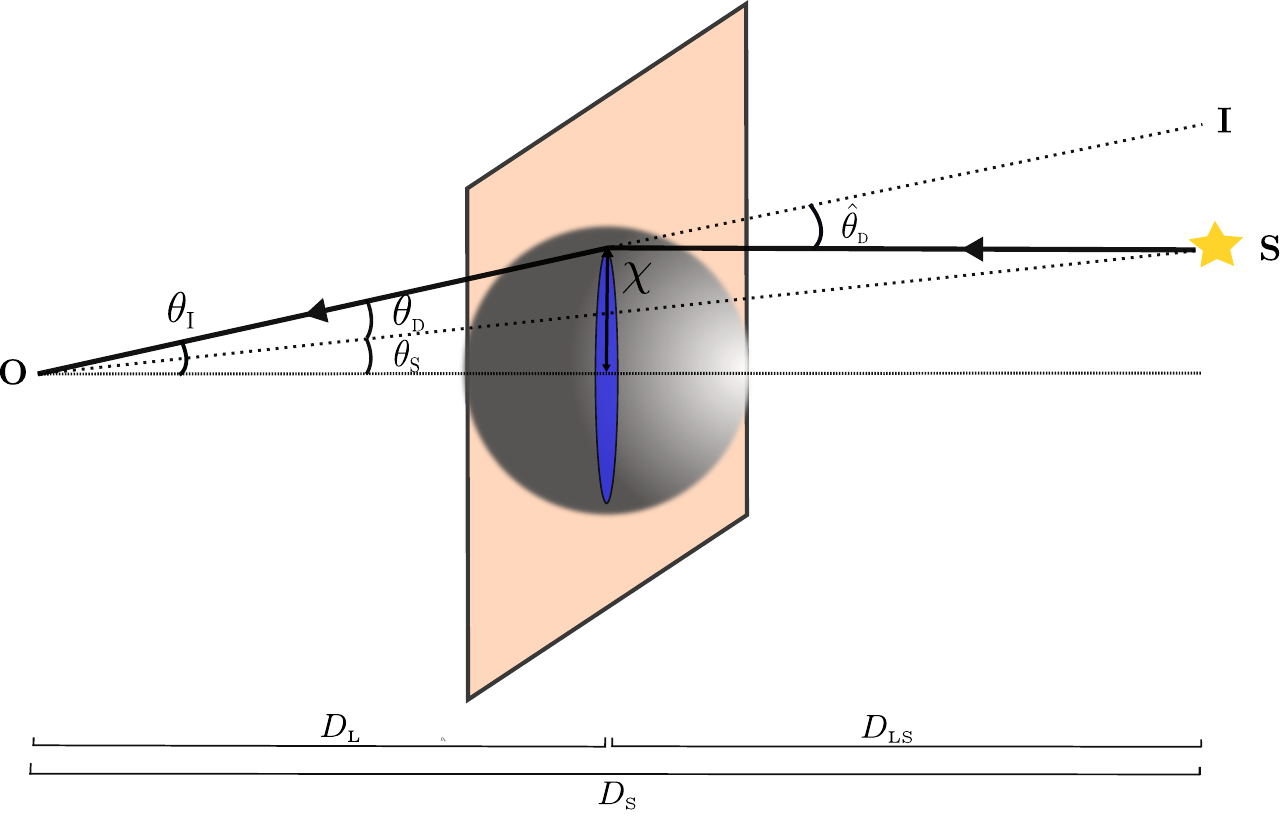}
			\end{center}
			\caption{ A schematic representation illustrates an extended lens, an observer (O), a point source (S), and image position (I) within a lensing scenario.   The center of a spherically symmetric extended lens with total mass $M_{\text{\tiny L}}$ is taken as the origin and is located at an angular diameter distance $D_{\text{\tiny L}}$ from an observer (O).  The source (S) is located at an angular diameter distance $D_{\text{\tiny S}}> D_{\text{\tiny L}}$ from the observer, and the angular diameter distance between the lens and the source is denoted as $D_{\text{\tiny LS}}$. Light rays illustrated by the solid line travel along the $z$ direction, with an impact parameter $\chi$ with the lens is deflected by an angle $\hat{\theta}_{\text{\tiny D}}$. The angular positions of the source and the image as seen by the observer are denoted by $\theta_{\text{\tiny S}},~\theta_{\text{\tiny I}}$, respectively and $\theta_{\text{\tiny D}}\equiv \theta_{\text{\tiny I}}-\theta_{\text{\tiny S}} $ . The lens plane is shown at the centre, passing through the lens. }
			\label{fig:gr_lens_fig}
		\end{figure}

The lensing configuration is shown in Fig.\,\ref{fig:gr_lens_fig}, where the origin is defined at the center of the spherically symmetric extended lens. The point source\footnote{ A source is considered as the point source when its angular size as seen by an observer is much less than the Einstein angle (see Eq.\,\eqref{eq:aml_rev_eins_ang}) of the lens.} (S) and the lens are located at the distances $D_{\text{\tiny S}}$ and $D_{\text{\tiny L}}$ from the observer ($O$), respectively. A light ray originates from the source S at the impact parameter $\chi$ relative to the center of the lens, is deflected by an angle $\hat{\theta}_{\text{\tiny D}}$ by the lens, and reaches the observer at $O$. The angular position of the source and image is represented by $\theta_{\text{\tiny S}}$ and $\theta_{\text{\tiny I}}$, respectively, and are defined as the angles subtended at the observer $O$ by the source and the image, respectively, with respect to the center of the lens (see Fig.\,\ref{fig:gr_lens_fig}). 

In most astrophysical contexts, the impact parameter is small compared to the distance between the source and the observer (i.e. $\chi\ll (D_{\text{\tiny S}},D_{\text{\tiny L}})$), and therefore we can write $D_{\text{\tiny LS}}\simeq D_{\text{\tiny S}}-D_{\text{\tiny L}}$\footnote{In a cosmological lensing setup, the distances $D_{\text{\tiny L}}$, $D_{\text{\tiny S}}$, and $D_{\text{\tiny LS}}$ are generally not simply additive (i.e. $D_{\text{\tiny S}}\ne D_{\text{\tiny LS}}+D_{\text{\tiny L}}$) due to spacetime curvature and cosmic expansion. However, for our context of lensing within a galaxy along with small impact parameters, space can be treated as approximately Euclidean. Therefore, these distances correspond to physical line-of-sight separations, so the angular diameter distances add linearly.
}. Additionally, the physical radius of the lens is much smaller than these distances (i.e. $R_{\text{\tiny L}}\ll (D_{\text{\tiny S}},D_{\text{\tiny L}})$). This enables us to treat the lens as a planar mass distribution. This is achieved by projecting the mass of the lens onto a plane that is orthogonal to the line of sight, commonly referred to as the lens plane (shown in Fig.\,\ref{fig:gr_lens_fig}). This allows us to relate the angular position of the source and image via a simple equation commonly referred to as the lens equation. For a spherically symmetric dark lens, it is given by (see Eq.\,\ref{eq:app_mlrev_lens_eq_eins_ang})
 \begin{equation}\label{eq:aml_rev_lens_eq_eins_ang}
    \theta_{\text{\tiny S}}=\theta_{\text{\tiny I}}-\frac{\tilde{M}_{\text{\tiny L}}(\theta_{\text{\tiny I}}) }{M_{\text{\tiny L}} }\frac{\theta_{\text{\tiny E}}^2}{\theta_{\text{\tiny I}}}\;.
 \end{equation}
Here $M_{\text{\tiny L}}$,  $\theta_{\text{\tiny E}}$ are the total lens mass and the Einstein angle, respectively. And  $\tilde{M}_{\text{\tiny L}}(\theta_{\text{\tiny I}})$ (see Eq.\,\eqref{eq:app_mlrev_mass_encl}) is the 2-D projected mass on the lens plane enclosed within impact parameter $\chi =\theta_{\text{\tiny I}}D_{\text{\tiny L}}$ and we refer to it as the mass function.

Notice that in Eq.\,\eqref{eq:aml_rev_lens_eq_eins_ang} there exists an intrinsic angular scale $\theta_{\text{\tiny E}}$, named the Einstein angle, and a corresponding length scale, the Einstein radius $R_{\text{\tiny E}}$. They are given by
 \begin{eqnarray}\label{eq:aml_rev_eins_ang}
    \theta_{\text{\tiny E}}&=&\left[\frac{4 G_{\text{\tiny N}} M_{\text{\tiny L}} }{c^2}\frac{D_{\text{\tiny S}}-D_{\text{\tiny L}}}{D_{\text{\tiny S}}D_{\text{\tiny L}}}\right]^{1/2}\;,\\
    &\simeq&2.85 \text{ mas}\,\left(\frac{D_{\text{\tiny S}}}{D_{\text{\tiny L}}}-1\right)^{1/2}\left(\frac{D_{\text{\tiny S}}}{\text{1 }  \text{kpc}}\right)^{-1/2}\left(\frac{M_{\text{\tiny L}}}{\text{1 } \solarmass}\right)^{1/2}\nonumber \;, \\    
    R_{\text{\tiny E}}&\equiv&D_{\text{\tiny L}}\theta_{\text{\tiny E}} \simeq 2.87\text{ AU}\,\left(1-\frac{D_{\text{\tiny L}}}{D_{\text{\tiny S}}}\right)^{1/2}\left(\frac{D_{\text{\tiny L}}}{\text{1 }  \text{kpc}}\right)^{1/2}\left(\frac{M}{\text{1 } \solarmass}\right)^{1/2}\;.
    \label{eq:aml_rev_eins_rad}
 \end{eqnarray}
  Physically, the Einstein angle represents the angular radius of a ring-shaped image that would form if the observer, the center of the point lens\footnote{A lens is considered as a point lens when its physical radius ($R_{\text{\tiny L}}$) is much smaller than the Einstein radius, or equivalently, when its angular size ($R_{\text{\tiny L}}/D_{\text{\tiny L}}$) is much smaller than the Einstein angle (see Eq.\,\eqref{eq:aml_rev_eins_ang}) of the lens.   }, and the source are in perfect alignment. Additionally, the radius of this ring-like image projected on the lens plane is the Einstein radius.

For a given mass distribution and source angular position, the lens equation Eq.\,\eqref{eq:aml_rev_lens_eq_eins_ang} determines the angular position of the image(s). For a generic, spatially bounded in extent, transparent, and finite mass and density distribution, a distant point source will always produce an odd number of lensed images\,\cite{Burke:1981zz}. The characteristic angular scale for image separation is the Einstein angle $( \theta_{\text{\tiny E}})$, which is typically of the order of milliarcseconds (mas) for stellar-mass lenses in our galaxy (see Eq.\,\eqref{eq:aml_rev_eins_ang}). Consequently, the multiple-lensed images are generally unresolved by conventional telescopes. Therefore, the detection of microlensing events often relies on observing the combined light from these unresolved images, leading to what is known as photometric microlensing.         
 
Gravitational lensing also alters the solid angles subtended by the images relative to the source. According to Liouville's theorem, the brightness per area of the apparent source  remains constant under gravitational lensing\,\cite{Schneider_1992}.  Therefore, this change in solid angle directly affects the observed flux, which can be quantified through the magnification of the image (see  Eq.\,\eqref{eq:app_mlrev_axial_mag}). 
 
In PML, the observable signal is the transient change in the total observed flux of light as the lens passes in the foreground, which is proportional to the sum of the magnifications of all lensed images
  \begin{equation}\label{eq:aml_rev_axial_mag_tot}
    \mathfrak{m}_{\text{\tiny T}} (t)=\sum_{I}\abs{\left(\frac{\theta_{\text{\tiny I}}}{\theta_{\text{\tiny S}}}\frac{d \theta_{\text{\tiny I}}}{d {\theta}_{\text{\tiny S}}}\right)}\;.
  \end{equation}

Moreover, an important concept in understanding gravitational lensing, particularly in microlensing, is that of critical curves and their corresponding caustics. Critical curves are loci in the lens plane where the magnification of the images of the moving sources formally becomes infinite\,\cite{Schneider_1992}.  The corresponding source locations on the source planes\footnote{The source plane is an imaginary 2-D plane, orthogonal to the observer's line of sight, where the unlensed background source is located.}  producing those infinite magnification image defines caustics, which are key regions where the number of lensed images changes abruptly as the source crosses them. This happens because such images are produced at the local extrema of the lens equation, Eq.\,\eqref{eq:aml_rev_lens_eq_eins_ang}. At these positions, the condition $d\theta_{\text{\tiny S}}/d\theta_{\text{\tiny I}}=0$ is met, which, as shown in Eq.\,\eqref{eq:aml_rev_axial_mag_tot}, leads to an infinite magnification for these images. 

This feature is also key to the detection of distinct microlensing events, as when a background source star crosses a caustic, there is a sharp increase in the observed flux known as a high magnification event\,\cite{Schneider_1992}. Also,  when a background source star crosses a caustic, a pair of images may either appear and disappear \,\cite{Schneider_1992}.  The location of these caustics is sensitive to the mass distribution of the lens, making them a powerful tool for inferring the properties of the lens.

In addition to photometric changes, microlensing also causes a subtle transient shift in the apparent centroid of the unresolved images, a phenomenon known as astrometric microlensing. This astrometric shift, while tiny, can be precisely measured with astrometric surveys like Gaia\,\cite{Gaia:2016zol}, allowing for the detection and detailed characterization of AML events. While PML provides information about the magnification, AML offers insights into the precise positions of the centroid of the lensed images and provides complementary information about the lens system.

The astrometric shift is precisely quantified by the displacement of the image(s) centroid from the unlensed source position\,\cite{Hog:1994xt}. The centroid of the lensed image system is formally defined as the magnification-weighted average of the individual image positions (see Eq.\,\eqref{eq:app_rev_rev_cent_pos},  and therefore AML is quantified by the shift in the centroid position, which is given by
\begin{equation}\label{eq:aml_rev_cent_shift}
    \Delta{\theta}_{\text{\tiny cent}}(t)=\frac{\sum_{I}{\theta}_{\text{\tiny I}} \mathfrak{m}_{\text{\tiny I}}}{\sum \mathfrak{m}_{\text{\tiny I}}}-\theta_{\text{\tiny S}}\;,
\end{equation}
where the sum is over all images.

 AML offers a significant advantage over PML for investigating  EADOs, mainly because it is sensitive to a broader range of impact parameters. Specifically, for large source positions,  the AML signals fall off as $(\theta_{\text{\tiny S}}/\theta_{\text{\tiny E}})^{-1}$, however the PML signal falls faster as $(\theta_{\text{\tiny S}}/\theta_{\text{\tiny E}})^{-4}$. This, combined with our current sensitivity\,\cite{Gaia:2016zol,Mroz:2022mfl} to detect these signals, enables the observation of AML signals at larger impact parameters relative to those detectable via PML.

The enhanced sensitivity to larger impact parameters makes AML particularly effective for studying  EADOs, like boson stars, Q-balls, and dark matter subhalos, compared to the point-like lenses like PBH. This advantage arises because extended EADOs produce larger astrometric shifts, as we shall discuss in the following sections. By exploring a greater volume of space for subtle, distributed lensing effects, AML presents a powerful and complementary approach for uncovering the nature and distribution of extended dark matter. In the next section, we will look at Gaia's ability to detect these AML events using its astrometric prowess.

\section{Astrometric microlensing due to  EADOs}\label{sec:aml_eds}
Having established the generic framework for astrometric microlensing in Sec.\,\ref{sec:aml_rev}, we will now
 apply it to investigate the AML characteristics of  EADOs described in Sec.\,\ref{sec:eds_rev}. Our investigation centers on two representative models--- a thin-wall Q-ball and a mini-boson star. The analysis focuses on how the lens's mass distribution and radius determine the astrometric shifts, and in some cases, are distinct from the point-lens case.

\subsection{Thin-wall Q-balls}\label{sec:aml_eds_uds}
We begin by analyzing the thin-wall Q-balls as our first model for an EADOs. As detailed in Sec.\,\ref{sec:eds_rev}, they are flat-space time solutions of EKG action, which exist for scalar potentials $V(\abs{\Phi}^2)$ satisfying certain conditions. In particular, the thin-wall Q-balls exist near the global minima of the $V(\abs{\Phi}^2)/\Phi^2$. These objects feature nearly uniform density throughout, with only a very thin crust; therefore, a uniform-density sphere provides a good approximation for modeling them. While this is an idealization, the UDS serves as a valuable approximation for studying other dark objects which have a constant density core like thick-wall Q-balls\,\cite{Heeck:2020bau}, gravastars\,\cite{Mazur:2004fk}, etc. 

The UDS model is characterized by a total mass $M_{\text{\tiny u}}$ distributed uniformly within a suitably defined radius $R_{\text{\tiny u}}$, providing a simple prototype for understanding how astrometric deflections differ for extended masses, in particular, differentiating them from point-mass lensing. The density of a UDS of radius $R_{\text{\tiny  u}}$ and total mass $M_{\text{\tiny  u}}$ is given by Eq.\,\eqref{eq:aml_eds_uds_dens_prof}.
Using Eqs.\,\eqref{eq:aml_eds_uds_dens_prof}, \eqref{eq:app_mlrev_mass_encl} and \eqref{eq:aml_rev_lens_eq_eins_ang},  we get the lens equation for the UDS lens as,
\begin{equation}\label{eq:aml_eds_uds_lens_eq_1}
    \theta_{\text{\tiny S}}=\begin{cases}
        \theta_{\text{\tiny I}}-\frac{\theta_{\text{\tiny E}}^2}{\theta_{\text{\tiny I}}}\left[1-\left(1-\frac{\theta_{\text{\tiny I}}^2/\theta_{\text{\tiny E}}^2}{R_{\text{\tiny  u}}^2/R_{\text{\tiny E}}^2}\right)^{3/2}\right]\,,& \theta_{\text{\tiny I}}/\theta_{\text{\tiny E}} <  R_{\text{\tiny  u}}/R_{\text{\tiny E}}\,, \\
        \theta_{\text{\tiny I}}-\frac{\theta_{\text{\tiny E}}^2}{\theta_{\text{\tiny I}}}\,,& \theta_{\text{\tiny I}}/\theta_{\text{\tiny E}} \ge  R_{\text{\tiny  u}}/R_{\text{\tiny E}}\,.
    \end{cases}
\end{equation}
		\begin{figure}[]
			\begin{center}
				\includegraphics[scale=0.25]{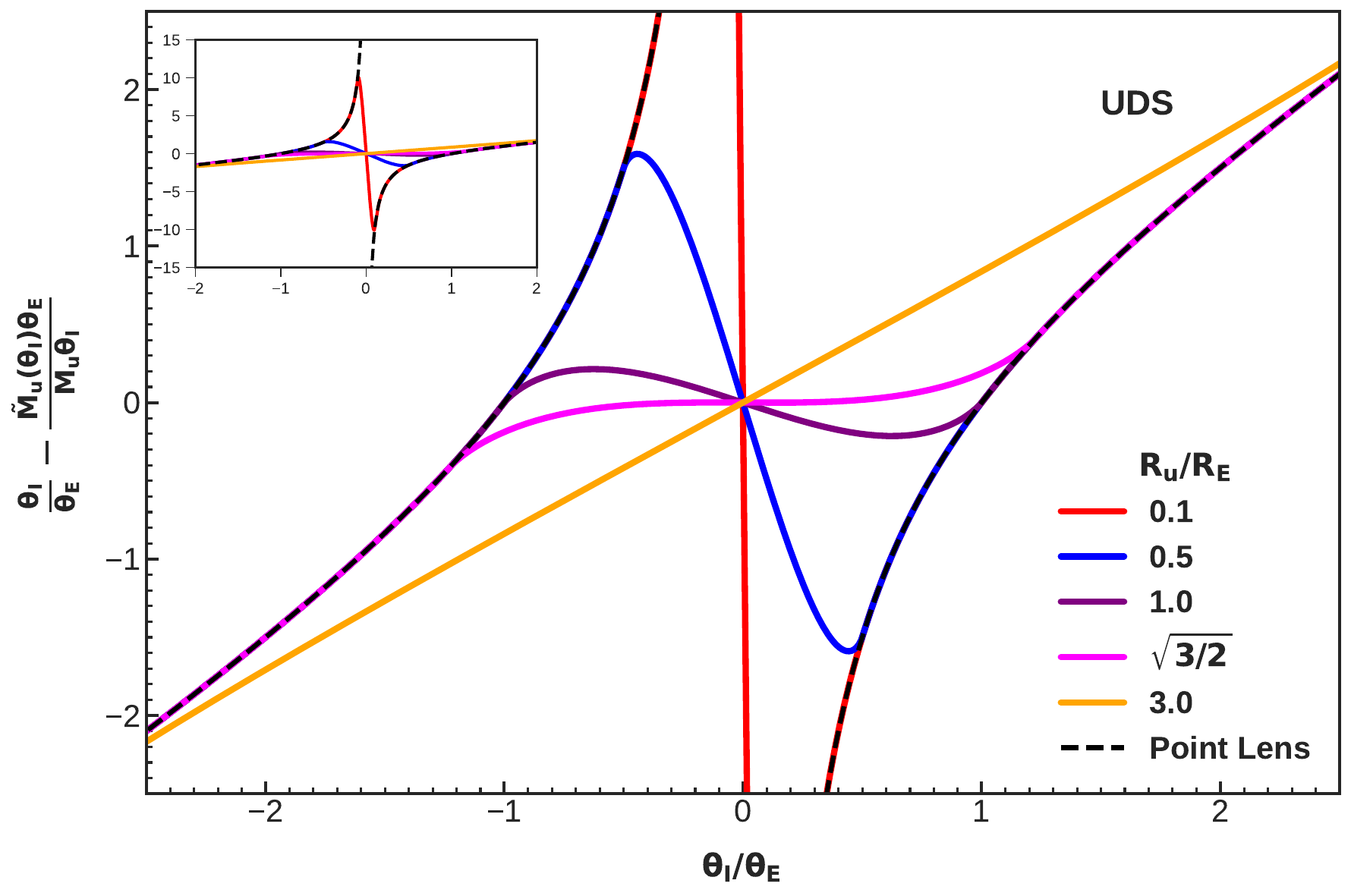}
                \includegraphics[scale=0.25]{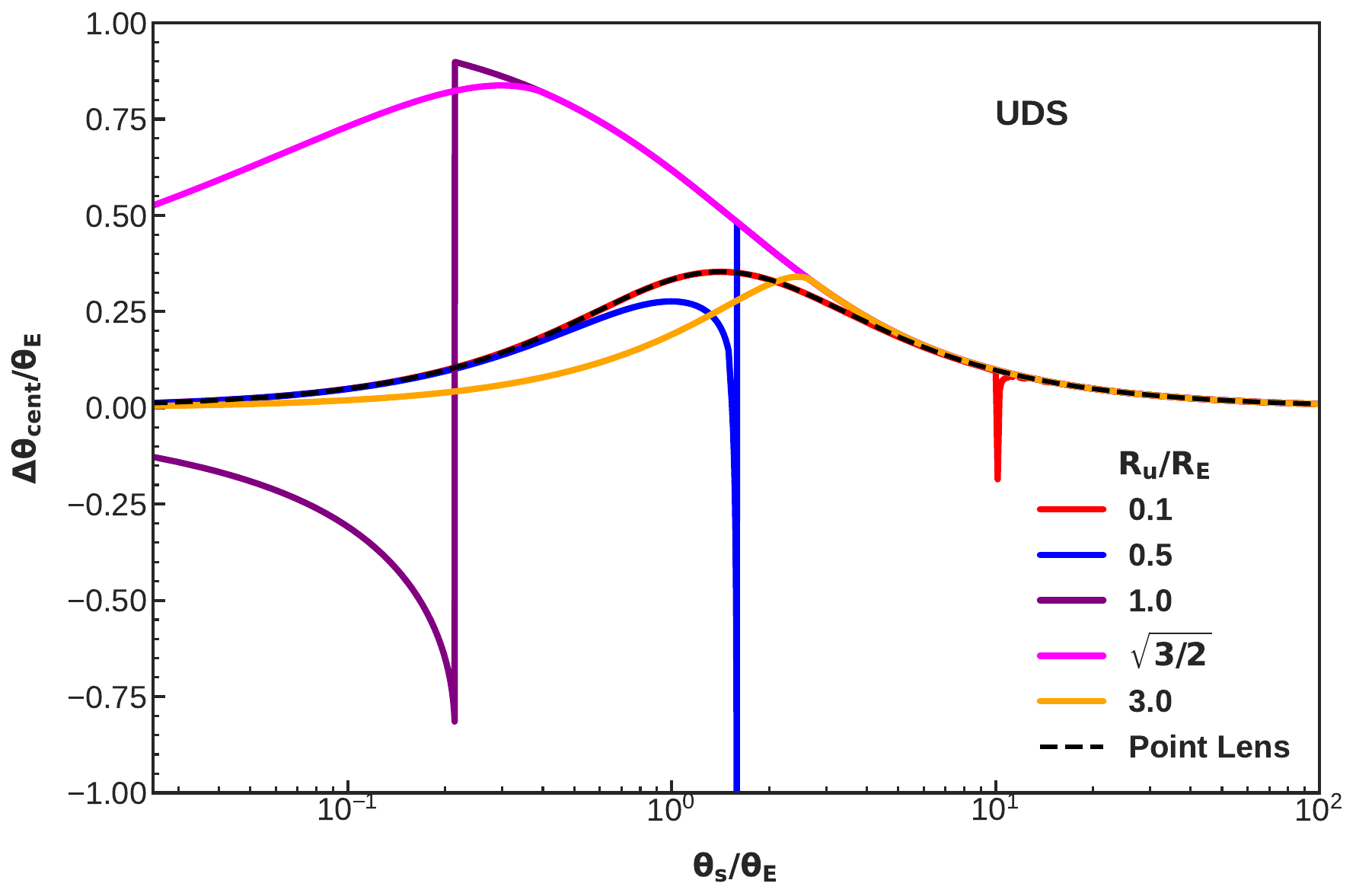}
			\end{center}
		
			\caption{ In the \textit{left panel}, we present the lens equation for the UDS lens (see Eq.\,\eqref{eq:aml_eds_uds_lens_eq_1}) as a function of the image position $\theta_{\text{\tiny I}}$. The \textit{right panel} shows the corresponding centroid shifts, $\Delta\theta_{\text{\tiny cent}}$ (see Eq.\,\eqref{eq:aml_rev_cent_shift}),  as a function of the source position $\theta_{\text{\tiny S}}$. The curves are plotted for several values of the scaled radius, $R_{\text{\tiny u}}/R_{\text{\tiny E}}=(0.1,0.5,1,\sqrt{3/2},3.0)$. The kink in the centroid shift for $R_{\text{\tiny u}}/R_{\text{\tiny E}}=(0.1, 0.5, 1)$  arises when the source crosses a caustic. For comparison, the point-lens case is indicated by the black dashed line. All angular quantities are normalised with respect to $\theta_{\text{\tiny E}}$.  See text for more details.}
			\label{fig:uds_massfunc_vs_ui}
		\end{figure}
 
The solutions to this equation\footnote{Note that Equation \eqref{eq:aml_eds_uds_lens_eq_1} is defined piecewise. Therefore, while each individual equation within the piecewise definition may yield multiple solutions, only the solution that meets its corresponding condition will be physically valid.} determines the angular positions of the images $\theta_{\text{\tiny I}}$, for a given source position $\theta_{\text{\tiny S}}$. The lens equation for the UDS model, Eq.\,\eqref{eq:aml_eds_uds_lens_eq_1}, is piecewise. The first case governs light rays passing inside the lens, resulting in a quintic polynomial that lacks an analytical solution. The second case governs light rays passing outside the lens and reduces to a quadratic equation, analogous to the point-lens case (see Appendix\,\ref{app:mlrev_UDS} for more details). As mentioned before,  the Einstein angle $\theta_{\text{\tiny E}}$ and the Einstein radius $R_{\text{\tiny E}}$, (see Eq.\,\eqref{eq:aml_rev_eins_ang}) inherently establish the characteristic scales for angles and distances within the system. Consequently, these quantities will be employed as normalization factors for their respective physical quantities.  

The number of images i.e. the number of solutions to the lens equation Eq.\,\eqref{eq:aml_eds_uds_lens_eq_1} depends upon the source position $\theta_{\text{\tiny S}}$, lens radius $R_{\text{\tiny u}}$, the Einstein angle $\theta_{\text{\tiny E}}$, and the Einstein radius $R_{\text{\tiny E}}$.  This dependence is illustrated in the left panel of Fig.\,\ref{fig:uds_massfunc_vs_ui}, where we display the right-hand side of the lens equation Eq.\,\eqref{eq:aml_eds_uds_lens_eq_1} (normalized by $\theta_{\text{\tiny E}}$) as a function of the source position $\theta_{\text{\tiny S}}/\theta_{\text{\tiny E}}$ for radii \(R_{\text{\tiny u}}/R_{\text{\tiny E}}=(0.1, 0.5, 1, \sqrt{3/2}, 3.0)\),  along with the point lens case.Notice that for $\theta_{\text{\tiny I}}/\theta_{\text{\tiny E}}>R_{\text{\tiny u}}/R_{\text{\tiny E}}$  the lens equation due to UDS lens matches with the point lens case as expected from Eq.\,\eqref{eq:aml_eds_uds_lens_eq_1}. 

For lenses characterized by radii such that
\begin{equation}\label{eq:aml_eds_uds_R_lim}
    R_{\text{\tiny u,Sch.}}<R_{\text{\tiny u}} <R_{\text{\tiny u,crit}} \equiv  \sqrt{3/2}~R_{\text{\tiny E}}\;,
\end{equation}
 the lens equation exhibits local maxima and minima, which may lead to the formation of multiple images. Here, we define $R_{\text{\tiny u,Sch.}}\equiv2 G_{\text{\tiny N}}M_{\text{\tiny u}}/c^2$, as the  Schwarzschild radius corresponding to the UDS lens with total mass $M_{\text{\tiny u}}$. In the regime given by Eq.\,\eqref{eq:aml_eds_uds_R_lim}, both critical curves ($\theta_{\text{\tiny I,crit}}$) and caustics ($\theta_{\text{\tiny S,caust}}$) are present and their angular position is given in Eq.\,\eqref{eq:app_eds_uds_crit_cur_caust}. It is important to note that these caustics and critical curves exist exclusively for lenses where $R_{\text{\tiny u,Sch.}}<R_{\text{\tiny u}} < \sqrt{3/2}~R_{\text{\tiny E}}$. Conversely, for $R_{\text{\tiny u}} > \sqrt{3/2}~R_{\text{\tiny E}}$, no local extrema are present, and consequently, only a single image is produced.

		\begin{figure}[]
			\begin{center}
				\includegraphics[scale=0.28]{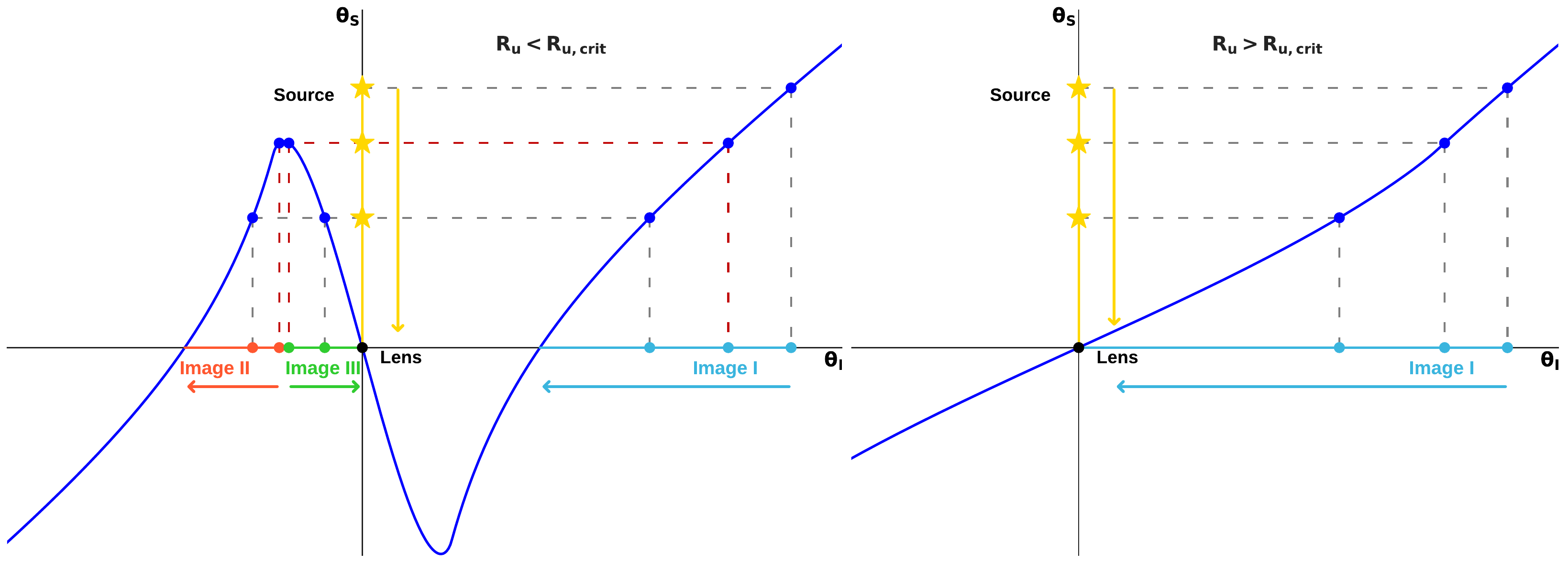}
			\end{center}
		
			\caption{Illustration for the solutions to the lens equation.
                    \textit{Left panel:} Case with $R_{\text{\tiny u}} < R_{\text{\tiny u,crit}}=\sqrt{3/2}~R_{\text{\tiny E}}$. 
                    \textit{Right panel:} Case with $R_{\text{\tiny u}} > R_{\text{\tiny u,crit}}=\sqrt{3/2}~R_{\text{\tiny E}}$. 
                    The lens is located at the center. 
                    The lens equation and source position are shown by the blue and yellow curves, respectively. 
                    The sky blue, green, and red solid lines represent the three image positions. 
                    For a given source position $\theta_{\text{\tiny S}}$, we draw a horizontal line whose intersection with the lens equation identifies the image positions along the $x$-axis. 
                    The red dashed horizontal line marks the caustic, $\theta_{\text{\tiny S}} = \theta_{\text{\tiny S,caust}}$. 
                    For $R_{\text{\tiny u}} < R_{\text{\tiny u,crit}}$, when the source lies beyond the caustic ($\theta_{\text{\tiny S}} > \theta_{\text{\tiny S,caust}}$), only one image (referred to as Image~I) is formed. 
                    However, for the source position near the caustic ($\theta_{\text{\tiny S}} \lesssim \theta_{\text{\tiny S,caust}}$, two additional images (referred to as Image~II and Image~III) appear. 
                    In contrast, for $R_{\text{\tiny u}} > R_{\text{\tiny u,crit}}$, no caustic exists, and only a single image (Image~I) is present for all source positions. See text for more details
 }
			\label{fig:uds_massfunc_vs_ui_ill}
		\end{figure}

Consider the illustration in Fig.\,\ref{fig:uds_massfunc_vs_ui_ill}, which depicts how the image positions are obtained from the lens equation and how their behaviour changes for the cases $R_{\text{\tiny u}} < R_{\text{\tiny u,crit}}$ (shown in the left panel) and $R_{\text{\tiny u}} > R_{\text{\tiny u,crit}}$ (shown in the right panel). The source and the lens equation are represented by the yellow and blue curves, respectively. The vertical direction represents the source position, and the horizontal direction represents the image position. For a given source position, we draw a horizontal line at  $\theta_{\text{\tiny S}}$ whose intersection with the lens equation (represented by blue dots) identifies the image positions, which are projected along the $x$-axis (represented by sky blue, red, and green dots). We use arrows to display how the image position changes when the source comes closer to the lens. We depict the caustics by the red dashed horizontal. 

As mentioned before, for lenses with radii \( R_{\text{\tiny u}} <R_{\text{\tiny u,crit}} \equiv \sqrt{3/2}~R_{\text{\tiny E}} \) there are local extrema in the lens equation (see for instance $R_{\text{\tiny u}}/R_{\text{\tiny E}}=(0.1, 0.5, 1)$ in the left panel of Fig.\,\ref{fig:uds_massfunc_vs_ui}).  We therefore expect distinct observational signatures due to the presence of caustics. In the left panel Fig.\,\ref{fig:uds_massfunc_vs_ui_ill}, when the source lies outside the caustic, i.e., at $\theta_{\text{\tiny S}} > \theta_{\text{\tiny S,caust}}$, only a single image is produced, located closer to the source position which we will call Image I (displayed in the sky blue dots). This feature contrasts with the point-lens case (see Eq.\,\ref{eq:rev_mlrev_point_lens_image_sol_1}, which invariably generates two images---one nearer to the source and another nearer to the lens. As the angular position of the source decreases for the UDS lens, increased deflection causes Image I to appear further from the actual source position. At the caustic crossing i.e $\theta_{\text{\tiny S}} = \theta_{\text{\tiny S,caust}}$, two additional images appear abruptly, raising the total image number to three. As displayed in the left panel of Fig.\,\ref{fig:uds_massfunc_vs_ui_ill}, these two new images (referred to as Image II (represented by red dots) and III (represented by green dots)) emerge near $-\abs{\theta_{\text{I,crit}}}$, indicating their formation in the direction opposite to the source, relative to the lens. As the source's angular position further decreases, Image II  moves away from the lens (towards -$\abs{\theta_{\text{\tiny E}}}$), while Image III moves closer to it (towards the lens center).

As shown in the right panel of Fig.\,\ref{fig:uds_massfunc_vs_ui_ill}, for lenses with $R_{\text{\tiny u}}\ge R_{\text{\tiny u,crit}} \equiv \sqrt{3/2}~R_{\text{\tiny E}}$, the lack of local extrema in the lens equation ensures that only a single image is produced (see for instance $R_{\text{\tiny u}}/R_{\text{\tiny E}}=(\sqrt{3/2},3.0)$ in the left panel of Fig.\,\ref{fig:uds_massfunc_vs_ui}).
 In this regime, as the angular position of the source decreases, the corresponding image position monotonically decreases, approaching zero when the source, lens, and observer become perfectly aligned. This behaviour arises because the mass distribution of the lens is too diffuse to generate the large deflection angles necessary for multiple imaging. This feature contrasts with that of a point lens, where perfect alignment of the source, lens, and observer leads to the formation of a circular ring-like image, known as the Einstein ring, at an angular position of $\theta_{\text{\tiny E}}$. 

 It is important to mention, that for all radii for large source positions, particularly $\theta_{\text{\tiny S}}\gg (\theta_{\text{\tiny S,caust}} \text{ and } R_{\text{\tiny u}}\theta_{\text{\tiny E}}/{R_{\text{\tiny E}}})$, the lens equation has only one solution determined by the second case in Eq.\,\ref{eq:aml_eds_uds_lens_eq_1}, which matches with the corresponding Image I with the point lens case.

\subsection*{Centroid shift due to UDS lens}
We numerically solve the lens equation, Eq.\,\eqref{eq:aml_eds_uds_lens_eq_1}, to determine the angular position of the images (${\theta}_{\text{\tiny I}}$) for a given source position ${\theta}_{\text{\tiny S}}$. Then the centroid shift for the UDS lens is given by
\begin{equation}\label{eq:aml_rev_cent_shift_uds}
    \Delta{\theta}_{\text{\tiny cent}}=\frac{\sum_{I}{\theta}_{\text{\tiny I}} \mathfrak{m}_{\text{\tiny I}}}{\sum \mathfrak{m}_{\text{\tiny I}}}-\theta_{\text{\tiny S}}\;,
\end{equation}
 Here the summation $(I)$ is over all images and $\mathfrak{m}_{\text{\tiny I}}$ is the magnification of the $I^{\text{\tiny th}}$ image, obtained by numerically solving  Eqs.\,\eqref{eq:app_mlrev_axial_mag}. The resulting dependence of the centroid shift on the source position for different lens radii is shown in the right panel of Fig.\,\ref{fig:uds_massfunc_vs_ui}, for $R_{\text{\tiny u}}/R_{\text{\tiny E}}=(0.1,0.5,1,\sqrt{3/2},3.0)$ and the point lens. Several distinctive features emerge in the UDS case, setting it apart from the point-lens behavior, which is indicated by the black dashed curve.

We begin our analysis with UDS lenses with radii $R_{\text{\tiny u}}/R_{\text{\tiny E}} < \sqrt{3/2}$. As noted earlier from the left panel of Fig.\,\ref{fig:uds_massfunc_vs_ui}, the number of images in this regime depends on the angular position of the source. For sources located outside the caustic, i.e., $\theta_{\text{\tiny S}} > \theta_{\text{\tiny S,caust}}$, only a single image (the one we call Image I) is produced, and therefore the centroid position coincides with Image I itself. However, a point lens always produces an additional, negatively-signed image on the opposite side of the source (see Eq.\,\ref{eq:rev_mlrev_point_lens_image_sol_1}). This secondary image pulls the magnification-weighted centroid (see Eq.\,\eqref{eq:app_rev_rev_cent_pos}) closer to the lens (the origin). Therefore, the centroid position (see Eq.\,\eqref{eq:app_rev_rev_cent_pos}) and centroid shift (see Eq.\,\eqref{eq:aml_rev_cent_shift}) due to a UDS lens are consequently larger than a point lens in this regime. This results in larger astrometric shift values for UDS lenses when $\theta_{\text{\tiny S}} > \theta_{\text{\tiny S,caust}}$ (see for instance $R_{\text{\tiny u}}/R_{\text{\tiny E}}=(0.1, 0.5, 1)$ in the right panel of Fig.\,\ref{fig:uds_massfunc_vs_ui}.

When the angular position of the source lies on a caustic \( \theta_{\text{\tiny S}} = \theta_{\text{\tiny S,caust}} \), two additional images emerge on opposite sides of the source with respect to UDS lens. The magnification of these images diverges at the caustic, leading to their predominant contribution to the overall centroid position. Consequently, the angular position of the centroid undergoes an abrupt change as the source crosses a caustic. This abrupt change in centroid position at the caustics is clearly reflected as kink in the right panel of Fig.\,\ref{fig:uds_massfunc_vs_ui} (see for radii \(R_{\text{\tiny u}}/R_{\text{\tiny E}}=(0.1, 0.5, 1)\)). When the source is located within the caustic (\( \theta_{\text{\tiny S}} < \theta_{\text{\tiny S,caust}} \)), three images are formed, with two forming on the opposite side to the source with respect to the lens, i.e. Image II and III (see Fig.\,\ref{fig:uds_massfunc_vs_ui_ill}). Near the caustics, these two images are highly magnified, and therefore, they provide the dominant contribution to the astrometric shift (see Eq.\,\eqref{eq:aml_rev_cent_shift}). Consequently, the centroid shift changes sign, a behaviour clearly distinct from that of the point-lens case.

As noted earlier, for UDS lenses with $R_{\text{\tiny u}}/R_{\text{\tiny E}} \ge \sqrt{3/2}$, the lensing configuration produces only a single image,  which naturally coincides with the centroid (see Eq.\,\eqref{eq:app_rev_rev_cent_pos}) since no secondary image is formed (see for radii \(R_{\text{\tiny u}}/R_{\text{\tiny E}}=(\sqrt{3/2}, 3.0)\) in the right panel of Fig.\,\ref{fig:uds_massfunc_vs_ui}). In instances where the source position dictates that Image I forms at an angular distance $\theta_{\text{\tiny I}}/\theta_{\text{\tiny E}}>R_{\text{\tiny u}}/R_{\text{\tiny E}}$, the corresponding light rays are exterior to the physical extent of the lens and consequently experience gravitational deflection attributable to the total mass of the lens. This scenario leads to a centroid shift that is greater than that predicted for a point lens. In this regime, the resulting centroid shift exceeds that of the point-lens case. Conversely, when the source position is such that the image is formed at angular position $\theta_{\text{\tiny I}}/\theta_{\text{\tiny E}} < R_{\text{\tiny u}}/R_{\text{\tiny E}}$, the light rays traverse the interior of the lens and experience a reduced effective deflecting mass. The centroid shift in this case exhibits a more intricate dependence on the lens radius and mass distribution. For example, in the right panel of the Fig.\,\ref{fig:uds_massfunc_vs_ui} shows that lenses with $R_{\text{\tiny u}}/R_{\text{\tiny E}} \sim \sqrt{3/2}$ produce a centroid shift significantly larger (up to $\sim 2.3$ times) than that of the point lens. However, as the lens radius increases further, the mass distribution becomes diluted, leading to weaker deflections. Consequently, for lenses with $R_{\text{\tiny u}}/R_{\text{\tiny E}} \gg \sqrt{3/2}$, the centroid shift is suppressed compared to the point lens. However, for source positions satisfying $\theta_{\text{\tiny S}}/\theta_{\text{\tiny E}} \gtrsim R_{\text{\tiny u}}/R_{\text{\tiny E}}$, it gives astrometric shift approximately similar to point lens which falls as $\Delta \theta_{\text{\tiny cent}}/\theta_{\text{\tiny E}}\sim \theta_{\text{\tiny E}}/\theta_{\text{\tiny S}}$.

In summary, our analysis demonstrates that the extended nature of a uniform density sphere introduces distinctive and rich features into the astrometric microlensing signal, which are absent in the conventional point-mass lens model. The key determinant of the lensing phenomenology is the ratio of the lens radius to its Einstein radius, \( R_{\text{\tiny u}}/R_{\text{\tiny E}}  \). We have shown that for the lenses with radii \(R_{\text{\tiny u,Sch.}}/R_{\text{\tiny E}}< R_{\text{\tiny u}}/R_{\text{\tiny E}} < \sqrt{3/2} \), the presence of caustics leads to discontinuous jumps in the astrometric centroid, causing the centroid shift to take negative values—a clear and potentially observable signature of a UDS lens. In contrast, lenses with radii \( R_{\text{\tiny u}}/R_{\text{\tiny E}}\ge \sqrt{3/2} \) always produce a single image, resulting in higher centroid shifts due to the absence of a negatively signed second image. This single-image scenario also arises for lenses with radii ($R_{\text{\tiny u}}/R_{\text{\tiny E}} < \sqrt{3/2}$), but only when the source is located outside the caustic ($\theta_{\text{\tiny S}} > \theta_{\text{\tiny S,caust}}$), again leading to astrometric shift larger than the point lens. These findings for the UDS lens model, which serves as a fundamental case study for an extended lens, provide a crucial baseline for understanding how mass distribution impacts astrometric observables. Building upon this foundation, we will now extend our investigation to a more physically motivated EDS with a non-uniform density profile---the mini-boson star.

\subsection{Boson Star (BS)}\label{sec:aml_eds_bs}
As described in Sec.\,\ref{sec:eds_rev}, boson stars are self-gravitating astrophysical objects composed of bosonic particles, stabilized against gravitational collapse by quantum pressure. We restrict ourselves to the case of mini-boson stars, which are described by a quadratic potential $V(\abs{\Phi}^2)=m^2\abs{\Phi}^2$, where $m$ is the mass of the bosonic particle. As mentioned, the mini-boson star profile is found by numerically solving the Schrödinger-Poisson equation. Additionally, a single solution can then be used to generate the entire family of solutions via the scaling relation in Eq.\,\eqref{eq:app_bs__scaling_final}. In this section, we will study the AML properties of mini-boson stars, utilizing the theoretical framework for spherically symmetric structures presented in Sec.\,\ref{sec:aml_rev} and the lensing analysis of the uniform density sphere from the previous section. This extension is motivated by the need to assess how a non-uniform density profile may further modify the AML signal. 

In contrast to the sharply bounded UDS case, BS exhibit a cored density profile that gradually decreases outward, as illustrated in Fig.\,\ref{fig:UDS_BS_dens_prof}, potentially giving rise to distinct observational signatures.
To quantify these effects, we employ the mini–boson star (BS) density (obtained numerically by solving Eqs.\,\eqref{eq:app_bs_einstein_tt_final}-\eqref{eq:app_bsmini_bos_star_density}) in the lens equation Eq.\,\eqref{eq:aml_rev_lens_eq_eins_ang} and compute the resulting astrometric centroid shifts. 
		\begin{figure}[]
			\begin{center}
				\includegraphics[scale=0.25]{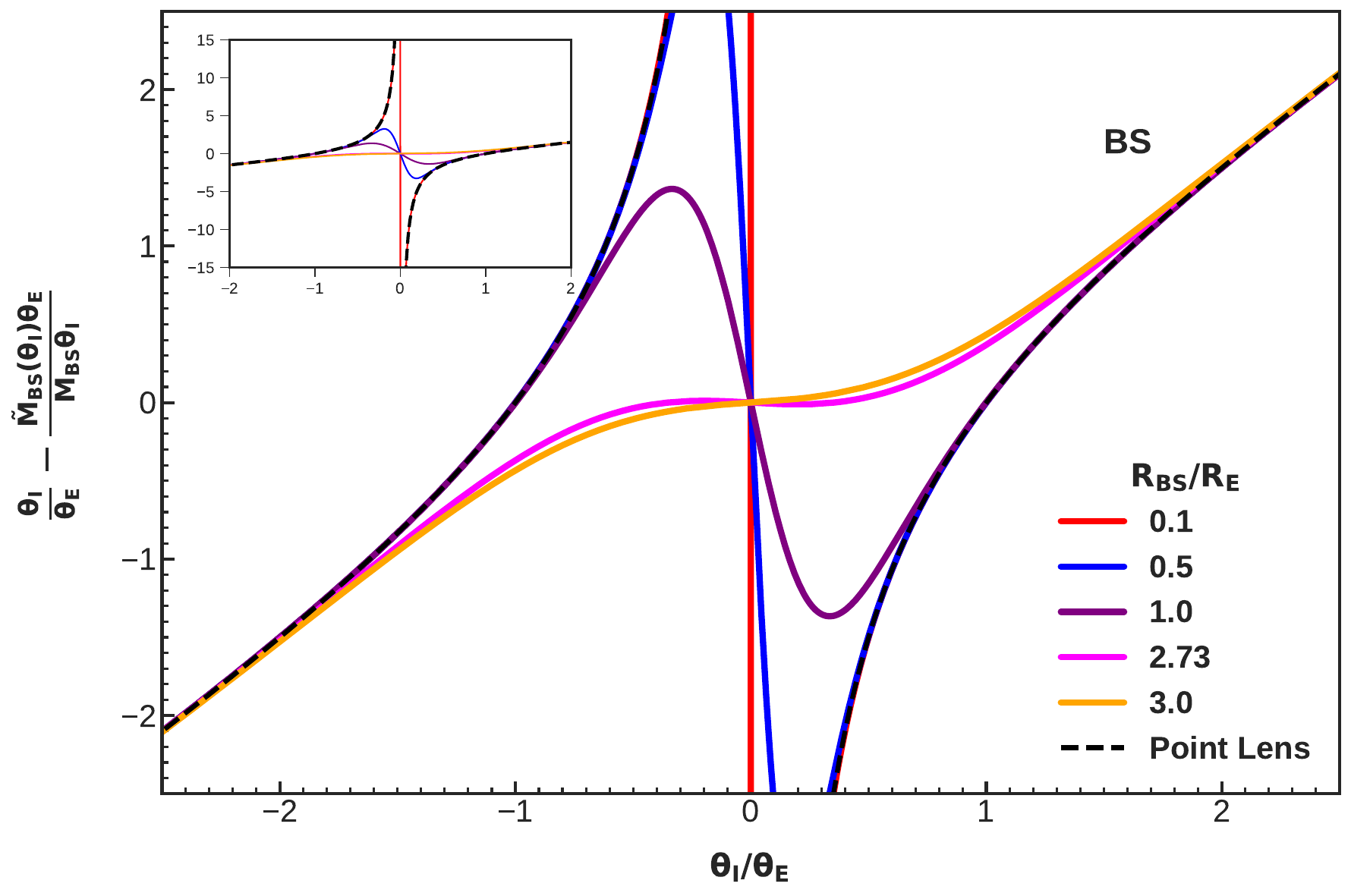}
                     \includegraphics[scale=0.25]{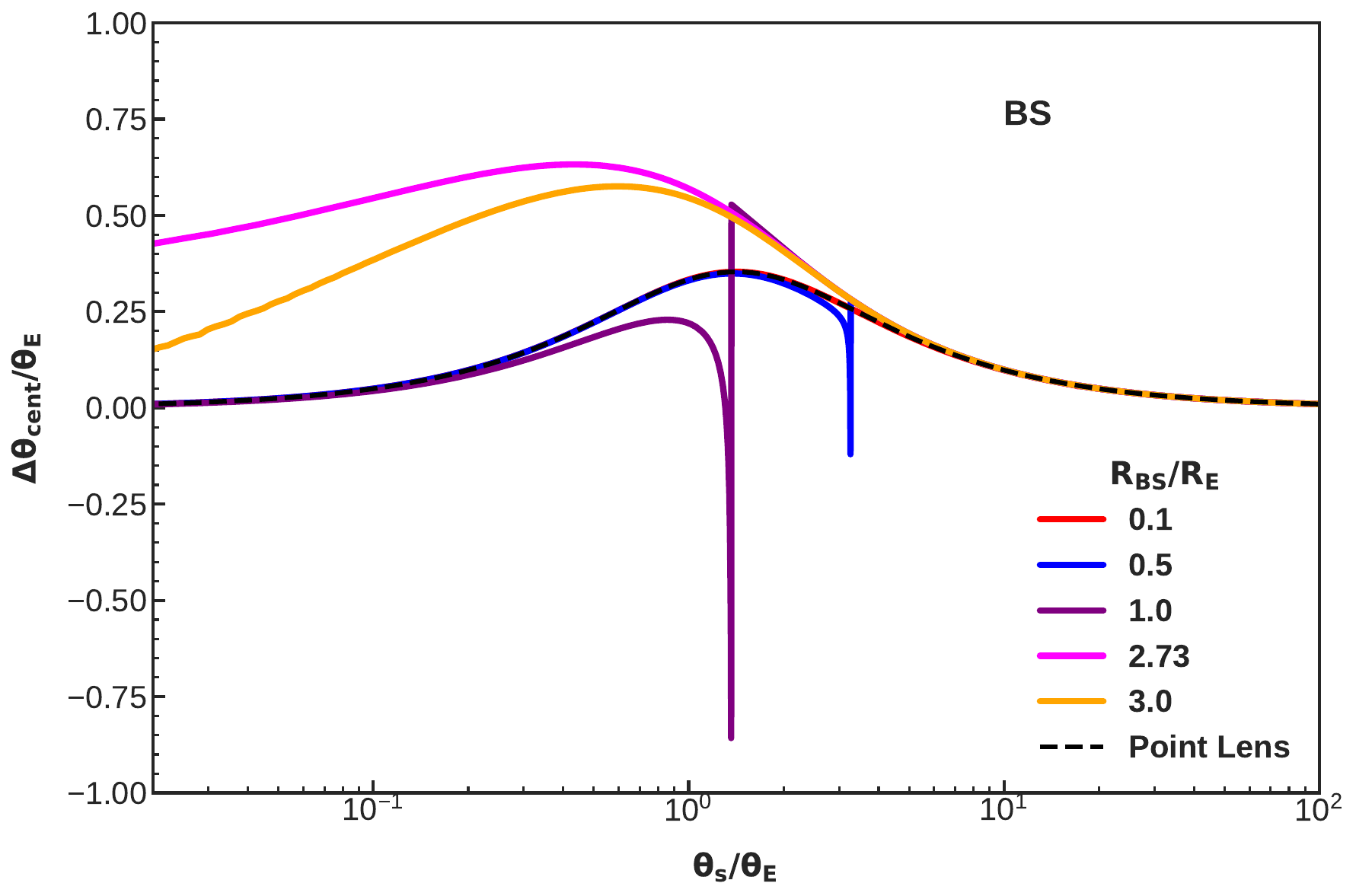}
			\end{center}
		
			\caption{ In the \textit{left panel}, the lens equation for the mini-boson star lens (see Eq.\,\eqref{eq:aml_eds_uds_lens_eq_1}) as a function of the image position $\theta_{\text{\tiny I}}$ is displayed. The corresponding centroid shifts, $\Delta\theta_{\text{\tiny cent}}$ (see Eq.\,\eqref{eq:aml_rev_cent_shift}),  as a function of the source position $\theta_{\text{\tiny S}}$ are illustrated in the \textit{right panel}. The curves are plotted for several values of the scaled radius, $R_{\text{\tiny  BS}}/R_{\text{\tiny E}}=(0.1,0.5,1,2.73,3.0)$. Again the kink in the centroid shift for $R_{\text{\tiny u}}/R_{\text{\tiny E}}=(0.5, 1)$  arises when the source crosses a caustic. For comparison, the point-lens case is indicated by the black dashed line. All angular quantities are normalised with respect to $\theta_{\text{\tiny E}}$. For details, see text.    }
			\label{fig:bs_massfunc_vs_ui}
		\end{figure}
The lens equation for the BS lens is formally given by 
 \begin{equation}\label{eq:aml_eds_bs_lens_eq_eins_ang}
    \theta_{\text{\tiny S}}=\theta_{\text{\tiny I}}-\frac{\tilde{M}_{\text{\tiny BS}}(\theta_{\text{\tiny I}}) }{M_{\text{\tiny BS}} }\frac{\theta_{\text{\tiny E}}^2}{\theta_{\text{\tiny I}}}\;.
 \end{equation}
 Here $\tilde{M}{\text{\tiny BS}}(\theta_{\text{\tiny I}})$ (see Eq.\,\eqref{eq:app_mlrev_mass_encl}) denotes the $2$-D mass function of the boson star lens projected onto the lens plane. In contrast to the UDS case, where a closed-form expression is available, $\tilde{M}{\text{\tiny BS}}(\theta_{\text{\tiny I}})$ must be evaluated numerically.
 
 Following the procedure outlined in Sec.\,\ref{sec:aml_eds_uds}, we numerically solve Eq.\,\eqref{eq:aml_eds_bs_lens_eq_eins_ang} to obtain the angular image positions ($\theta_{\text{\tiny I}}$) for different lens radii ($R_{\text{\tiny BS}}\equiv R_{\text{\tiny 99,BS}}$), total mass ($M_{\text{\tiny BS}}$), density distribution $\rho_{\text{\tiny BS}}(r)$ (see Eq.\,\eqref{eq:app_bsmini_bos_star_density}), and angular source positions ($\theta_{\text{\tiny S}}$). It is crucial to reiterate that the image multiplicity, determined by the number of solutions to the lens equation, again exhibits a direct dependence on source position $\theta_{\text{\tiny S}}$, BS lens radius $R_{\text{\tiny BS}}$, the Einstein angle $\theta_{\text{\tiny E}}$ and the Einstein radius $R_{\text{\tiny E}}$.  Fig. \ref{fig:bs_massfunc_vs_ui}\,(left panel) serves to illustrate this dependence, presenting the right-hand side of Eq.\,\eqref{eq:aml_eds_bs_lens_eq_eins_ang} (normalized by $\theta_{\text{\tiny E}}$) as a function of the normalized source position $\theta_{\text{\tiny S}}/\theta_{\text{\tiny E}}$ for radii \(R_{\text{\tiny BS}}/R_{\text{\tiny E}}=(0.1, 0.5, 1,2.73,3.0)\) and the point lens. Again similar to the UDS case, the lens equation due to BS lens matches with the point lens case for  $\theta_{\text{\tiny I}}/\theta_{\text{\tiny E}}\gtrsim R_{\text{\tiny u}}/R_{\text{\tiny E}}$.

For BS lenses with radii
 \begin{equation}\label{eq:aml_eds_bs_R_lim}
    R_{\text{\tiny BS,Sch.}}< R_{\text{\tiny BS}} <R_{\text{\tiny BS,crit}} \approx2.73 R_{\text{\tiny E}}\;,
\end{equation}
 the non-monotonic nature of the lens equation allows the formation of up to three images. Here, again $R_{\text{\tiny BS,Sch.}}\equiv2 G_{\text{\tiny N}}M_{\text{\tiny u}}/c^2$ is the  Schwarzschild radius corresponding to the BS lens with total mass $M_{\text{\tiny BS}}$. As before, the existence of multiple images makes BSs distinct from that of a point lens, which produces only two images. Moreover, it contrasts with the UDS case,  where multiple images are possible only for  $R_{\text{\tiny u,Sch.}}<R_{\text{\tiny u}}< \sqrt{3/2}~R_{\text{\tiny E}}$. Similar to the UDS case, here again the caustics ($\theta_{\text{\tiny S,caust}}$) coincide with the local extrema of the RHS of the lens equation. A key observation for BS lenses is that caustics emerge at a larger source position than in the UDS case. This is significant as it implies a potentially larger impact parameter (scaled with respect to $\theta_{\text{\tiny E}}$) for caustic crossing events for a given total mass and radius.

Given the qualitative similarity between the BS lens equation and the UDS case, Fig.\,\ref{fig:uds_massfunc_vs_ui_ill} can again be employed to qualitatively understand the behavior of the images. As in the UDS case, when the source position lies outside the caustic, $\theta_{\text{\tiny S}} > \theta_{\text{\tiny S,caust}}$, only a single image (Image I) is formed, located close to the source. At the caustic boundary, $\theta_{\text{\tiny S}} = \theta_{\text{\tiny S,caust}}$, two additional images appear on the opposite side of the source with respect to the lens. Since, for the given mass and radius, the mass distribution of the BS lens is more centrally concentrated than that of the UDS lens, BS lenses induce greater deflection for light rays passing through the lens. Consequently, Images II and III form closer to the lens for BS lenses than for UDS lenses (see, for instance $R_{\text{\tiny  BS}}=1~R_{\text{\tiny E}} $ in the left panel of Fig.\,\ref{fig:bs_massfunc_vs_ui} and corresponding curve in Fig.\,\ref{fig:uds_massfunc_vs_ui}).  As before, when the source’s angular position further decreases, one image (Image II) moves away from the lens (towards -$\abs{\theta_{\text{\tiny E}}}$), while the other image (Image III) moves closer to it (towards the lens center) as displayed in Fig.\,\ref{fig:uds_massfunc_vs_ui_ill}. 

For BS lenses with $R_{\text{\tiny BS}}/R_{\text{\tiny E}} \gtrsim 2.73$ (see for $R_{\text{\tiny BS}}/R_{\text{\tiny E}}=(2.73,3.0)$ in the left panel of Fig.\,\ref{fig:bs_massfunc_vs_ui}), the right-hand side of the lens equation is a monotonically increasing function, resulting in the formation of only a single image (Image I). As in the UDS scenario, the image position decreases monotonically with the source’s angular position, ultimately approaching zero at perfect alignment. As before, this behavior arises because, unlike the point-lens case, the diffuse mass distribution (see Fig.\,\ref{fig:UDS_BS_dens_prof}) of the BS lens is incapable of producing the large deflection angles required for multiple imaging. Since the BS lens has a more centrally concentrated mass distribution than the UDS lens, small source positions for the BS lens lead to relatively larger deflections compared to the UDS lens. Consequently, the centroid shift for the diffuse BS lens at small source positions is greater than the UDS case, though still smaller than the point lens case.

\subsection*{Centroid shift due to BS lens}
Following the methodology applied to the UDS lens, the centroid shift $\Delta{\theta}_{\text{\tiny cent}}$ which is again given by
\begin{equation}\label{eq:aml_rev_cent_shift_BS}
   \Delta{\theta}_{\text{\tiny cent}}=\frac{\sum_{I}{\theta}_{\text{\tiny I}} \mathfrak{m}_{\text{\tiny I}}}{\sum \mathfrak{m}_{\text{\tiny I}}}-\theta_{\text{\tiny S}}\;,
\end{equation}
will be derived from the solution of the lens equation, specifically Eq.\eqref{eq:aml_eds_bs_lens_eq_eins_ang}. Here again the summation $(\text{\tiny I})$ is over all images produced by BS lens and $\mathfrak{m}_{\text{\tiny I}}$ is the magnification of the $\text{\tiny I}^{\text{\tiny th}}$ and obtained by numerically solving  Eqs.\,\eqref{eq:app_mlrev_axial_mag}. In the right panel of Fig.\,\ref{fig:bs_massfunc_vs_ui}, we illustrate the functional dependence of angular position of centroid shift on the angular source position across a range of BS lens radii, specifically \(R_{\text{\tiny BS}}/R_{\text{\tiny E}}=(0.1, 0.5, 1,2.73,3.0)\) and the point lens. 

We begin our analysis of the centroid shift $\Delta\theta_{\text{\tiny cent}}$ for BS lenses with radii $R_{\text{\tiny BS}}/R_{\text{\tiny E}} <2.73$. Consistent with prior discussions, when sources lies beyond the caustic (\( \theta_{\text{\tiny S}} > \theta_{\text{\tiny S,caust}} \)), only one image (Image I) is present, and the centroid coincides with the image's center.  In this regime, the centroid shift is comparable to that of a UDS lens, although it is consistently larger than the point-lens prediction.  Again, this enhancement occurs because, unlike a point lens, a BS lens does not produce an image on the side opposite to the source relative to the lens, which would otherwise reduce the centroid position. Once the source crosses the caustic, however, two more images, which we call Images II and III, appear, producing a sudden negative shift in the centroid. At the caustics, the centroid position is dominated by the two newly formed, highly magnified images. As established, the higher central concentration of the BS lens pulls these negatively signed images closer to the lens center than in the UDS case. Because these dominant images are less separated from the lens at the origin, their (negative) contribution to the magnification-weighted centroid is smaller, resulting in a less pronounced "kink" or jump in the centroid trajectory. Finally, it is worth emphasizing that for BS lenses, the caustic occurs at larger source position (scaled with respect to $\theta_{\text{\tiny E}}$) compared to UDS lenses; therefore, these caustic crossing features can be observed for larger source positions for fixed values of $M_{\text{\tiny BS}}$ and $R_{\text{\tiny BS}}$.

Considering lenses with radii $ R_{\text{\tiny BS}}/R_{\text{\tiny E}} \gtrsim2.73$, the lensing configuration yields only Image I. As discussed earlier, in instances where Image I forms at an angular distance $\theta_{\text{\tiny I}}/\theta_{\text{\tiny E}}>R_{\text{\tiny BS}}/R_{\text{\tiny E}}$ the light rays propagate outside the physical extent of the lens, and the resulting centroid shift is greater than that predicted for a point lens. Conversely, when the source position is such that the image forms inside the physical extent of the lens, i.e. $\theta_{\text{\tiny I}}/\theta_{\text{\tiny E}}<R_{\text{\tiny BS}}/R_{\text{\tiny E}}$, the light rays forming Image I pass through the BS lens, experiencing a reduced effective deflecting mass, and the behavior of the centroid shift becomes more complicated.  For instance, Fig.\,\ref{fig:bs_massfunc_vs_ui} demonstrates that lenses with $R_{\text{\tiny BS}}/R_{\text{\tiny E}} \sim 2.73$ exhibit a larger centroid shift compared to the point lens in this scenario. In this scenario, compared to the UDS case,  the more centrally concentrated mass distribution of the BS lens causes light rays to experience a larger effective deflecting mass. This leads to a larger centroid shift for the BS lens compared to the UDS lens. However, as the lens radius increases further, the mass distribution becomes increasingly diluted, reducing the deflection. In the limit $R_{\text{\tiny BS}}/R_{\text{\tiny E}} \gg2.73$, this results in centroid shifts that can fall below the point-lens expectation.

 		\begin{figure}[]
 			\begin{center}
 				\includegraphics[width=\linewidth]{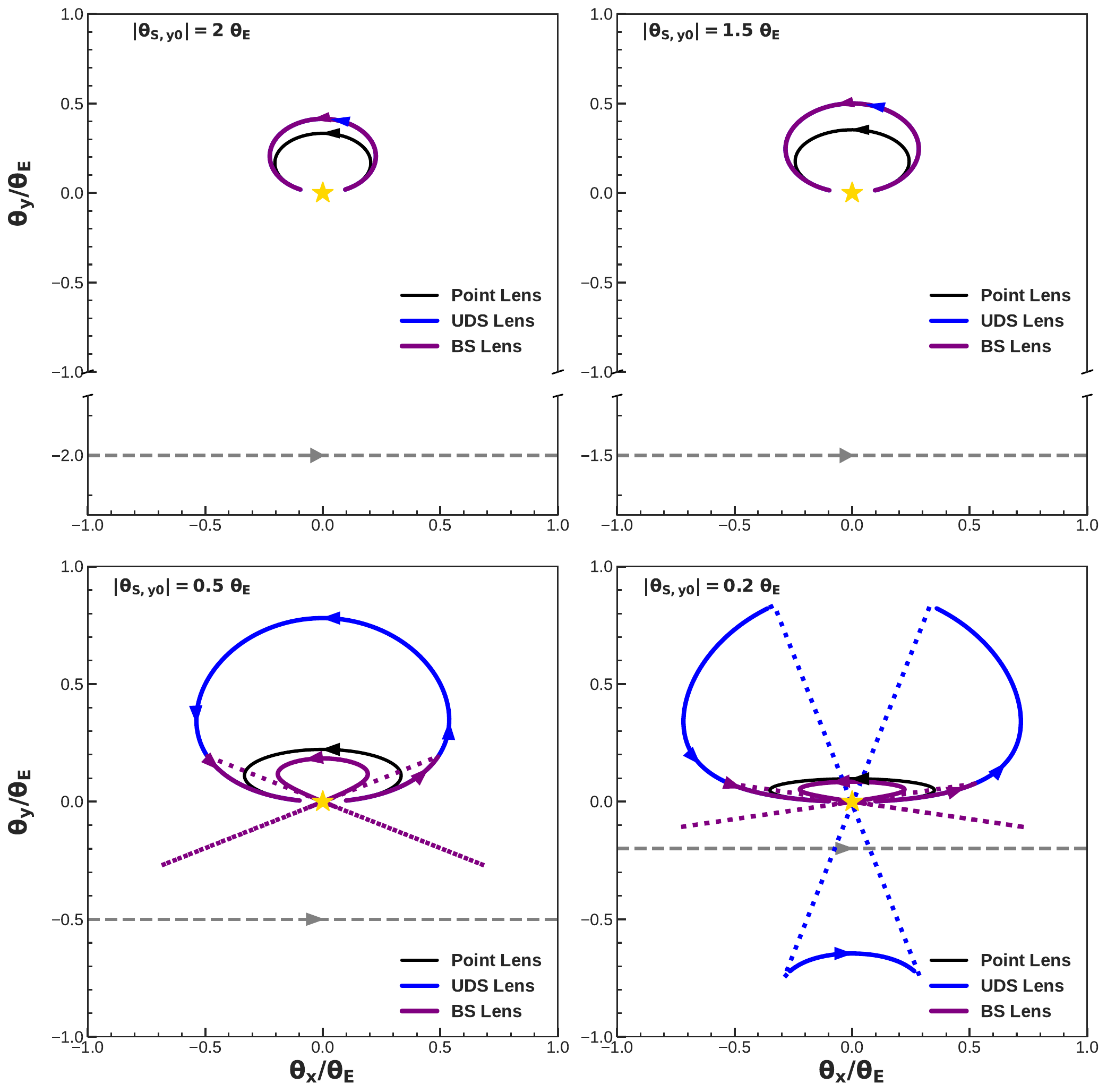}
 			\end{center}
 			\caption{Comparison of AML signals due to UDS (blue) and BS (purple) lenses with radius $R_{\text{\tiny L}}/R_{\text{\tiny E}}=1$ during lens motion. The blue and purple curves trace the centroid-shift trajectories of the source induced by UDS and BS lenses, respectively, for lenses moving along the $x$-axis. Four cases are shown with minimum impact parameters $\theta_{\text{\tiny S,y0}}/\theta_{\text{\tiny E}} = 2$ (\textit{top left}), $1.5$ (\textit{top right}), $0.5$ (\textit{bottom left}), and $0.2$ (\textit{bottom right}). The gray line marks the lens trajectory, while black curves indicate the corresponding centroid shifts for a point lens.  The dotted lines represent the abrupt shift in the centroid position due to caustic crossing. The source star is located at the origin. All axes indicate angular positions on the sky, normalized by $\theta_{\text{\tiny E}}$. }
 			\label{fig:uds_centshift_vs_us_2d}
 		\end{figure}

In the following section, we translate these characteristic astrometric features of UDS and BS lenses into the corresponding observable AML events. We further examine how such signals can be detected by Gaia and classify the events according to their characteristic durations.

\section{AML signal due to  EADOs and Gaia detectability}
\label{sec:AML_sig_Gaia_det}
This section outlines the observability prospects of AML signals produced by  EADOs. We begin by characterizing the distinctive observable AML signals generated by the UDS and BS lens models in Sec.\,\ref{sec:AML_UDS_BS}. We then turn to the observational capabilities of Gaia in Sec.\,\ref{sec:gaia}, reviewing its astrometric precision and detailing the contents of the Gaia DR3 catalogue\,\cite{Gaia:2023fqm}. This discussion provides a basis for assessing Gaia’s sensitivity to the astrometric shifts induced by EADO. Building on this, we establish the detection criteria for identifying AML events produced by such lenses in Sec.\,\ref{sec:AML_class}. We also discuss the classification of AML events according to their characteristic durations. 

\subsection{Astrometric lensing signal due to a UDS and BS}
\label{sec:AML_UDS_BS}
We have discussed the behavior of centroid shift for UDS and BS lenses in the preceding sections. We now extend this analysis to examine how these behaviors manifest in the full astrometric microlensing signal as would be observed. For this purpose, we consider a lens moving along a rectilinear trajectory in the positive $x$-direction with a minimum impact parameter $\abs{\theta_{\text{\tiny S,y0}}}$ and compute the resulting AML signal. The trajectory of the astrometric centroid ${\theta}_{\text{\tiny cent}}$ is then traced for a background source located at the origin, allowing us to directly compare the  EADOs like UDS and BS, with the point-lens case.\footnote{In this configuration, since the source is placed at the origin, the centroid coordinates coincide directly with the centroid shift.}

This comparison is presented in Fig.\,\ref{fig:uds_centshift_vs_us_2d}, which provides a two-dimensional sky view (normalized by $\theta_{\text{\tiny E}}$) of the AML signal generated by a moving UDS lens (blue) and a BS lens (purple), each with a radius  $R_{\text{\tiny L}}/R_{\text{\tiny E}}=1$. The figure illustrates the trajectory of the astrometric centroid $\vec{\theta}_{\text{\tiny cent}}$\footnote{Here, $\vec{\theta}_{\text{\tiny cent}} =(\theta_{\text{\tiny cent,x}},\theta_{\text{\tiny cent,y}})$, where $\theta_{\text{\tiny cent,x}}$ and $\theta_{\text{\tiny cent,y}}$ are the components of centroid position projected along $x$ and $y$ direction respectively for the moving lens (see Appendix\,\ref{app:mlrev}) for details. } as the lens passes in front of a stationary background source located at the origin. The analysis is presented across four panels, each corresponding to a different minimum impact parameter: $\theta_{\text{\tiny S,y0}}/\theta_{\text{\tiny E}}=(0.2,0.5,1.5,2.0)$. For reference, the lens's path is shown as a solid gray line, and the corresponding signal for a standard point lens is shown in black.  A similar comparison plot for the $R_{\text{\tiny L}}/R_{\text{\tiny E}}=3$ is presented in Fig.\,\ref{fig:uds_centshift_vs_us_2d_1} in App.\,\ref{app:sup_result}.

For large-impact parameters ($\abs{\theta_{\text{\tiny S,y0}}/\theta_{\text{\tiny E}}}=2.0$ and $1.5$), both UDS and BS lenses produce similar, smooth, and closed centroid trajectories. These loops are significantly larger than those generated by a point lens, reflecting the stronger deflection caused by extended mass distributions of the same mass and radii. This behavior arises from the fact that, for such source positions, only a single image is formed, and therefore the centroid directly follows its distorted path. As expected, the size of this closed loop is smaller for the case when $\theta_{\text{\tiny S,y0}}/\theta_{\text{\tiny E}}=2.0$ (top left panel of Fig.\,\ref{fig:uds_centshift_vs_us_2d}), as compared to the case when $\theta_{\text{\tiny S,y0}}/\theta_{\text{\tiny E}}=1.5$ (top right panel of Fig.\,\ref{fig:uds_centshift_vs_us_2d}). This feature is consistent with the diminishing lensing effect at larger angular separations.

For the intermediate impact parameter ($\theta_{\text{\tiny S,y0}}/\theta_{\text{\tiny E}}=0.5$), the centroid trajectories of UDS and BS lenses remain similar when the source lies outside the caustics. However, in the BS case, the centroid trajectory exhibits abrupt deviations due to caustic crossing. Specifically, for the BS trajectory, the source intersects its corresponding caustic twice, leading to the sudden appearance and subsequent disappearance of additional images, which causes the centroid to jump between separate regions of the plot. By contrast, the UDS trajectory remains smooth and loop-like, as no caustic crossing occurs. In both cases, the centroid shifts are larger than for a point lens.

At the smallest impact parameter ($\theta_{\text{\tiny S,y0}}/\theta_{\text{\tiny E}}=0.2$), both UDS and BS lenses undergo caustic crossings, but the signature of the centroid shift differs markedly. This is a consequence of the fact that the onset of caustics happens at smaller source positions for the UDS lens. It is again important to mention that UDS produces a larger centroid shift compared to a BS lens for this configuration. After the second caustic crossing for the UDS and BS lens, the centroid resumes a continuous path until returning to its original position.

The analysis highlights that the astrometric signatures of  EADOs are closely tied to their internal mass distributions. Although, we have displayed the centroid trajectories for $R_{\text{\tiny L}}/R_{\text{\tiny E}}=1$, however the presented qualitative behaviour of centroid trajectories remains same for the cases when the lens have caustics (i.e. for radii satisfied by Eqs.\,\eqref{eq:aml_eds_uds_R_lim} and \eqref{eq:aml_eds_bs_R_lim}). In summary, the Fig.\,\ref{fig:uds_centshift_vs_us_2d} highlights that while both UDS and BS lenses can produce enhanced centroid shifts relative to a point lens, their behavior differs significantly in the vicinity of caustics. For fixed mass and radii, the UDS lenses tend to exhibit larger centroid shifts, whereas BS lenses reveal caustic crossings at a larger source position. This increases the impact parameter over which discontinuous features are observable, providing a promising pathway to distinguish between different extended lens models through astrometric microlensing.

For non-caustic crossing events, the characteristics of the astrometric shifts depend sensitively on the lens radius, in particular, on the ratio $R_{\text{\tiny L}}/R_{\text{\tiny E}}$. Specifically, for dilute lenses where $R_{\text{\tiny  L}}/R_{\text{\tiny E}} \gg 1$ (see for instance Fig.\,\ref{fig:uds_centshift_vs_us_2d_1} in App.\,\ref{app:sup_result} which compares AML signals for $R_{\text{\tiny  L}}/R_{\text{\tiny E}} = 3$), the BS lens generally produces larger centroid shifts at small source positions ($\theta_{\text{\tiny S}}/\theta_{\text{\tiny E}} < R_{\text{\tiny L}}/R_{\text{\tiny E}}$), a direct consequence of its more compact mass distribution compared to a UDS lens of the same mass and radius. Fig.\,\ref{fig:uds_centshift_vs_us_2d_1} demonstrates that even for non-caustic-crossing events, the AML signal remains quantitatively distinct for the BS model compared to the UDS model. This highlights a crucial point: the ability to distinguish between UDS and BS lenses is not limited to the dramatic signatures of caustic crossings. A distinction also exists in the dilute-lens regime, where their different internal mass profiles produce quantitatively different astrometric shifts. However, for $\theta_{\text{\tiny S}}/\theta_{\text{\tiny E}} > R_{\text{\tiny L}}/R_{\text{\tiny E}}$, the centroid shifts from both BS and UDS models become quantitatively similar. In the asymptotic limit $\theta_{\text{\tiny S}}/\theta_{\text{\tiny E}} \gg R_{\text{\tiny L}}/R_{\text{\tiny E}}$, the results from both extended lenses converge to the point-lens case, as expected.

We have displayed in Figs.\,\ref{fig:uds_centshift_vs_us_2d} and \,\ref{fig:uds_centshift_vs_us_2d_1} that the AML signatures of EADOs can be significantly distinct from those of point-mass lenses like PBHs.  Furthermore, depending on the impact parameter and the ratio $R_{\text{\tiny L}}/R_{\text{\tiny E}}$, the UDS and BS models themselves exhibit distinguishable signals, offering a potential probe of the lens's internal structure.  While these phenomenological differences are promising, a dedicated simulation is required to rigorously quantify the observational distinguishability of these models with Gaia. 

Having established the distinctive astrometric features of UDS and BS lenses, we now turn to quantifying their observability. In the following section, we address the question of whether these signals are detectable by Gaia. Therefore, we will start by discussing the instrument’s characteristics, with particular emphasis on its astrometric sensitivity.

\subsection{Gaia}\label{sec:gaia}

In the previous section, we showed that the typical AML shift in the centroid of a source star due to an exotic astrophysical dark object can be of the order of a milliarcsecond\footnote{A typical shift can be of order of Einstein angle, which for a $1 \solarmass$ lens inside the galaxy can be of order of milliarcsecond (mas) (see Eq.\,\ref{eq:aml_rev_eins_ang})}. To test the hypothesis that such objects contribute to the dark matter content of the Milky Way, we require a telescope with the ability to measure stellar positions with mas-level precision over a long temporal baseline. This is precisely the capability of the European Space Agency's Gaia mission.

Gaia\,\cite{Gaia:2016zol,Lindegren_2012,Gaia:2023fqm} is a space-based optical telescope at the second Lagrange point $L_2$ of the Earth-Sun system. Its principal goal is to chart a 3D map of the Milky Way by measuring the positions, proper motion, and parallaxes of more than a billion stars with unprecedented accuracy. Launched in December 2013, Gaia began its scientific observations in mid-2014 and has since been continuously surveying the sky. Over more than a decade of operations which has produced successive data releases (i.e. DR1\,\cite{Gaia:2016zol}, DR2\,\cite{Gaia_dr2}, and DR3\,\cite{Gaia:2023fqm}), Gaia has observed approximately 1.47 billion stars, covering the entire sky with sensitivity brighter than about 21~mag. 

Gaia\,\cite{Gaia:2016zol} employs two telescopes simultaneously observing two distinct fields of view , separated by a constant angle of $106^\circ$. Each telescope has a rectangular mirror of size 1.49 m $\times$ 0.54 m. This wide angular separation between the fields of view is essential for achieving a global astrometric solution (for more details, check ref.~\cite{Lindegren_2012}). The detector collects light in three broad photometric bands---a broad $G$ band (330-1050 nm), and two narrower bands - the blue photometer (BP: 330-680 nm) and the red photometer (RP: 640-1050 nm). The instrument's point spread function (PSF) has a typical size of $\sim$100 mas. This implies that in most microlensing events induced by  EADOs, for which the angular separation ($\sim \theta_E$) is typically smaller than the PSF, Gaia will detect an unresolved blend of the lensed images. This results in a small shift in the centroid of the stellar image rather than the resolution of distinct lensed images, thereby manifesting as an AML signal. 

It is important to note that the centroid precision is limited by the signal-to-noise ratio of the measurement for a given PSF size. For a well-sampled PSF, the centroid can be determined to a precision much smaller than the PSF width. The centroid uncertainty approximately varies as\footnote{ This is because the PSF represents the probability distribution of a single photon, and by collecting $N_{\rm photons}$, the uncertainty on the center of that distribution (the centroid) is reduced by $\sqrt{N_{\rm photons}}$} $\sigma_a \sim \sigma_{\rm psf}/\sqrt{N_{\rm photons}}$, where $\sigma_{\rm psf}$ is the size of the PSF and $N_{\rm photons}$ are the number of photons detected from the source~\cite{1983PASP...95..163K}. Each stellar transit on the focal plane of the detector of Gaia provides typically 9 CCD observations from each field of view. These $2\times9$ measurements are then combined to obtain about a milliarcsecond level of astrometric precision.

A key advantage of Gaia is its astrometric time-series data. For stars brighter than 14 magnitude (mag)\footnote{The apparent magnitude ($m$) of a celestial object is a measure of its observed brightness as seen from Earth. It is related to the observed flux ($F$) as $m_{\text{\tiny G}}=-2.5~\text{Log}_{10}(F/F_0)$, where $F_0$  is a reference flux corresponding to a magnitude of $0$.}in the $G$ band, Gaia achieves an astrometric precision of about 1 mas per observational epoch. Owing to its scanning law—a predefined, non-uniform sky-scanning strategy optimized for global astrometry—each point on the sky is observed roughly once per month by the combined fields of view, resulting in about 140 measurements per star (70 in each field of view) over the 10-year mission lifetime\,\cite{Gaia:2016zol}. For simplicity, we treat the two successive observations of a star within a combined field of view passage as a single effective pass. Assuming these passes are uniformly distributed, the corresponding cadence is $t_{\text{\tiny s}}=10~\text{years}/70\approx 52.2~\text{days}$. This cadence and mission duration provide a rich time-domain dataset for identifying the subtle astrometric shifts induced by gravitational microlensing.
		\begin{figure}[]
			\begin{center}
				\includegraphics[scale=0.4]{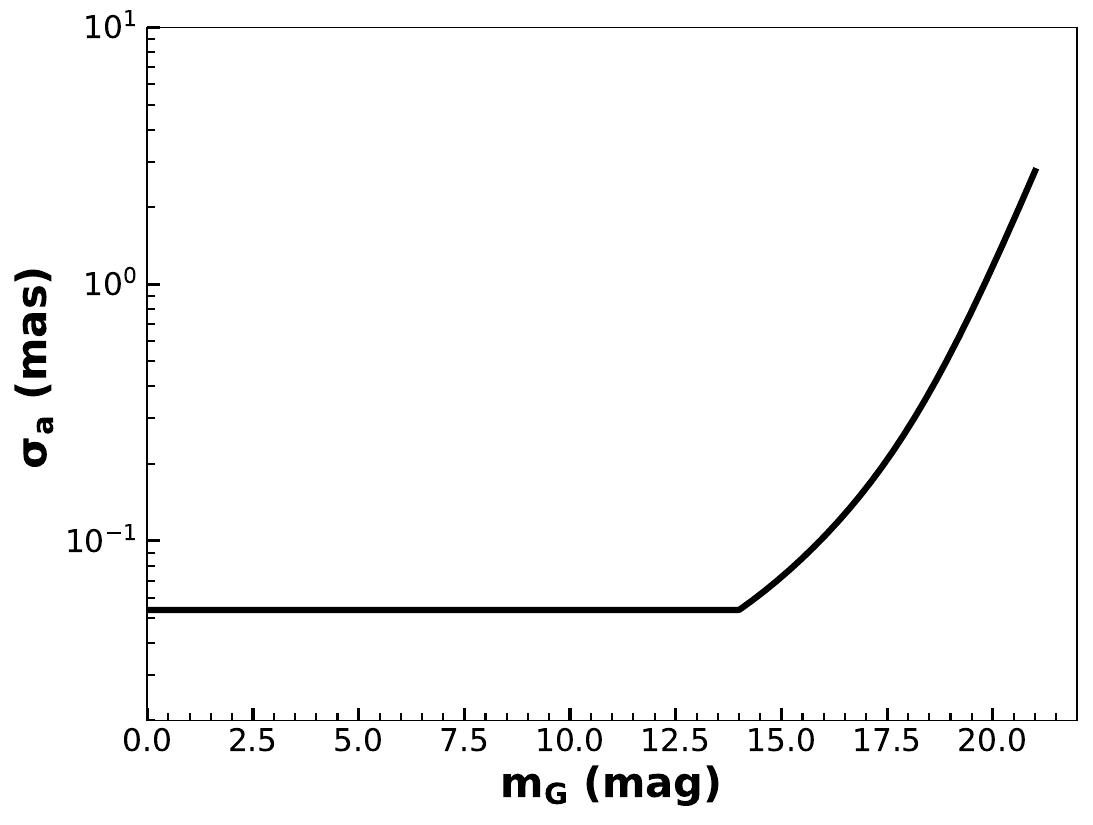}
                \includegraphics[scale=0.375]{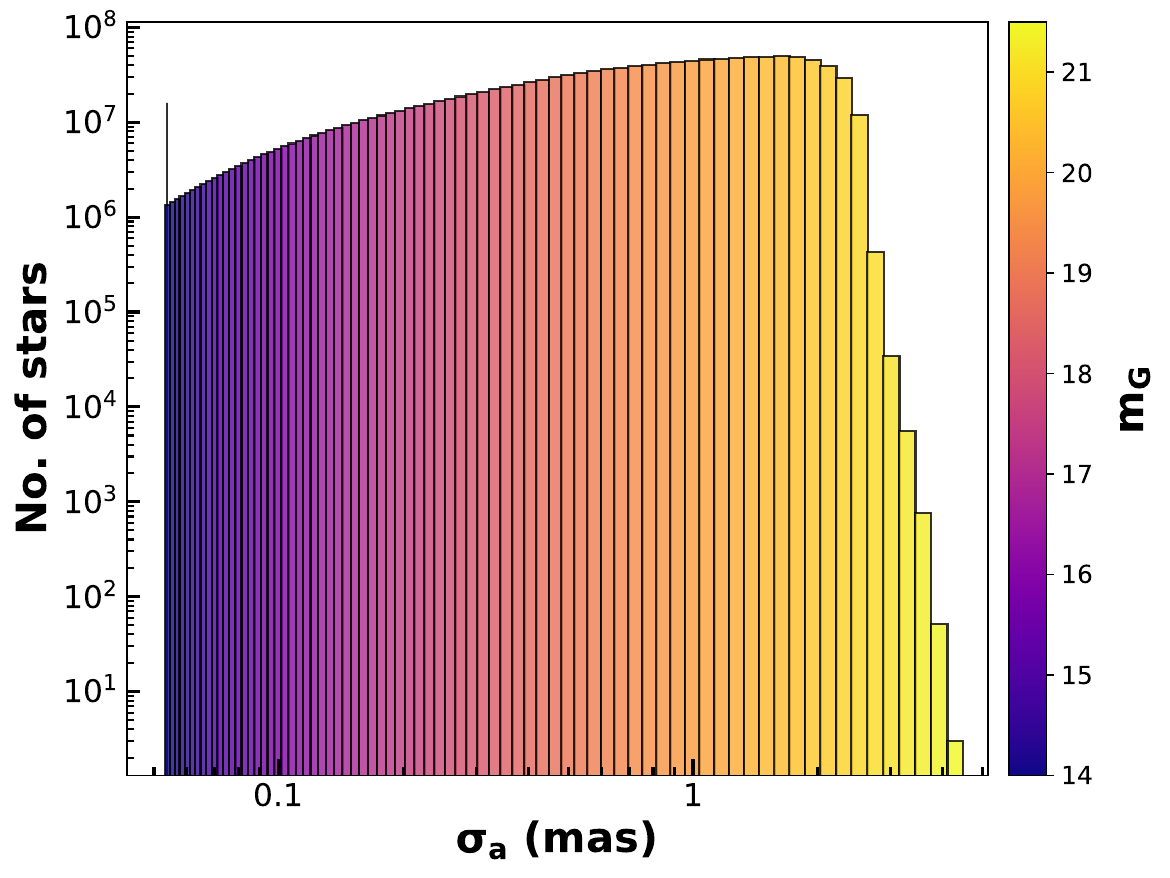}
			\end{center}
			\caption{In the \textit{left plot}, we show the empirical relation between astrometric uncertainty per epoch along the scanning direction and G-band apparent magnitude of a star for Gaia DR3\,\cite{Gaia:2023fqm}. This will be used to obtain the astrometric precision for a star of a given magnitude in the catalogue and to determine the impact parameter of the AML event for the star. In the \textit{right plot}, the distribution of Gaia DR3 stars over the typical astrometric uncertainty per epoch is shown.}
			\label{fig:sigma_a}
		\end{figure}

The per-epoch astrometric uncertainty of Gaia depends on the $G$-band magnitude of the source and is empirically modeled as~\cite{2018A&A...616A...2L, 2020A&A...640A..83K},
\begin{align}
\label{eq:Asigma}
    \sigma_{a} = & \frac{\sqrt{-2.031 + 680.766 z + 32.732 z^2}\times7.75+100}{\sqrt{N_\textrm{CCD}}} \,\mu\textrm{as},
\end{align}
where
\begin{equation}
    z = 10^{(0.4(\textrm{max}(m_G,14) - 15))}.
\end{equation}
where $N_\textrm{CCD} = 18$ is the number of CCD transits per epoch as the star crosses the focal plane along the scanning direction. The above expression gives us the expected uncertainty in the centroid position per epoch for a given $G$-band magnitude and will be instrumental in defining the detection threshold for AML signals from  EADOs.

The left panel of Fig.\,\ref {fig:sigma_a} shows this empirical relation of per-epoch astrometric uncertainty along the scanning direction and G-band magnitude for Gaia DR3 catalogue\,\cite{Gaia:2023fqm}. This relationship shows how astrometric precision systematically degrades for fainter objects, where fainter stars (large $m_G$) exhibit larger positional uncertainties along the scan direction. This empirical relation is used to determine the AML accessible impact parameter for a star with a given magnitude present in the Gaia DR3 catalogue\,\cite{Gaia:2023fqm}. 

In the right panel of the Fig.\,\ref {fig:sigma_a}, we show the distribution of the stars in the Gaia DR3\,\cite{Gaia:2023fqm,Lindegren_2021} catalogue over the astrometric precision per epoch. This distribution peaks around $\sim$1 mas, representing the typical accuracy attained for the majority of sources. This per-epoch precision is crucial to determine the detectability threshold for AML events, as it sets the minimum deflection amplitude required for reliable signal detection above the measured noise.

Although Gaia's time-series astrometric measurements have not yet been made publicly available, their analysis has yielded the derivation of a five-parameter astrometric solution for over 1.5 billion stars in the most recent data release (Gaia DR3)\,\cite{Gaia:2023fqm,Lindegren_2021}. These five parameters include the reference position of the sources at epoch $t_{\rm ref}=2016.0$ ($\alpha_{\rm ref}, \delta_{\rm ref}$), its proper motion ($\mu_\alpha, \mu_\delta$), and annual parallax ($\omega$). These parameters were derived using approximately 34 months (July 25, 2014 to May 28, 2017) of data and are publicly available, enabling us to access a high-precision "snapshot" of the Milky Way. We utilize this Gaia DR3 catalogue as the most precise static map of the stellar content of the Galaxy to estimate the rate and distribution of AML signals induced by  EADOs.

		\begin{figure}[]
			\begin{center}\includegraphics[scale=0.29]{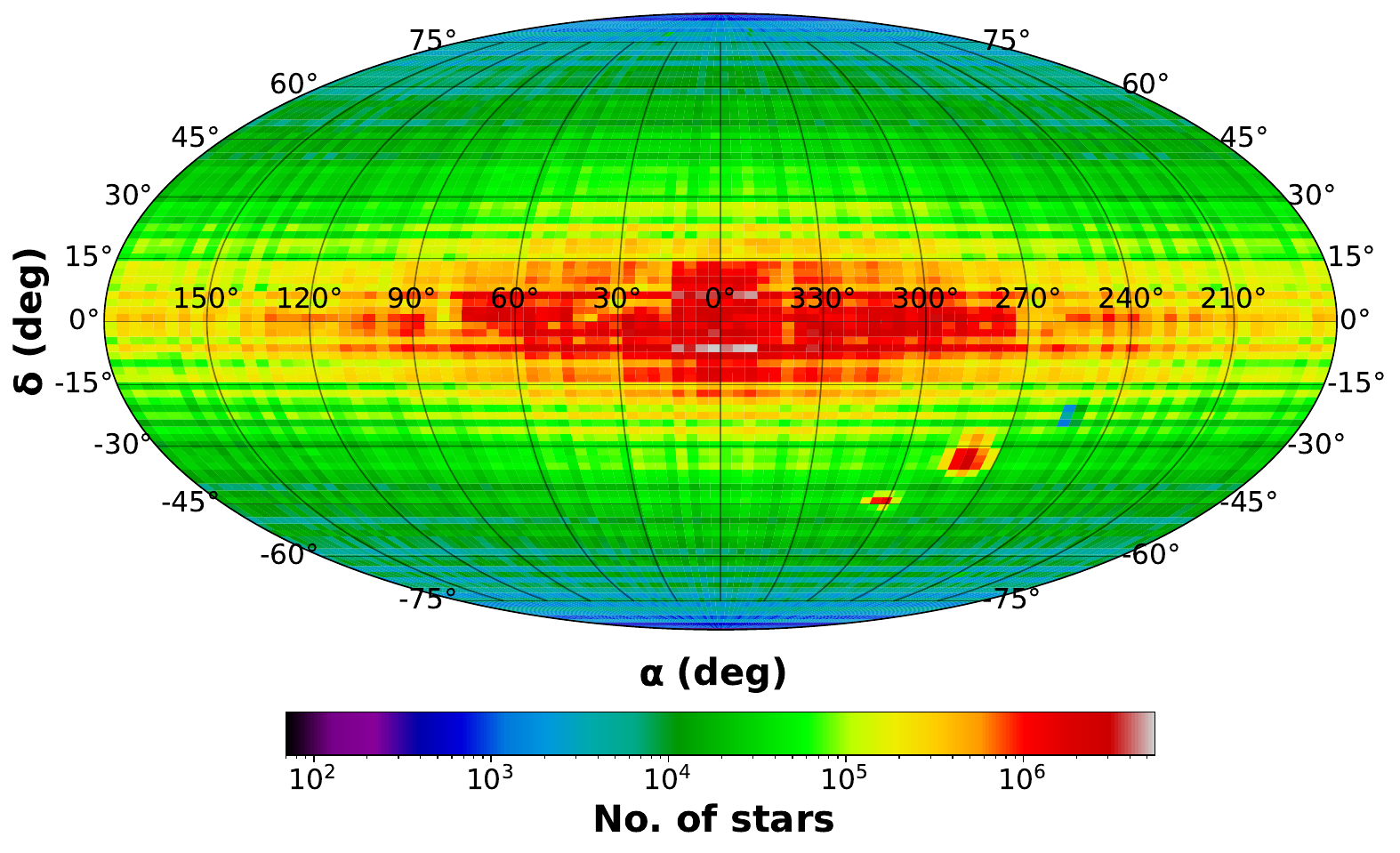}
                \includegraphics[scale=0.33]{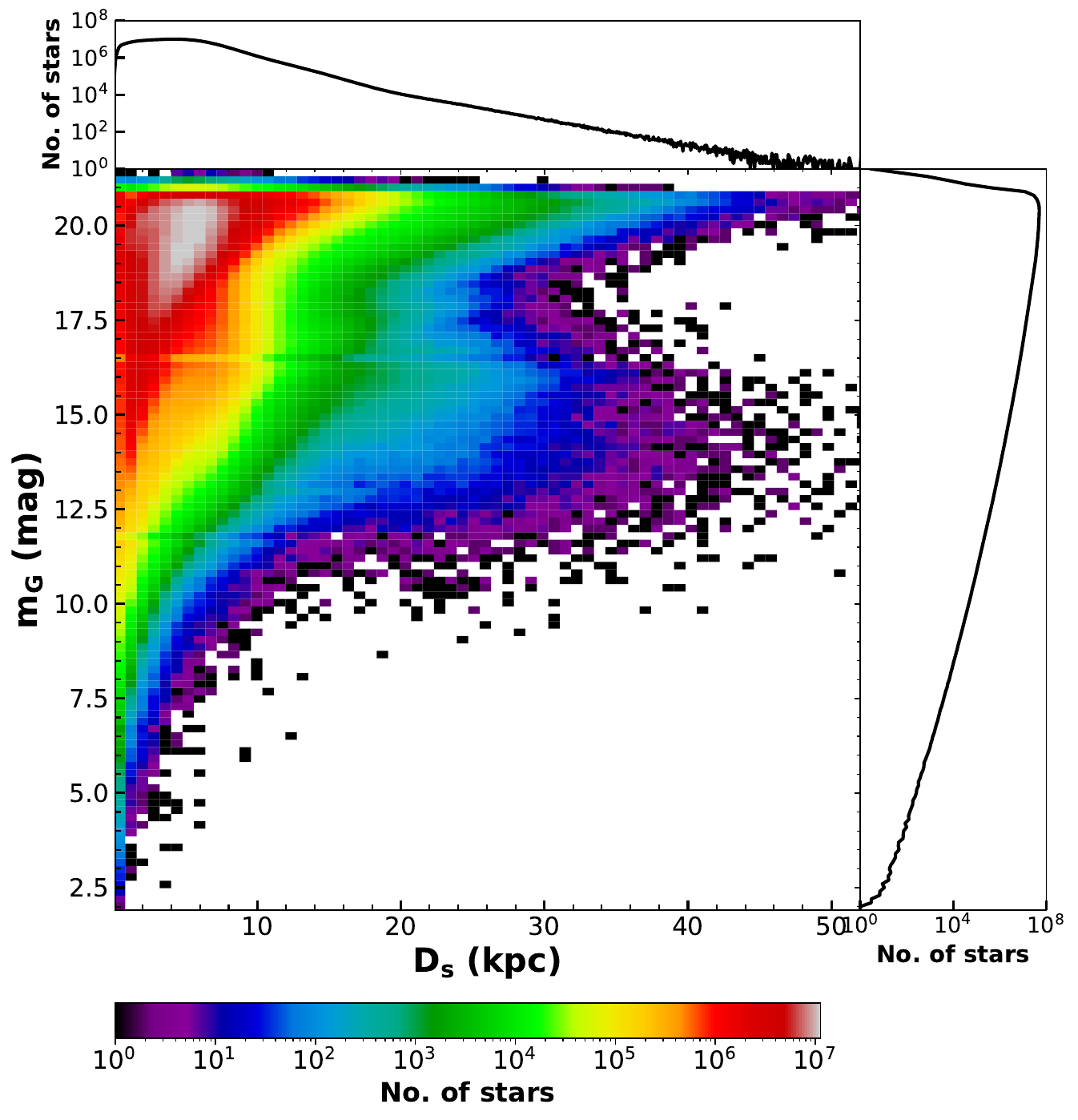}                 
			\end{center}
			\caption{In the \textit{left panel}, it shows the distribution of star in the Gaia DR3\,\cite{Gaia:2023fqm} catalogue over Galactic longitude ($\alpha$) and latitude ($\delta$). It clearly reveals that the structure of the Milky Way consists of the Galactic disk and the bulge. The Large Magellanic Cloud and Small Magellanic Cloud can also be seen in the south-east quadrant. We use this gridded catalogue as the most precise model of the Milky Way to estimate the AML event rate due to EADOs  potentially observable at Gaia. In the \textit{right panel}, it shows the distribution of Gaia DR3 sources in the distance-magnitude plane. The 1D projection of the distribution is also shown on the two axes.}
			\label{fig:griddedGDR3}
		\end{figure}
Gaia does not directly measure stellar distances for all sources; however, it infers them through parallax measurements. For stars with high signal-to-noise parallaxes, the distances estimated are robust. However, for fainter or more distant stars where parallax error becomes significant, Bayesian distance estimation methods can be applied~\cite{2015PASP..127..994B, 2016ApJ...832..137A}, incorporating priors based on Galactic structure models. In our analysis, we will use these distances estimated by Bailer-Jones et al.~\cite{2021AJ....161..147B} when calculating the AML signal rate due to EADOs.

The left panel of Fig.~\ref{fig:griddedGDR3}, shows the all sky 2D projection of the distribution of stars as a function of Galactic longitude ($\alpha$) and latitude ($\delta$) in Gaia DR3 catalogue\,\cite{Gaia:2023fqm}. It displays the distribution of stars across the sky in $100\times100$ degree bins, clearly showing the Galactic disk, bulge, Large Magellanic Cloud, and Small Magellanic Cloud components. This gridded stellar catalogue serves as the Galactic model for our AML event rate calculation.  

The right panel of the figure presents the 2D distribution of the stars in the catalogue on a distance-magnitude plane, and it also shows a 1D projection of the distribution on the two axes. The distribution reveals a concentration of stars at distances less than 8~kpc and represents the local Galactic disk population that dominates the high-precision astrometric sample. The sharp cut-off at the faint magnitudes ($>20.5$~mag) corresponds to the detection limit of Gaia, while the decrease in stellar density at large distances reflects both the decrease in stellar density with Galactocentric radius and the magnitude-limited nature of the survey. We use this catalogue to weigh the probability of lensing for each bin to obtain the final observable event rate with Gaia.

It is important to note that Gaia's ability to observe stars in the Galactic center is limited due to extreme source crowding and significant extinction in this direction. These effects reduce Gaia's completeness and astrometric accuracy towards the Galactic bulge. Thus, the Galactic bulge is not a promising place to search for AML signals of  EADOs using Gaia's astrometry. However, we will include it in our analysis and see this degradation of the AML signal toward the Galactic bulge in the lensing rate all-sky map.

\subsection{AML signal detection criteria and classifications}
\label{sec:AML_class}

In Sec.\,\ref{sec:AML_UDS_BS}, we demonstrated that exotic astrophysical dark objects, such as thin-wall Q-balls (modeled as UDS) and boson stars, generate distinctive AML signatures. While these signatures are theoretically well characterized, their astrophysical relevance depends critically on the observational capabilities of Gaia, as discussed in Sec.\,\ref{sec:gaia}. Building upon the insights gained in those sections, we now establish the criteria for detecting such AML events.

Foreground  EADOs can, in principle, be detected through astrometric deviations, assuming a simplified scenario of rectilinear stellar motion. The ability of Gaia to resolve such astrometric distortions is determined by three primary factors---(a) observation time of Gaia $t_{\text{\tiny obs}}$; (b) Gaia's intrinsic astrometric uncertainty, ($\sigma_{\text{\tiny a} }(m_{\text{\tiny G} })$) in the position measurement of a star;  (c)  observation cadence, or the rate at which Gaia collects data for a specific stellar trajectory $t_{\text{\tiny s}}$.

In our analysis, we consider two representative observation times for Gaia, $t_{\text{\tiny obs}}=\{5,10\}$ years. These observation times (e.g., 5 and 10 years) were chosen to align with the expected observational baselines of the time-series astrometric solutions that will be released in the forthcoming Gaia DR4 and DR5, respectively.\footnote{The forthcoming Gaia DR4 (expected late 2026) and DR5 (expected  after 2030) will provide the full time-series astrometric solutions based on 66 months and the full mission data ($\sim 10$ years), respectively. For more details: \url{https://www.cosmos.esa.int/web/gaia/release}} As described in Sec.\,\ref{sec:gaia}, Gaia's operational strategy involves scanning every region of the sky approximately every $t_{\text{\tiny s}} \sim 52.2$ days\,\cite{Verma:2022pym}. For each scan, the instrument's precision in measuring source positions is given by $\sigma_{\text{\tiny a} }(m_{\text{\tiny G} })$, where $m_{\text{\tiny G} }$ represents the star's G-band magnitude (see Sec.\,\ref{sec:gaia} for details).

Consequently, for Gaia to effectively detect a deviation in a star's observed trajectory from its unlensed rectilinear path, the absolute value of the astrometric shift must surpass the instrument's astrometric precision, expressed as,
\begin{equation}\label{eq:AML_lens_class_lens_crit}
   \abs{\Delta \theta_{\text{\tiny cent}}}\ge \sigma_{\text{\tiny a} }(m_{\text{\tiny G} })\,.
\end{equation}

Based on the astrometric shifts for uniform density sphere (UDS) and boson star (BS) lenses shown in Figs.\,\ref{fig:uds_centshift_vs_us_2d}, this detection criterion will be satisfied within an annulus $ \theta_{\text{\tiny S-}}<\theta_{\text{\tiny S}}<\theta_{\text{\tiny S+}}$ around the lens. The radii $\theta_{\text{\tiny S+}}$ and $\theta_{\text{\tiny S-}}$, delineate the boundaries where $\abs{\Delta \theta_{\text{\tiny cent}}(\theta_{\text{\tiny S+}})}=\abs{\Delta \theta_{\text{\tiny cent}}(\theta_{\text{\tiny S-}})}=\sigma_{\text{\tiny a} }(m_{\text{\tiny G} })$\,\cite{Verma:2022pym}.\footnote{Importantly, for certain extended lenses, no such annular region may exist where Eq.\,\ref{eq:AML_lens_class_lens_crit} is met. These lenses are therefore unable to induce sufficient deflection for Gaia detectability, resulting in no astrometric microlensing events.}

The outer radius of the annulus, $\theta_{\text{\tiny S+}}$, is a critical quantity for observational prospects of AML. As will be discussed, it sets the threshold impact parameter for the detection of an AML event and is therefore a key component of the event rate calculation. This threshold impact parameter, $\theta_{\text{\tiny S+}}$, is a function of the lens's internal structure ($\rho_{\text{\tiny L}}$, $M_{\text{\tiny L}}$, $R_{\text{\tiny L}}$), the overall lensing geometry (which sets the Einstein angle $\theta_{\text{\tiny E}}$), and the astrometric precision of Gaia. We numerically determine this quantity by solving the detection criterion, Eq.\,\eqref{eq:AML_lens_class_lens_crit}, using the specific AML signatures for EADOs derived in Sec.\,\ref{sec:aml_eds}. 

\begin{figure}
    \centering
    \includegraphics[scale=0.45]{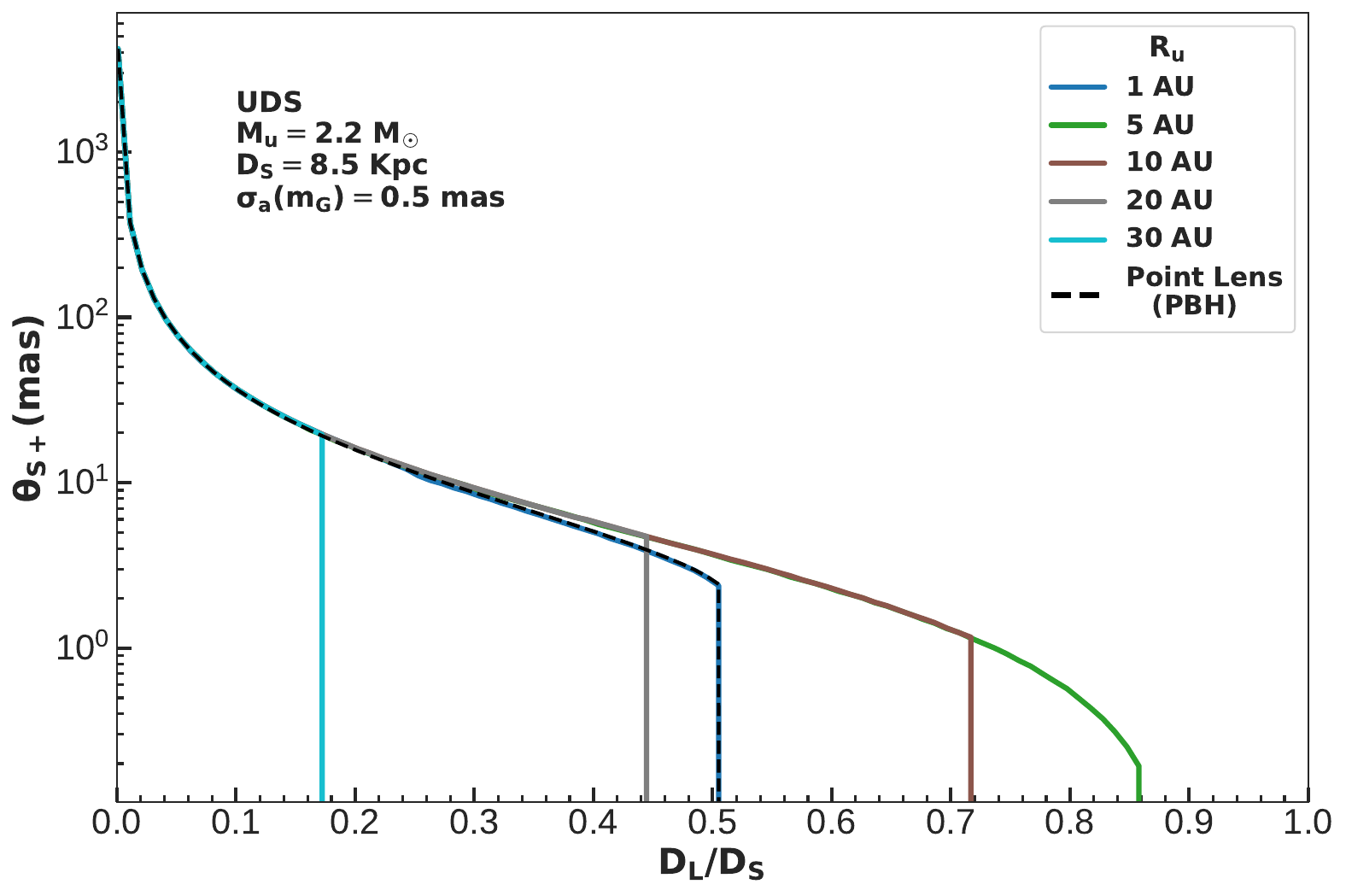}
    \caption{The plot shows variation of $\theta_{\text{S+}}$ with the lens distance $D_{\text{\tiny L}}/D_{\text{\tiny S}}$ for UDS lens for different radii. The results are shown for several radii, $R_{\text{\tiny u}}=\{1,5,10,20,30\}~\text{AU}$, represented by different curves. A lens mass of $M_{\text{\tiny u}}=2.2~\solarmass $, source distance of  $D_{\text{\tiny S}}=8.5~\text{kpc}$ and astrometric precision of $\sigma_{\text{\tiny a}}(m_{\text{\tiny G}})=0.5~\text{mas}$ are assumed. For chosen parameters, we find that unlike the point lens, the UDS lens with radii $R_{\text{\tiny u}}=(5,10)~\text{AU}$ can produce a detectable astrometric shift even when the lens is closer to the source. The point-lens approximation is included as a black dotted line. See text for details.}
    \label{fig:theta_illustration_0}
\end{figure}

In Fig.\,\ref{fig:theta_illustration_0}, we display how the threshold impact parameter, $\theta_{\text{S+}}$, changes as the function of lens distance $D_{\text{\tiny L}}/D_{\text{\tiny S}}$ for a UDS lens. We assume a lens mass of $M_{\text{\tiny u}} = 2.2~\solarmass$, a source distance of $D_{\text{\tiny S}} = 8.5~\text{kpc}$, and an astrometric precision of $\sigma_{\text{\tiny a}}(m_{\text{\tiny G}}) = 0.5~\text{mas}$ . The results are presented for several lens radii, $R_{\text{\tiny u}} = \{1,5,10,20,30\}~\text{AU}$ and a point lens, indicated by different curves. For these parameters, the point-lens case only produces a detectable shift ($ 0.5~\text{mas}$) when the lens is relatively close ($D_{\text{\tiny L}}\lesssim 0.5~D_{\text{\tiny S}}$). This is because for $D_{\text{\tiny L}} \gtrsim 0.5 ~D_{\text{\tiny S}}$, the maximum possible astrometric shift for a point lens, $\Delta \theta_{\text{C,max}} = \theta_{\text{\tiny E }}/(2\sqrt{2})$ (see Sec.\,\ref{app:mlrev}), falls below the $0.5~\text{mas}$ detection threshold. For smaller radii,  i.e., for $R_{\text{\tiny u}}=1~\text{AU}$, the curve matches the point lens case.

It is interesting to note that for UDS lens with radii $R_{\text{\tiny u}}=(5,10)~\text{AU}$, can still produce $\sigma_{\text{\tiny a}}(m_{\text{\tiny G}})=0.5~\text{mas}$ astrometric shift even for lens positions $D_{\text{\tiny L}}\gtrsim 0.5~D_{\text{\tiny S}}$. This is in stark contrast to the point-lens case, where the detectable signal vanishes in this regime. This arises because, since for the same mass, a UDS lens with radii $R_{\text{\tiny u}}/R_{\text{\tiny E}}\sim \mathcal{O}(1)$ can produce a larger centroid shift compared to a point lens (see, for instance, $R_{\text{\tiny u}}/R_{\text{\tiny E}}=(0.5,1,\sqrt{3/2})$ in the right panel of Fig.\,\ref{fig:uds_massfunc_vs_ui}).  This will lead to potentially more events detectable for Gaia due to EADOs than a point lens. However, this trend reverses as the radius is further increased. As shown in Fig.\,\ref{fig:uds_massfunc_vs_ui}, as the radius increases (see, for instance, $R_{\text{\tiny u}}/R_{\text{\tiny E}}=3$), the UDS lens becomes more dilute and therefore produces a smaller shift than the point lens. This is reflected in Fig.\,\ref{fig:theta_illustration_0}, as UDSs with radii $R_{\text{\tiny u}}=(20,30)~\text{AU}$  are unable to produce the $0.5~\text{mas}$ astrometric shift for  $D_{\text{\tiny L}}/D_{\text{\tiny S}}\gtrsim (0.44, 0.18)$ respectively.  As will be seen in subsequent sections, this dependence plays a crucial role in determining the event rate, leading to a decrease in the number of detectable events for larger EADOs.

Another key observable is the duration of the AML event, $t_{\text{\tiny e}}$, which is defined as the time the source spends within this detectable annular region for a given trajectory.  This event duration, $t_{\text{\tiny e}}$ is contingent upon the specific impact parameter of the trajectory. For convenience, we may define the mean event duration as the average of $t_{\text{\tiny e}}$ over all trajectories crossing the annulus\,\cite{Verma:2022pym}. This mean event duration is given by $\bar{t}_{\text{\tiny e}} = \bar{l} D_{\text{\tiny S}}/v_{\text{\tiny S,T}}$, where $\bar{l}$ is the average angular chord length within the annular region and $v_{\text{\tiny S,T}}$ is the tangential velocity of source along the lens plane. Hence, the average AML event timescale\footnote{This is different from the photometric event duration, which is characterized by the Einstein  time, i.e.   $t_{\text{\tiny E}}\sim \theta_{\text{\tiny E}}D_{\text{\tiny S}}/v_{\text{\tiny S,T}}$} is given by

\begin{equation}\label{eq:AML_lens_t_e}
  \bar{t}_{\text{\tiny e}}=\frac{\pi\left(\theta_{\text{\tiny S+}}^2-\theta_{\text{\tiny S-}}^2\right)}{2\theta_{\text{\tiny S+}}v_{\text{\tiny S,T}}}D_{\text{\tiny S}}\;.
\end{equation}
For $\theta_{\text{\tiny S+}}\gg \theta_{\text{\tiny S-}}$, one may simply take the mean event duration as $\bar{t}_{\text{\tiny e}}\sim \pi\theta_{\text{\tiny S+}}D_{\text{\tiny S}}/(2 v_{\text{\tiny S,T}})$.
Based on this event duration, AML events can be classified into three distinct categories,\cite{Verma:2022pym}:

\begin{itemize}
    \item \textbf{Short duration lensing events (SDLEs)} -- These events have durations ${t}_{\text{\tiny e}}$ shorter than Gaia’s sampling interval $t_{\text{\tiny S}}$ i.e. ${t}_{\text{\tiny e}}<t_{\text{\tiny S}}$. As a result, Gaia is likely to miss the majority of SDLEs, since the detectable deflection above the threshold typically occurs between successive observations. Nevertheless, in rare cases, a single observation may coincidentally capture a significant deviation if the star happens to be lensed at the exact time of measurement.
    \item \textbf{Intermediate duration lensing events (IDLEs)} -- These events are determined by event durations that fall between Gaia’s sampling interval and the total observational time $t_{\text{\tiny S}}< {t}_{\text{\tiny e}}  < t_{\text{\tiny obs}}$. IDLEs are typically characterized by trajectories that, for the most part, resemble rectilinear motion. However, at a few sampled epochs, the trajectory exhibits detectable deviations with $\Delta \theta_{\text{\tiny cent}} > \sigma_{\text{\tiny a} }(m_{\text{\tiny G} })$.
    
    \item \textbf{Long duration lensing events (LDLEs)} --   These events are characterized by durations longer than the observational timescale, i.e., $t_{\text{ \tiny e}} > t_{\text{\tiny obs}}$. In the case of long-duration lensing events (LDLEs), almost the entire observed trajectory is affected by lensing distortions. In the extreme limit of very long durations, Gaia may only capture a partial segment of the lensed path.
\end{itemize}

Building on our findings, Fig.\,\ref{fig:Gaia_sens} provides a heuristic estimate for mass and size ranges of exotic astrophysical dark objects that can be probed with Gaia\,\cite{Gaia:2016zol}. The blue-shaded region marks Gaia’s sensitivity to such EADOs through astrometric microlensing. We estimate that Gaia can access  EADOs with masses in the range $M_{\text{\tiny L}} \sim 10^{-2}-10^7\,\solarmass$ for radii $R_{\text{\tiny L}} \lesssim 10^8\,\text{AU}$. For comparison, the PML sensitivities of EROS\,\cite{Croon:2018ybs}, OGLE\,\cite{Croon:2018ybs}, and HSC-Subaru\,\cite{Croon:2020ouk} are also shown. These surveys probe masses $M_{\text{\tiny L}} \lesssim 10^2\,\solarmass$ for radii $R_{\text{\tiny L}} \lesssim 10\,\text{AU}$, highlighting their complementarity with Gaia in constraining the EADO's population.

In Fig.\,\ref{fig:Gaia_sens}, the lower bound on the EADO's radius for  a given mass is set by the corresponding Schwarzschild radius, $R_{\text{\tiny Sch}}$, which scales as $\sim M_{\text{\tiny L}}$. The upper bound on the radius is determined by Gaia’s threshold for detecting dilute lenses. For $R_{\text{\tiny L}}/R_{\text{\tiny E}} \gg 1$, the outer radius of annulur region scales with astrometric shifts as $\theta_{\text{\tiny S+}}\sim \theta_{\text{\tiny E}}^2/\Delta \theta_{\text{\tiny cent}}$. Also, the maximum centroid shift occurs just outside the physical extent of the lens, i.e., when $\theta_{\text{\tiny S,+}}/\theta_{\text{\tiny E}} \approx R_{\text{\tiny L}}/R_{\text{\tiny E}}$, giving $\Delta \theta_{\text{\tiny cent,max}} \approx R_{\text{\tiny E}}^2/( D_{\text{\tiny L}}R_{\text{\tiny L}})$. Using Eq.\,\ref{eq:AML_lens_class_lens_crit}, the Gaia's sensitivity to extended EADOs is limited to lenses that can produce a maximum astrometric shift greater than the instrument's minimum uncertainty $\sigma_{\text{\tiny a,min}}$. The maximum lens radius that satisfies this detection threshold, can be estimated as $R_{\text{\tiny L,max}} \approx R_{\text{\tiny E}}^2/(D_{\text{\tiny L}}  \sigma_{\text{\tiny a,min}})$ which scales as $\sim M_{\text{\tiny L}}^2$. 

We take $D_{\text{\tiny S}}=8.5$ kpc, and $D_{\text{\tiny L}}=D_{\text{\tiny S}}/2$ and $\sigma_{\text{\tiny a,min}}=0.05~\text{mas}$ for estimating maximum radius $R_{\text{\tiny L,max}}$ that can be probed by Gaia for a given lens mass $R_{\text{\tiny L}}$. Additionally, the minimum mass that Gaia can probe (for IDLE type events) is estimated by setting the event duration equal to the minimum sampling time, $\bar{t}_{\text{\tiny e}}=t_{\text{\tiny s}}$. Substituting this condition into Eq.\,\eqref{eq:AML_lens_t_e} (assuming  $\theta_{\text{\tiny S+}}\gg\theta_{\text{\tiny S-}}$), we find $M_{\text{\tiny L,min}}\sim c^2 t_{\text{\tiny s}}v_{\text{\tiny S,t}} \sigma_{\text{\tiny a,min}}/(2\pi G (D_{\text{\tiny S}}-D_{\text{\tiny L}})/D_{\text{\tiny S}})\approx 10^{-2}~\solarmass$. The existence of the LDLE class ($\bar{t}_{\text{\tiny e}} > t_{\text{\tiny obs}}$) implies that there is no intrinsic upper mass limit for Gaia's sensitivity. A lens with a mass exceeding this boundary would simply be observed as a long-duration event.  The PML sensitivity estimates for EROS, OGLE, and HSC-Subaru are based on the results of the EADOs analysis presented in \cite{Ansari:2023cay}.

In summary, this section established the detectability criteria for AML signals induced by  EADOs and introduced a classification scheme for such events based on their characteristic durations.  We also discussed the astrometric precision of Gaia and utilized the Gaia DR3 catalogue\,\cite{Gaia:2023fqm} to account for the distribution and properties of the stellar population. The detectability criteria link the theoretical lensing signatures of UDS and BS models to Gaia’s instrumental capabilities. Together, these elements set the foundation for quantifying the actual detection potential of dark lenses. In the following section, we build upon this framework to evaluate the lensing probability and estimate the expected number of AML events detectable by Gaia.

\section{AML lensing probability and event number due to  EADOs }\label{sec:aml_lens_prob_no}
To quantify Gaia’s sensitivity to exotic astrophysical dark objects, we now turn to calculating the probability that any given source star produces a detectable astrometric microlensing signal. This probability depends on both the intrinsic properties of the lens and its spatial distribution, along with the location and the apparent brightness of the given source star in the Milky Way under consideration. Specifically, we will define the total probability, $P_{\text{\tiny Star}}$, that a given source star undergoes a detectable AML event as the sum of the individual probabilities for the three event classes—short, intermediate, and long duration—introduced above. We then calculate this $P_{\text{\tiny Star}}$ for each star in the Gaia DR3 catalogue\,\cite{Gaia_2021} to estimate the number of AML events observed by Gaia.  

\subsection{Lensing probability  for extended lens }\label{sec:aml_lens_prob_no_lens_prob}

\begin{figure}
    \centering
    
    \includegraphics[scale=1]{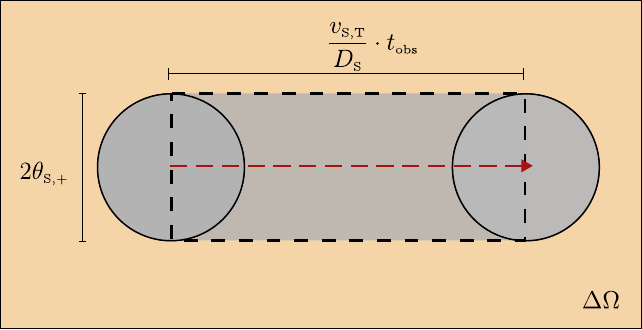}
    \caption{Illustration of the estimation of the conditional probability for a star in the Gaia DR3 catalogue to undergo AML due to a foreground lens. A source star (origin of red dashed line) moves with relative transverse velocity $v_{\text{\tiny S,T}}$ for an observation time $t_{\text{\tiny obs}}$, traversing an angular distance $v_{\text{\tiny S,T}}t_{\text{\tiny obs}}/D_{\text{\tiny S}}$. A detectable AML event occurs if the source passes within an angular distance $\theta_{\text{S+}}$ of a foreground lens.  This condition is equivalent to the lens lying within the gray rectangle swept out by the source's detection radius. This rectangular area has an angular width of $2\theta_{\text{\tiny {S+}}}$ and a length of $v_{\text{\tiny S,T}}t_{\text{\tiny obs}}/D_{\text{\tiny S}}$. The probability $P_{\text{\tiny Star}}$ is proportional to this area multiplied by the number density of lenses.}
    \label{fig:prob_illustration}
\end{figure}

We will now estimate the probability of a source star undergoing a Gaia-detectable astrometric microlensing event, thin-wall Q-balls, and boson stars. The astrometric shifts induced by these lenses were discussed in Sec.\,\ref{sec:aml_eds}, while the corresponding Gaia detectability criteria were specified in Eq.\,\eqref{eq:AML_lens_class_lens_crit}. We adapt the methodology outlined in\,\cite{Verma:2022pym} for our analysis.

For each lens model, the system is characterized by its total mass, radius, and dark matter fraction, denoted as $M_{\text{\tiny L}}$, $R_{\text{\tiny L}}$, and $f_{\text{\tiny L}}$, respectively. Our goal is then to compute $P_{\text{\tiny Star}}$, the probability that a given source star in Gaia DR3 (see sec.\,\ref{sec:gaia} for details)---specified by its apparent G-band magnitude $m_{\text{\tiny G}}$ and Galactic coordinates $(D_{\text{\tiny S}}, \alpha, \delta)$—undergoes a detectable lensing event caused by the distribution of such extended lenses in the Milky Way halo.

The total probability $P_{\text{\tiny Star}}$ is obtained by first evaluating the conditional probability $p_{\text{\tiny AML}}^{\text{\tiny c}}$ for a given star located within a patch of the sky subtending a solid angle $\Delta \Omega$, to produce detectable AML event due to a lens situated at a distance $D_\text{\tiny L}$ along the line of sight.  For simplicity, we assume that the source star is moving in a rectilinear motion with the velocity $v_{\text{\tiny S,T}}$ tangential to the lens plane as illustrated in Fig.\,\ref{fig:prob_illustration}.   As the source star (origin of red dashed line) moves across the patch $\Delta \Omega$ over the total observation time $t_{\text{\tiny obs}}$, it traverses an angular distance $v_{\text{\tiny S,T}}t_{\text{\tiny obs}}/D_{\text{\tiny S}}$ on the sky. A detectable AML event occurs if the source's path passes within an angular distance $\theta_{\text{\tiny S+}}$ of a lens, as this is the region where the astrometric shift exceeds the detection threshold, $\sigma_{\text{\tiny a}}(m_{\text{\tiny G}})$.

For IDLEs, events where the entire lensed trajectory can be observed, this conditional probability is simply the ratio of the area of an imaginary rectangle (shown as a gray rectangular region\footnote{  We neglect the semicircular end-caps of this detection region. This is a valid approximation because the angular length of the path ($v_{\text{\tiny S,T}}t_{\text{\tiny obs}}/D_{\text{\tiny S}}$) is typically much larger than its width ($2\theta_{\text{\tiny S+}}$), making the contribution from the end-caps negligible.} in Fig.\,\ref{fig:prob_illustration} with angular width $2\theta_{\text{\tiny S+}}$ and angular length $v_{\text{\tiny S,T}}t_{\text{\tiny obs}}/D_{\text{\tiny S}}$. Therefore, for IDLEs events the conditional probability $p_{\text{\tiny AML}}^{\text{\tiny c}} = 2\theta_{\text{\tiny S+}} v_{\text{\tiny S,T}}t_{\text{\tiny obs}}/(D_{\text{\tiny S}}\Delta \Omega)$. However, since SDLEs and LDLEs have different event durations and sampling characteristics, their conditional probability expressions are correspondingly different. Additionally, when the lenses have a velocity distribution, the conditional probability $p_{\text{\tiny AML}}^{\text{\tiny c}}$ must be averaged over that distribution. A detailed derivation of this calculation is provided in App.\,\ref{app:lens_prob}, where the averaged conditional probability $\expval{p_{\text{\tiny AML}}^{\text{\tiny c}}}$ is obtained as Eq.\,\ref{eq:app_AML_lens_cond_prob_bolt}.

This conditional probability is then integrated over all lenses along the line of sight. At each lens distance interval $(D_{\text{\tiny L}}, D_{\text{\tiny L}}+\Delta D_{\text{\tiny L}})$, the contribution is weighted by the number of lenses contained within the angular window $\Delta \Omega$. Summing these contributions over all possible lens positions from $0$ to $D_{\text{\tiny S}}$ yields
\begin{equation}\label{eq:AML_lens_tot_prob}
		P_{\text{\tiny Star,L}} =  \int_{0}^{D_{\text{\tiny S}}} dD_{\text{\tiny L}} D_{\text{\tiny L}}^2  \Delta \Omega \frac{f_{\text{\tiny L }}\rho_{\text{\tiny DM}}(D_{\text{\tiny L}},\alpha,\delta)}{M_{\text{\tiny L }}} \expval{p_{\text{\tiny AML}}^{\text{\tiny c}}}\;.
\end{equation}

\indent We consider that the dark matter (DM) density within the Milky Way halo follows the standard spherically symmetric Navarro–Frenk–White (NFW) profile\,\cite{Navarro:1995iw}, (see Eq.\,\eqref{eq:app_AML_lens_NFW}). Furthermore, the lens is assumed to follow a Maxwell-Boltzmann velocity distribution (see Eq.\,\eqref{eq:app_max_dist}).

\begin{table}
    	\centering
    	\begin{tabular}{ |p{3cm}|p{2.5cm}|p{2.5cm}| p{1.5cm}| }
    		\hline
    		Parameters& min value &max value& grid size\\
    		\hline
    		$\alpha$ & 0 deg & 360 deg&50\\
    		$\delta$ & -90 deg & 90 deg&50\\
    		$D_s$& 0.5 kpc& 20.5 kpc&50\\
    		$m_G$ & 11.4 & 21.4&50\\
    		\hline
    	\end{tabular}
    	\caption{This table specifies the parameter ranges and the total number of sampling points that define a four-dimensional grid in the source parameter space $(D_{\text{\tiny S}},\alpha,\delta,~m_{\text{\tiny G}})$. We evaluate $P_{\text{\tiny Star}}$ at each of the $50^4$ uniformly distributed grid points along respective axes.}
    	\label{tab:gaia_data_binned}
    \end{table}

The total probability that a given star undergoes an AML event, $P_{\text{\tiny Star}}$ (see Eq.\,\eqref{eq:AML_lens_tot_prob}), is sensitive to the lens model through its dependence on $\expval{p_{\text{\tiny AML}}^{\text{\tiny c}}}$. In particular, $\expval{p_{\text{\tiny AML}}^{\text{\tiny c}}}$ depends on the total lens mass $M_{\text{\tiny  L }}$, radius $R_{\text{\tiny L }}$, and its density distribution via the AML impact parameter $\theta_{\text{\tiny S,+}}$. Additionally, $P_{\text{\tiny Star}}$ also depends upon the parameters of the source star i.e. $(D_{\text{\tiny S}},\alpha,\delta, m_{\text{\tiny G}})$, and the population of lens along the line of sight $f_{\text{\tiny L }}\rho_{\text{\tiny DM}}(D_{\text{\tiny L}},\alpha,\delta)/M_{\text{\tiny L}} $.

\begin{figure}
    \centering
    \includegraphics[scale=0.375]{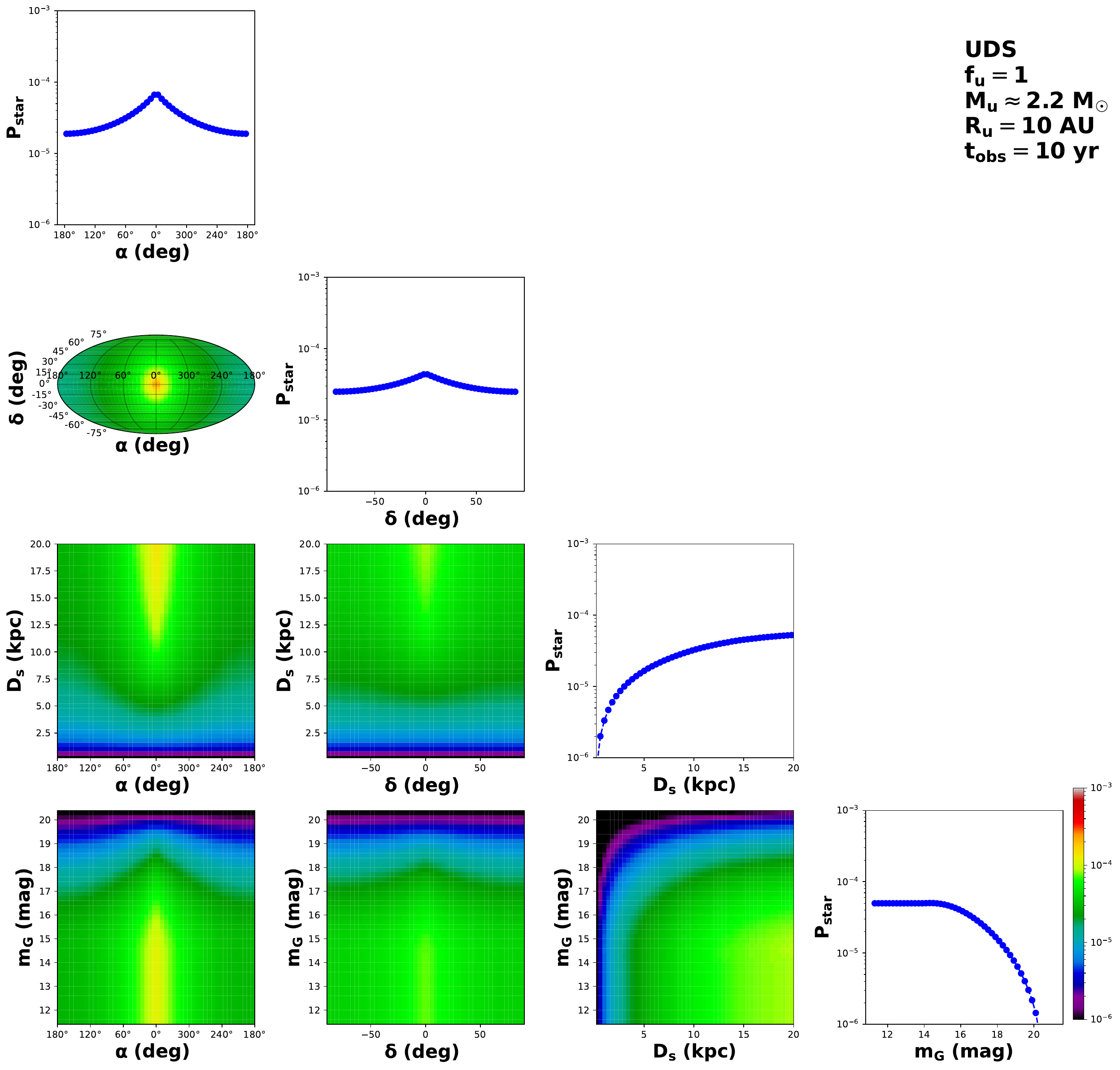}
    \caption{  The corner plot shows differential distribution of probability $P_{\text{\tiny Star}}$ for a star at position $(D_s,\alpha,\delta)$ with apparent magnitude $m_G$ to undergo a detectable AML event from a UDS lens by Gaia over an observational time ($t_\textrm{obs}=10~\text{yr}$). The example shown corresponds to a UDS lens with parameters $(f_{\text{\tiny u}},M_{\text{\tiny u}},R_{\text{\tiny u}})=(1,2.2~\solarmass,10~\text{AU})$. As we shall see later, this parameter combination for EADOs produces the maximum number of  AML events. The 2D histogram plots (located at off-diagonal position) display $P_{\text{\tiny Star}}$ as a function of two source star parameters chosen from $(D_s,\alpha,\delta,m_G)$, while averaging uniformly over the remaining two parameters. The source parameters are obtained on the grid defined in Tab.\,\ref{tab:gaia_data_binned}. Additionally, the 1D histogram (shown at diagonal position) plots depict the dependence of probability $P_{\text{\tiny Star}}$ on a single parameter, averaged over the other three. In Sec.\,\ref{sec:aml_lens_prob_no_no}, this probability distribution is convolved with the Gaia stellar catalogue to estimate the expected number of detectable events. Further details are provided in the text.}
    \label{fig:uds_prob_dist_plot}
\end{figure}

Given these stellar parameters i.e. $(D_{\text{\tiny S}},\alpha,\delta, m_{\text{\tiny G}})$, the probability $P_{\text{\tiny Star}}$ is obtained by numerically evaluating the line-of-sight integral in Eq.\,\eqref{eq:AML_lens_tot_prob}, which incorporates the conditional probability (see Eq.\,\eqref{eq:app_AML_lens_cond_prob_bolt} and \eqref{eq:app_AML_lens_cond_prob_bolt_cut}) together with the NFW dark matter density profile (see Eq.\,\eqref{eq:app_AML_lens_NFW}). This procedure yields the lensing probability for any individual star in the Gaia catalogue. In principle, one could repeat this calculation for each star in the Gaia catalogue to estimate the total expected number of AML events that Gaia will observe.  However, performing such an exhaustive star-by-star computation would be prohibitively expensive in terms of computational resources.

To make the computation of $P_{\text{\tiny Star}}$ across the Gaia catalogue tractable, we adopt a grid-based approach in the source parameter space $(D_{\text{\tiny S}}, \alpha, \delta, m_{\text{\tiny G}})$. Specifically, we pre-compute $P_{\text{\tiny Star}}$ on a four-dimensional grid spanning the relevant parameter ranges (see Table \ref{tab:gaia_data_binned}) with a resolution of $50^4$ uniformly spaced points. This grid-based strategy enables us to efficiently evaluate $P_{\text{\tiny Star}}$ for different lens hypotheses, characterized by $(M_{\text{\tiny L}}, R_{\text{\tiny L}}, f_{\text{\tiny L}})$. For each star in the Gaia catalogue, the probability is then obtained by interpolating from the nearest grid point. This framework allows us to systematically and efficiently explore a broad range of UDS and BS lens parameters without too much loss of accuracy.

We have compiled the results for the total $P_{\text{\tiny Star}}$ by summing contributions from SDLEs, IDLEs, and LDLEs across our parameter grid, and for UDS lens it is displayed in Fig.\,\ref{fig:uds_prob_dist_plot} (The corresponding plot for the BS lens is shown later in Fig. \ref{fig:bs_prob_dist_plot} in App.\,\ref{app:sup_result}). These figures specifically showcase results for lens mass $M_{\text{\tiny L }}=2.2\,\solarmass$, radius $R_{\text{\tiny L }}=10~\text{AU}$, and dark matter density fraction of $f_{\text{\tiny L }}=1$. As we shall later see, EADOs with these parameters produce the maximum number of  AML events for both lens models. To better visualize this probability, We have presented its distribution over the source property parameters given in Table \ref{tab:gaia_data_binned}. For instance, Fig.\,\ref{fig:uds_prob_dist_plot} presents 2D histogram plots (shown at the off-diagonal positions) illustrating the distribution of $P_{\text{\tiny Star}}$ as a function of two source star parameters selected from $(D_s,\alpha,\delta,m_G)$, while averaging uniformly over the remaining two parameters.  Furthermore, the 1D histograms (displayed at diagonal position) show how $P_{\text{\tiny Star}}$ varies with a single parameter, averaged over the remaining three. 

In Figs.\,\ref{fig:uds_prob_dist_plot}, and \ref{fig:bs_prob_dist_plot}, we find that both UDS and BS cases exhibit qualitatively similar trends, with approximately similar values of the probability $P_{\text{\tiny Star}}$. As seen in Sec.\,\ref{sec:AML_UDS_BS}, this behavior stems from the fact that UDS and BS lenses produce nearly identical astrometric shifts over most impact parameters, differing appreciably only in caustic crossing events at small impact parameters. Although their total probabilities are comparable, the distinction becomes evident for low-impact-parameter AML events (see Fig.\,\ref{fig:uds_centshift_vs_us_2d}).

Focusing on the 2D histogram plot depicting $P_{\text{\tiny Star}}$ as a function of galactic coordinate $\alpha$ and $\delta$, we find that $P_{\text{\tiny Star}}$ reaches its maximum toward the Galactic Center (i.e., $\alpha=\delta=0^\circ$) and decreases with increasing angular separation from the galactic center. This is also observed in 1D histogram plot showing $P_{\text{\tiny Star}}$ vs $\alpha$ and $P_{\text{\tiny Star}} $ vs $\delta$. This is expected, as the line of sight in this direction contains a higher population of potential lenses, and this population decreases as the line of sight moves away from the galactic center. Moreover, the lensing probability remains relatively flat with respect to the apparent stellar magnitude $m_{\text{\tiny G}}$ up to about 14 (see, for instance, 1D histogram plot showing $P_{\text{\tiny Star}}$ as a function of $m_{\text{\tiny G}}$). Beyond this threshold, the probability drops, reflecting Gaia’s astrometric uncertainty (see Fig.\,\ref{fig:sigma_a}), which increases substantially for fainter sources with $m_{\text{\tiny G}} \gtrsim 14$. 

Additionally, looking at 1D histogram showing the variation of  $P_{\text{\tiny Star}}$ with source distance $D_{\text{\tiny S }}$, we observe that  $P_{\text{\tiny Star}}$  initially increases steeply with the source distance $D_{\text{\tiny S}}$. However, the increase $P_{\text{\tiny Star}}$ becomes smaller after $D_{\text{\tiny S }}\sim 10$ kpc, near the galactic center. This happens because, initially, more lenses are encountered along longer lines of sight, but this increase gradually slows once we extend past the region where the lens population significantly contributes. Moreover, we have checked that $P_{\text{\tiny Star}}$ flattens around $D_{\text{\tiny S }}\sim 20$ kpc, corresponding to the distance approximately equal to the sum of the NFW scale radius and the distance between the Earth and the Galactic Center---at which the NFW profile begins to flatten and the lens population changes only weakly.

The calculated $P_{\text{\tiny Star}}$ values across different stellar and lens parameters are important. In the next section, we will weight  $P_{\text{\tiny Star}}$ by the stellar population distributed over the predefined grid (see Tab.\,\ref{tab:gaia_data_binned}). This computation yields the total number of AML events that Gaia is expected to detect for each lens model, thereby enabling quantitative predictions of Gaia’s observational yield.

\subsection{Number of lensing events observable by Gaia}\label{sec:aml_lens_prob_no_no}
\begin{figure}
    \centering
    \includegraphics[scale=0.375]{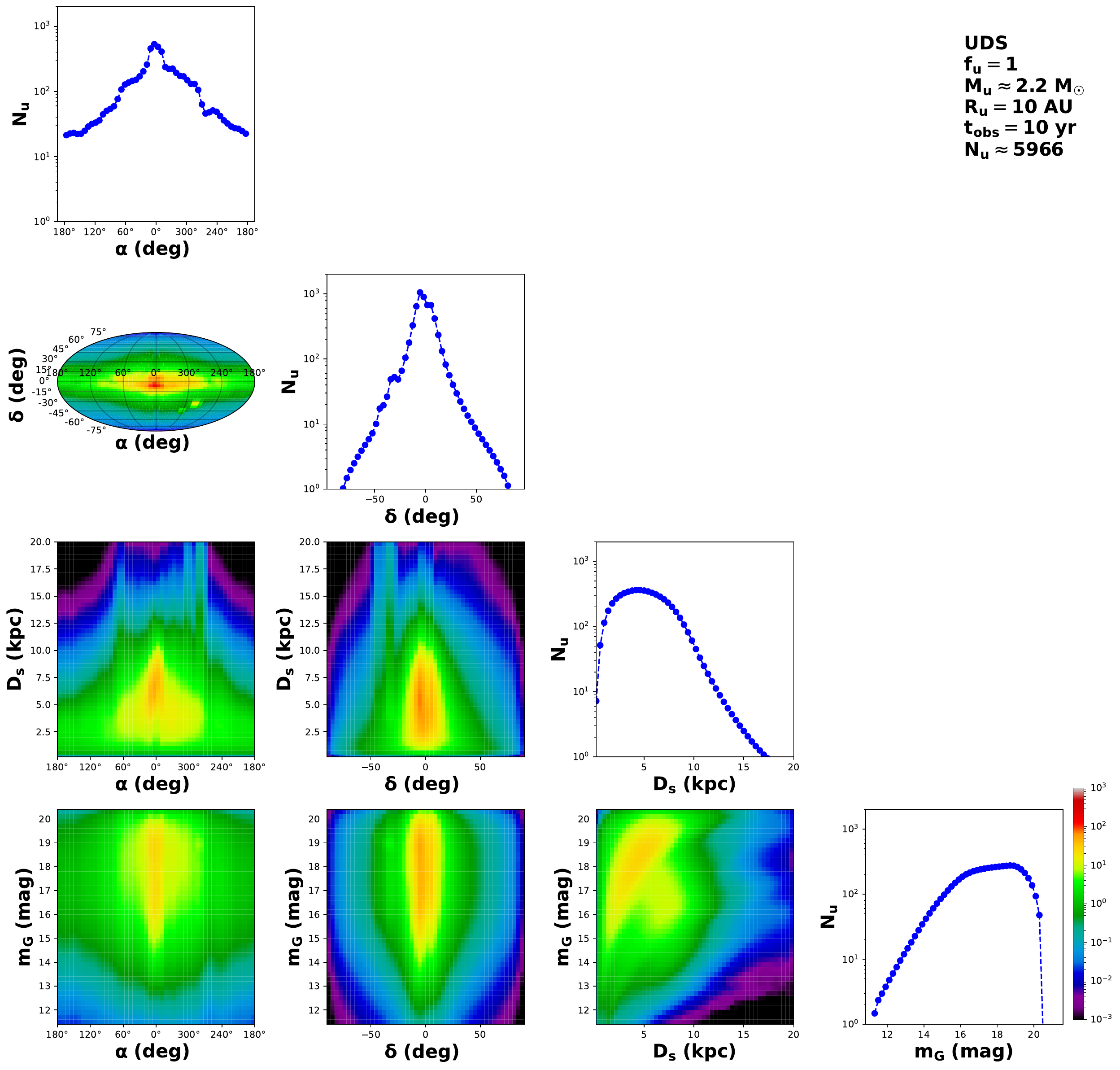}
    \caption{Distribution of expected number of AML events $N_{\text{\tiny u}}$ for stars in Gaia catalogue with parameter $(D_s,\alpha,\delta,m_G)$ (course-grained over grid data defined by Tab.\,\ref{tab:gaia_data_binned}) due to UDS lens with observational time $t_\textrm{obs} = 10~\text{yr}$. The results correspond to a uniform density sphere lens with parameters $(f_{\text{\tiny u}}, M_{\text{\tiny u}}, R_{\text{\tiny u}}) = (1, 2.2~\solarmass, 10~\text{AU})$. Each off-diagonal 2D histogram shows $N_{\text{\tiny u}}$ as a function of two source star parameters from $(D_s,\alpha,\delta,m_G)$, having been summed over the remaining two parameters. The diagonal 1D histogram plots present $N_{\text{\tiny u}}$ as a function of a single parameter, summed over the other three.  We expect roughly $6000$ AML events due to the UDS lens that Gaia may observe over a 10-year mission. }
    \label{fig:uds_ANL_dist_plot}
\end{figure}

Following the calculation of $P_{\text{\tiny Star}}$ for the stars in the Gaia catalogue and for both lens models with varying masses and radii, we now proceed to compute the total number of expected astrometric microlensing events, $N_{\text{\tiny L}}$, that may be potentially observed by Gaia. This calculation will be carried out for observation periods of $t_{\text{\tiny obs}} = 5~\text{yr}$ and $10~\text{yr}$ (as discussed in Sec.\,\ref{sec:AML_class}), and will be performed for both the uniform density sphere (UDS) and boson star (BS) lens models, specified by their respective parameters ($M_{\text{\tiny L}}, R_{\text{\tiny L}}, f_{\text{\tiny L}}$).

To estimate $N_{\text{\tiny L}}$, we first evaluate the expected number of AML events—covering short, intermediate, and long-duration lensing events (SDLEs, IDLEs, and LDLEs)—for each star in the Gaia catalogue. This is achieved by computing the individual $P_{\text{\tiny Star}}$ values, as outlined in Sec.\,\ref{sec:aml_lens_prob_no_lens_prob}, for a star characterized by $(D_{\text{\tiny S}}, \alpha, \delta, m_{\text{\tiny G}})$. The total number of events $N_{\text{\tiny L}}$ is then obtained by summing these probabilities across all Gaia stars, weighted by the fractional abundance of the chosen lens model.

As mentioned earlier, directly evaluating $P_{\text{\tiny Star}}$ for every star in the Gaia catalogue is computationally intensive, and to overcome this, we have adopted a coarse-graining strategy, which is expected to still retain a high degree of reliability. Specifically, as we have already discussed the stellar parameters $(D_{\text{\tiny S}}, \alpha, \delta, m_{\text{\tiny G}})$ of each star are mapped to the nearest point on a predefined grid as detailed in Tab.\,\ref{tab:gaia_data_binned}. Since $P_{\text{\tiny Star}}$ has been pre-computed for each of these grid points, this approach significantly reduces the computational time without sacrificing accuracy. 

The total number of AML events, $N_{\text{\tiny L}}$, is then obtained by summing over all Gaia stars, weighting each star by its corresponding grid-based probability $P_{\text{\tiny Star}}$. Equivalently, this procedure amounts to weighting $P_{\text{\tiny Star}}$ by the distribution of stellar positions and magnitudes across the Galaxy, as represented in the binned grid data. Thus, we have

\begin{equation}\label{eq:aml_lens_no_AML}
N_{\text{\tiny L}} = \sum_{i} N_{\text{\tiny Star,i}} P_{\text{\tiny Star}}(D_{\text{\tiny S},i}, \alpha_i, \delta_i, m_{\text{\tiny G},i})\;,
\end{equation}
where $N_{\text{\tiny Star,i}}$ is the number of Gaia stars in the $i$-th bin of the parameter grid, and $P_{\text{\tiny Star}}$ is the pre-computed probability corresponding to that bin.

\begin{figure}
    \centering
    \includegraphics[scale=0.45]{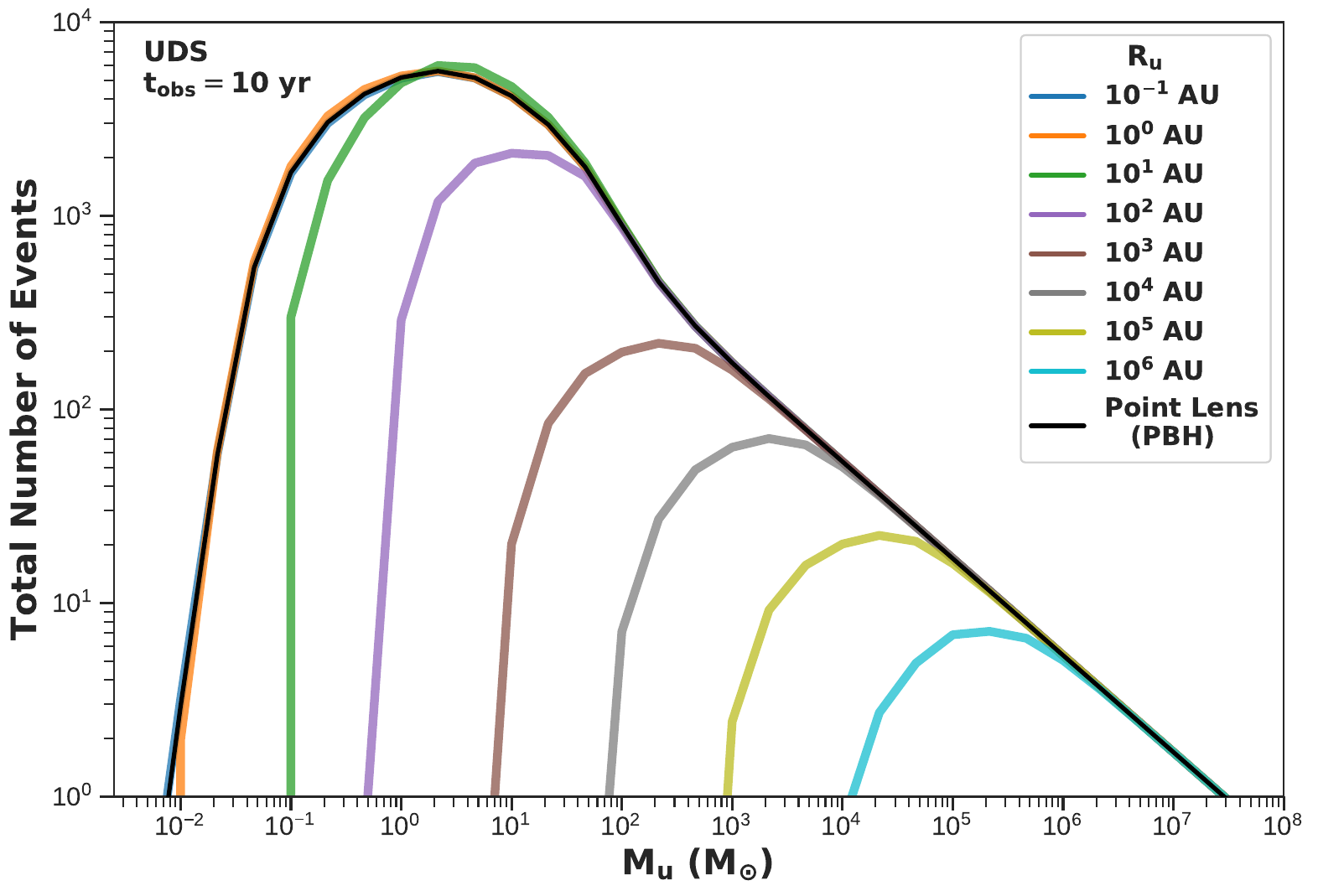}
    \includegraphics[scale=0.45]{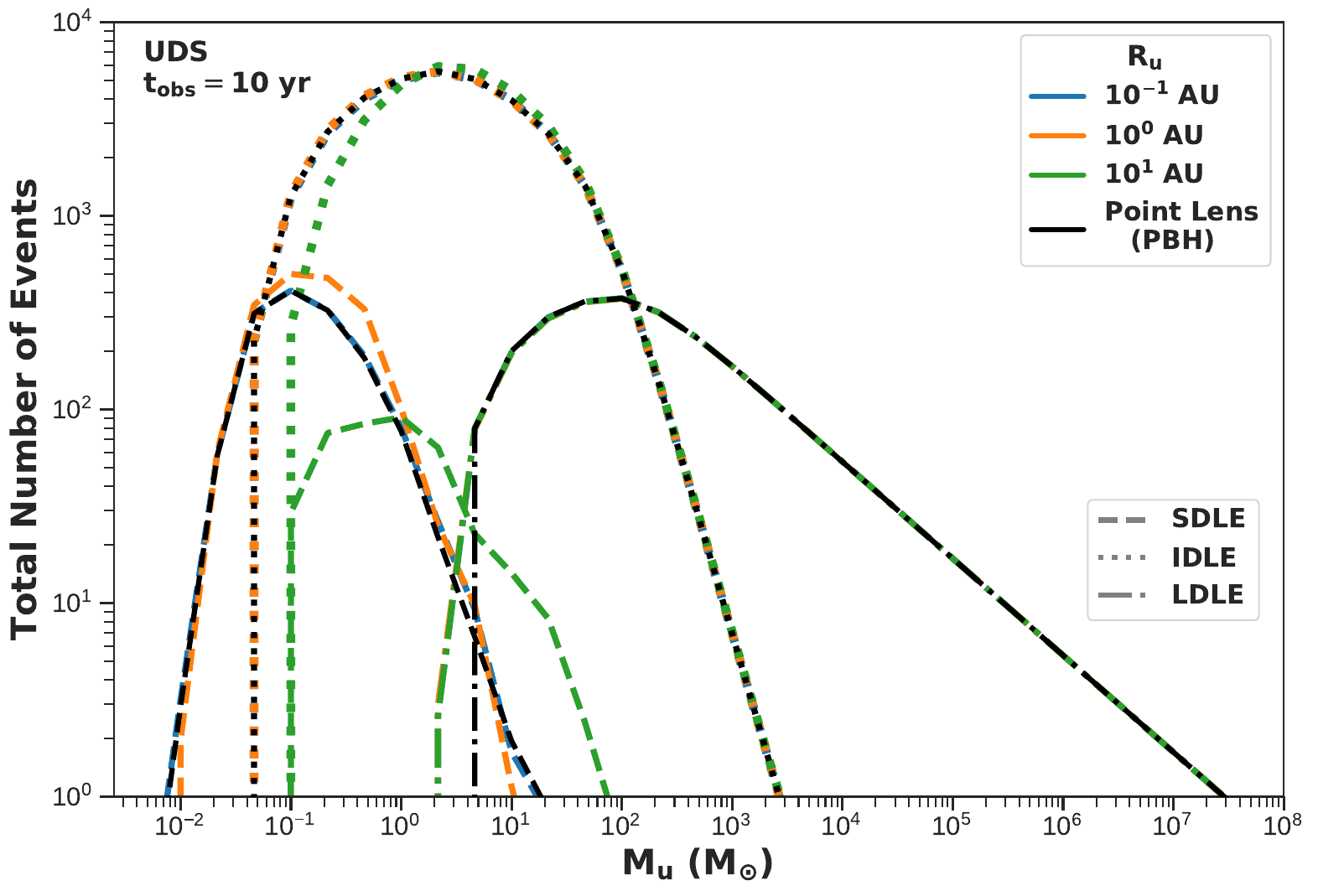}
    \caption{The \textit{top panel} shows the total number of expected microlensing events, $N_{\text{\tiny u}}$, produced by the UDS lens is shown as a function of lens mass for Gaia's observation period of  $t_{\text{\tiny obs}} = 10$ years assuming $f_{\text{\tiny u}}=1$. The plots display results for lenses with radii $R_{\text{\tiny u}} =\{0.1,1,10,10^2,10^3,10^4,10^5,10^6\}~\text{AU}$, with each radius represented by a different colored curve. The \textit{bottom panel} displays the contribution of SDLEs, IDLEs, and LDLEs events for radii $R_{\text{\tiny u}} =(0.1,1,10)~\text{AU}$.  The black line shows the point-lens approximation for comparison. As expected, the UDS results converge to the point-lens approximation (black line) whenever the physical radius is negligible compared to the Einstein radius ($R_{\text{\tiny u}}\ll R_{\text{\tiny E}}$). This limit is reached in two regimes: (a) for intrinsically small radii (e.g., the $R_{\text{\tiny u}} = 0.1$ AU curve) and (b) for any fixed radius at sufficiently high mass, where the Einstein radius ($R_{\text{\tiny E}} \propto \sqrt{M_{\text{\tiny u}}}$) becomes large. See text for more details. }
    \label{fig:uds_bs_Nu_vs_M}
\end{figure}

In the Fig.\,\ref{fig:uds_ANL_dist_plot} we present the distributions of the expected AML events $N_{\text{\tiny u}}$ for UDS lens model, while the corresponding plot for the BS lens is shown later in Fig.\,\ref{fig:bs_ANL_dist_plot} in Sec.\,\ref{app:sup_result}. Theses figures show results for lens parameters $M_{\text{\tiny L}} = 2.2~\solarmass$, $R_{\text{\tiny L}} = 10~\text{AU}$, and a dark matter fraction $f_{\text{\tiny L}} = 1$ for the respective lens models. The off-diagonal 2D histogram plots (in Fig.\,\ref{fig:uds_ANL_dist_plot}) show $N_{\text{\tiny u}}$ for the UDS lens as a function of two source star parameters taken from  $(D_s,\alpha,\delta,m_G)$, while summing over the remaining two parameters. Additionally, the diagonal 1D histogram plots show the corresponding distribution for a single parameter, while summing over the other three. These distribution plots are obtained by weighting the $P_{\text{\tiny Star}}$ values (as shown in Fig.\,\ref{fig:uds_prob_dist_plot} ) with the stellar distribution closest to the predefined grid data as given by Eq.\,\eqref{eq:aml_lens_no_AML}.  The total expected number of events accounts for all event types: SDLEs, IDLEs, and LDLEs.

From this analysis, we estimate that Gaia could detect approximately $6000$ AML events for both lens models with the above parameters, with the BS lens yielding slightly more events (by about 200 events) over a 10-year observational period. Again, we reiterate that while the UDS and BS models predict a comparable total number of AML events, their observational signatures are not degenerate. As discussed in Sec.\,\ref{sec:AML_UDS_BS}, the most pronounced differences arise during the caustic-crossing phases of low-impact-parameter events.

The 2D histogram showing the distribution of $N_{\text{\tiny u}}$ with galactic coordinate ($\alpha$, $\delta$)  in Fig.\,\ref{fig:uds_ANL_dist_plot}, reveals a higher number of events coming from the  Galactic center. This is the anticipated result, as this line of sight combines the highest density of potential source stars with the maximum concentration of dark matter lenses. Additionally, a substantial number of events occur within the galactic plane, consistent with the high stellar density of the Milky Way disk. The 1D histogram showing the distribution of  $N_{\text{\tiny u}}$ with the source distance $D_{\text{\tiny S}}$ reveals that the expected number of events increases with source distance $D_{\text{\tiny S}}$, peaking near the Galactic Center $D_{\text{\tiny S}}$ due to the dense stellar population, and subsequently declines beyond this region as the stellar density decreases.

Examining the 1D histogram distribution of $N_{\text{\tiny u}}$ as a function of G-band magnitude $(m_{\text{\tiny G}})$ in Fig.\,\ref{fig:uds_ANL_dist_plot}, we observe an initial increase in the expected number of events, driven by the larger stellar population (see the right panel of Fig.\,\ref{fig:sigma_a}), while $P_{\text{\tiny Star}}$ remained relatively constant in Fig.\,\ref{fig:uds_prob_dist_plot}. Beyond $m_{\text{\tiny G}} \gtrsim 15$, $N_{\text{\tiny u}}$ still increases though $P_{\text{\tiny Star}}$ decreases due to increase in stellar population. This happens till $m_{\text{\tiny G}} \sim 20$ after which $N_{\text{\tiny u}}$ declines sharply due to the decreasing astrometric sensitivity at fainter magnitudes, which also reduces $P_{\text{\tiny Star}}$ as shown in Fig.\,\ref{fig:uds_prob_dist_plot}.

We now examine how the mass and radius of the EADO lens affect the number of AML events. In the top panel of Fig.\,\ref{fig:uds_bs_Nu_vs_M}, we show the expected number of AML events $N_{\text{\tiny u}}$, for the UDS lens model as a function of UDS lens mass $M_{\text{\tiny u}}$, assuming a 10-year Gaia observation period. The corresponding plots for a 5-year observation period and BS lens model are presented later in Fig.\,\ref{fig:uds_bs_Nu_vs_M_1}, which shows that the BS lens produces quantitatively similar events as produced by the UDS lens. The plot is obtained assuming  $f_{\text{\tiny u}}=1$ and shows curves for different radii $R_{\text{\tiny u}} =\{0.1, 1, 10, 10^2, 10^3, 10^4, 10^5, 10^6\}~\text{AU}$, each represented by a distinct colored curve. Solid lines indicate the total $N_{\text{\tiny u}}$, obtained by summing contributions from SDLEs, IDLEs, and LDLEs, which are shown separately. The black curve corresponds to the point-lens approximation for reference. For $R_{\text{\tiny u}}=  0.1~\text{AU}$, the expected number of events coincides with that of the point-lens case. Furthermore, we have verified that for UDS lenses with radii  $R_{\text{\tiny u}} < 0.1~\text{AU}$, the expected event number $N_{\text{\tiny u}}$ is indistinguishable from the point-lens approximation.

We find in Fig.\,\ref{fig:uds_bs_Nu_vs_M} that Gaia is sensitive to extended UDS lenses with masses in the range $10^{-2} - 10^{7}\,\solarmass$ and radii $R_{\text{\tiny u}} \lesssim 10^6~\text{AU}\sim 5~\text{pc}$. The UDS lens model maximum number of AML events occurs at $R_{\text{\tiny u}} = 10~\text{AU}$ and $M_{\text{\tiny u}} \simeq 2.2,\solarmass$, yielding up to $\sim 6000$ events for $t_{\text{\tiny obs}} = 10$ years. As shown in Fig.\,\ref{fig:theta_illustration_0}, this is because EADOs with radii $R_{\text{\tiny u}} \sim 10~\text{AU}$ can produce an observable astrometric shift even when the lens is close to the source. However, this detectable signal is lost when lens is closer to the source for other radii---for $R_{\text{\tiny u}} \lesssim 1~\text{AU}$, the lens acts like a point lens (whose signal vanishes in this regime), and for $R_{\text{\tiny u}} \gtrsim 10~\text{AU}$, the lens becomes too diffuse to produce a detectable shift.

Additionally, for $t_{\text{\tiny obs}} = 5$ years, Gaia can observe up to $\sim 2500$ events as shown in the top left panel of Fig.\,\ref{fig:uds_bs_Nu_vs_M_1}. It is again important to note from Fig. \ref{fig:uds_bs_Nu_vs_M_1} that while the UDS and BS models predict a comparable total event rate, both are distinct from the point-lens case. We reemphasize that this behavior arises because, for larger impact parameters, the UDS and BS lenses produce similar astrometric shifts. The distinguishing features, however, are confined to the caustic-crossing signatures of low-impact-parameter events, meaning the models are not degenerate.

We observe that for large physical radii ($R_{\text{\tiny L}}\gtrsim100~\text{AU}$), both lens models predict a significantly lower event rate compared to the point-lens case.  This suppression occurs in the mass range where the lens is extended,  as its diffuse mass distribution produces a much weaker astrometric shift. For instance, the $R_{\text{\tiny u}}=100~\text{AU}$ curve  in Fig.\,\ref{fig:uds_bs_Nu_vs_M} falls well below the point-lens curve for for $M_{\text{\tiny u}}\sim 1-100~\solarmass$.

For sufficiently large masses or smaller radii, the number of events $N_{\text{\tiny u}}$ converges to the point-lens prediction, as expected from the discussion in Sec.\,\ref{sec:aml_eds}. In particular, the high-mass regime across all radii, as well as the low-mass regime for lenses with $R_{\text{\tiny u}} \lesssim 1~\text{AU}$, closely follows the point-lens approximation for the same reason. The bottom panel of Fig.\,\ref{fig:uds_bs_Nu_vs_M} displays the contribution of SDLEs, IDLEs, and LDLEs events for radii $R_{\text{\tiny u}} =(0.1,1,10)~\text{AU}$ and the point lens.
It is found that SDLEs (displayed in dashed lines in Fig.\,\ref{fig:uds_bs_Nu_vs_M}) contribute significantly for lenses with mass $M_{\text{\tiny u}} \lesssim 10^2~\solarmass$ and radii $R_{\text{\tiny u}} \lesssim 10~\text{AU}$. In particular, for compact lenses with radii $R_{\text{\tiny u}} \lesssim 1~\text{AU}$, SDLE events dominate around $M_{\text{\tiny u}} \sim 10^{-2}~\solarmass$, while for $R_{\text{\tiny u}} \sim 10~\text{AU}$, they dominate at $M_{\text{\tiny u}} \sim 10^{-1}~\solarmass$. Within these ranges of lens mass and radius, Gaia is expected to detect on the order of $10$–$100$ AML events.

However, IDLEs (displayed in dashed-dot lines in Fig.\,\ref{fig:uds_bs_Nu_vs_M}) span a wider parameter space, contributing across the mass range $M_{\text{\tiny u}} \sim 10^{-1} - 10^{3}~\solarmass$. Their dominant contribution arises in the intermediate mass regime, $M_{\text{\tiny u}} \sim 10^{-1} - 10^{2}~\solarmass$, where they predict up to $10^{2} - 10^{3}$ AML events. For lenses with radii $R_{\text{\tiny u}} \lesssim 10^{2}~\text{AU}$, the predictions for $N_{\text{\tiny u}}$ are predominantly driven by IDLE-type events.

For LDLEs (shown as dotted curves in Fig.\,\ref{fig:uds_bs_Nu_vs_M}), Gaia is sensitive to lenses across a broad mass range, $M_{\text{\tiny u}} \sim 10^{-1} - 10^{7}~\solarmass$, for all considered radii. However, LDLEs dominate only in the high-mass regime, $M_{\text{\tiny u}} \gtrsim 10^{2}~\solarmass$, where they predict up to $\sim 100$ AML events. 

The predicted AML event numbers account for both smooth trajectory distortions and caustic crossing events. These results provide insight into the rates and characteristics of events expected from  EADOs. More importantly, the future non-observation of anomalous events—those not explainable by standard astrophysical models—will be key to estimating robust constraints on the populations of thin-wall Q-balls and boson star lenses. We'll explore the detailed methodology for deriving these bounds in Sec.\,\ref{sec:gaia_detectibility}.

\section{Constraint on  exotic astrophysical dark objects}\label{sec:gaia_detectibility}

Having estimated the expected number of AML events detectable by Gaia for  EADOs with parameters ($M_{\text{\tiny L}}, R_{\text{\tiny L}}, f_{\text{\tiny L}}=1$) in the previous section, we now proceed to place constraints on these models. However, it should be noted that, over a broad range of impact parameters, various sources of background can produce signatures that mimic AML events produced from  EADOs. Since these backgrounds can lower the statistical significance of a potential AML signal from EADOs, deriving robust exclusion limits is contingent upon a careful modeling of the background event rate. In the following subsection, we analyse a specific background scenario that poses a significant challenge to identifying SDLE-type events in the Gaia data.

\subsection*{Backgrounds for AML due to EADOs}
One can categorize the potential backgrounds into three primary types: astrophysical, systematic, and statistical. The astrophysical backgrounds are genuine signals that mimic a lensing event produced by EADO.  This category of backgrounds mainly contains two classes: (i) lensing by other, non-exotic compact objects, and (ii) intrinsic non-rectilinear motion of the source star, most notably the astrometric perturbation caused by an unresolved binary companion. The systematic backgrounds include non-random errors originating from the instrument or data processing, such as uncorrected Point Spread Function (PSF) variations or calibration errors that could mimic a coherent astrometric drift.
Finally,  statistical background arises from the inherent, random uncertainty in centroiding the image of a given star, $\sigma_a(m_G)$, which sets the fundamental noise floor for a single measurement. 

While some aspects of astrophysical backgrounds have been explored (e.g., for primordial black holes in\,\cite{Verma:2022pym}), a comprehensive treatment of all background rates remains a topic for future study. Additionally, as detailed in Sec.\,\ref{sec:AML_UDS_BS}, the unique AML signals produced during caustic-crossing events are a key characteristic of EADOs; therefore, these events can be readily distinguished from the potential backgrounds. Additionally, the simultaneous observation of a photometric microlensing signal—especially for high-magnification events—can serve as a powerful discriminant to eliminate random astrometric backgrounds. In the current context, we will focus our analysis on addressing the statistical background. To assess the statistical backgrounds for AML due to EADOs, we adopt the methodology outlined in\,\cite{Verma:2022pym}. 

For an unlensed trajectory,  the statistical background arises from the inherent, random uncertainty in estimating the centroid of a star, due to astrometric uncertainty $\sigma_a(m_G)$ of Gaia's instruments. A sequence of such random fluctuations could mimic a "genuine" AML event due to EADOs. For such fluctuations to fake an AML event of a given duration $\bar{t}_{\text{\tiny e}}$, it would require producing a specific sequence of astrometric shifts such that each of these fluctuations is greater than $1\text{-}\sigma$ away from the true unlensed trajectory of the star.   Therefore, a fake event of duration $\bar{t}_{\text{\tiny e}}$ would require $N_s \approx \bar{t}_{\text{\tiny e}}/t_{\text{\tiny s}}$ (remember $t_{\text{\tiny s}}\approx 52.2\text{ days}$ is Gaia's sampling rate ) consecutive measurements to all fluctuate by more than $1\text{-}\sigma$ from the star's true path, and all in the same general direction. 

 We assume the centroiding fluctuations are Gaussian with standard deviation $\sigma_a(m_G)$, meaning the probability of a single, one-sided $1\text{-}\sigma$ fluctuation is $1-\text{erf}(1)\approx 0.16$. For a 10-year mission with $N_{\text{tot}} \approx 70$ total observations, the probability for a star to exhibit a fake event of $N_s = \bar{t}_{\text{\tiny e}}/t_{\text{\tiny s}}$ consecutive $1\text{-}\sigma$ fluctuations is the product of these individual probabilities (i.e. $(0.16)^{N_{\text{\tiny s}}}$), multiplied by the number of possible trial periods (i.e. $70-N_{\text{\tiny s}}$) and is given by $P^{\text{\tiny stat}}_{\text{\tiny star}}=2\times(0.16)^{N_{\text{\tiny s}}}\left(70-N_{\text{\tiny s}}\right)$. Here, the factor of 2 is included to account for the two possible directions of the coherent statistical fluctuation, as a fake event could be produced by deviations on either side of the true path. Since $P^{\text{\tiny stat}}_{\text{\tiny star}}$ is independent of individual stellar parameters (such as magnitude or distance), the total number of statistical background events, $N^{\text{\tiny stat}}$, can be estimated by multiplying this probability by the total number of stars in the Gaia Catalogue (i.e., $1.47\times 10^9$). Therefore, we can write  the total statistical  background as $N^{\text{\tiny stat}}= 1.47\times 10^9 P^{\text{\tiny stat}}_{\text{\tiny star}}$ and can be written as
\begin{equation}
    N^{\text{\tiny stat}}=2.94\times10^9\times (0.16)^{N_{\text{\tiny s}}}\left(70-N_{\text{\tiny s}}\right)\;.
\end{equation}

We estimate that for SDLEs, which rely on a single anomalous observation ($N_s=1$), the background rate is overwhelming, scaling to $N^{\text{\tiny stat}}\sim \mathcal{O}(10^{10})$ events. This rate drops exponentially with event duration; however, even for a 1-year duration ($N_s \approx 7$), the background remains significant at $\sim \mathcal{O}(10^{5})$. A background-free regime ($N^{\text{\tiny stat}}\lesssim 1$) is achieved only for durations $\bar{t}_{\text{\tiny e}} \gtrsim 2$ years. Therefore, only IDLEs with event durations $\bar{t}_{\text{\tiny e}} \gtrsim 2$ years and LDLEs are free from any statistical backgrounds.  

Therefore, a simple and effective method to remove this statistical background is to apply a conservative cut on the event duration. As mentioned before, the total number of fake background events, $N^{\text{\tiny stat}}$, falls to a negligible level (e.g., $N^{\text{\tiny stat}}\lesssim 1$) for event durations $\bar{t}_{\text{\tiny e}} \gtrsim 2$ years.  Therefore, by demanding $\bar{t}_{\text{\tiny e}} \gtrsim 2$ years, we can conservatively eliminate this background from AML event observations in Gaia. This analysis demonstrates that the statistical background for SDLEs, where $\bar{t}_{\text{\tiny e}}\lesssim t_{\text{\tiny s}}$, is overwhelmingly large and therefore SDLEs will be completely ignored while deriving constraints. Hence, we will derive our final exclusion limits on EADOs parameters considering only events with $\bar{t}_{\text{\tiny e}}\gtrsim 2~\text{years}$. This is achieved by using the conditional probability from Eq.\,\eqref{eq:app_AML_lens_cond_prob_bolt_cut} to evaluate the lensing probability $P_{\text{\tiny Star,L}}$ in Eq.\,\eqref{eq:AML_lens_tot_prob}, which is then used to calculate the total number of events via Eq.\,\eqref{eq:aml_lens_no_AML}.

 While a detailed treatment of event rates from such astrophysical backgrounds is an important topic, it is left for a separate analysis. This work remains focused on the specific signatures of extended EADOs. However, as detailed in Sec.\,\ref{sec:AML_UDS_BS}, the unique AML signals produced during caustic-crossing events are a key characteristic of EADOs. We posit that these distinct signatures are sufficiently anomalous to be readily distinguished from stellar binaries or other astrophysical mimics.

 We note that statistical backgrounds can be significant for event durations $t_{\text{\tiny e}}\lesssim 2$ years \cite{Verma:2022pym}. While a conservative cut on this timescale would reduce contamination, it would also introduce a false negative rate by discarding genuine AML events. Caustic-crossing events, in particular, would manifest as "louder" astrometric jumps and must produce a simultaneous photometric brightening—a feature that random background would lack.  While a full simulation is needed for quantitative justification, but a significant population of these shorter-duration events is still recoverable. Therefore, to provide a comprehensive estimate, we do not apply any cuts on the event duration in our analysis. 

\subsection*{Exclusion Limits on EADO Parameters}

\begin{figure}
    \centering
    \includegraphics[scale=0.5]{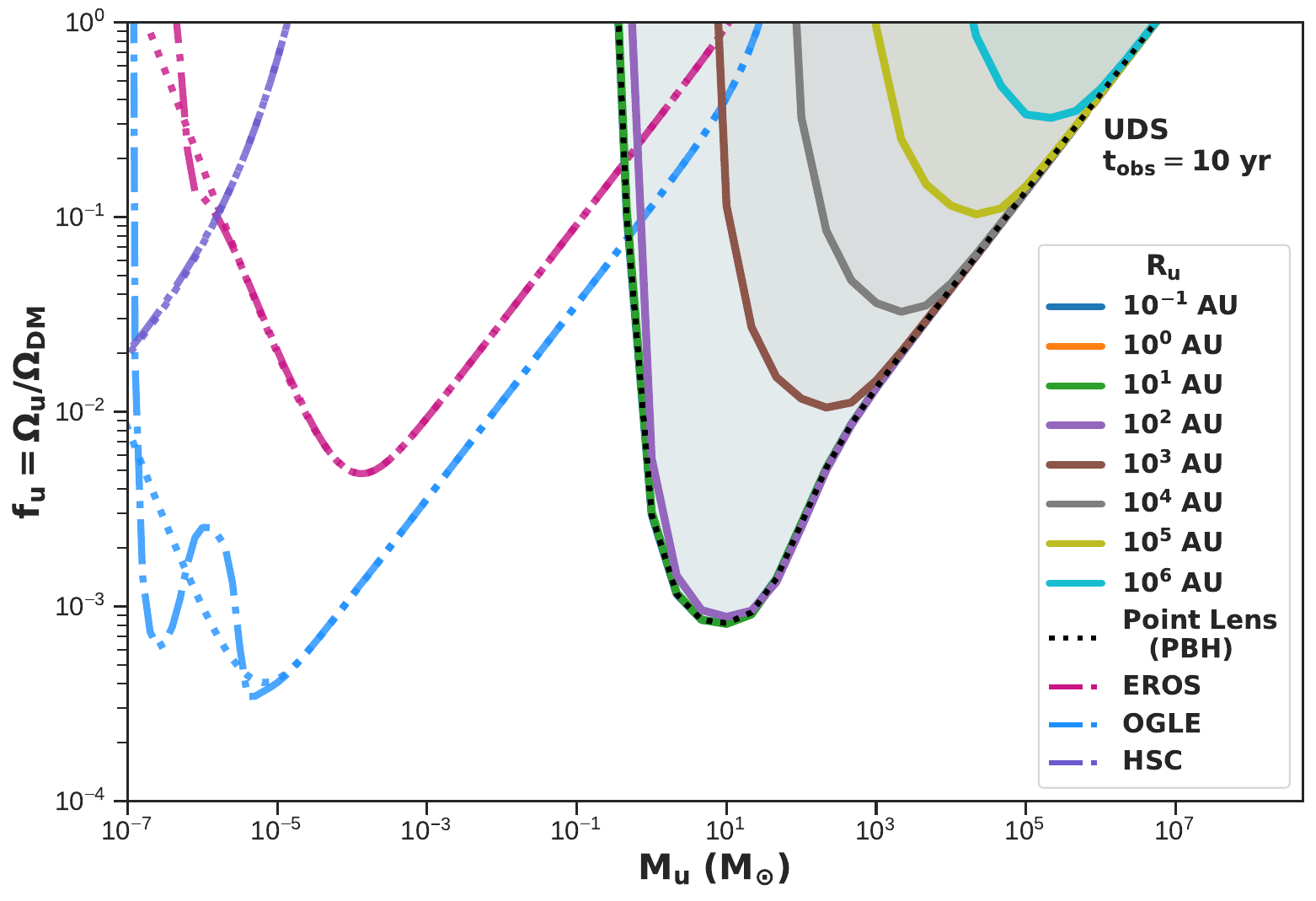}
    \includegraphics[scale=0.5]{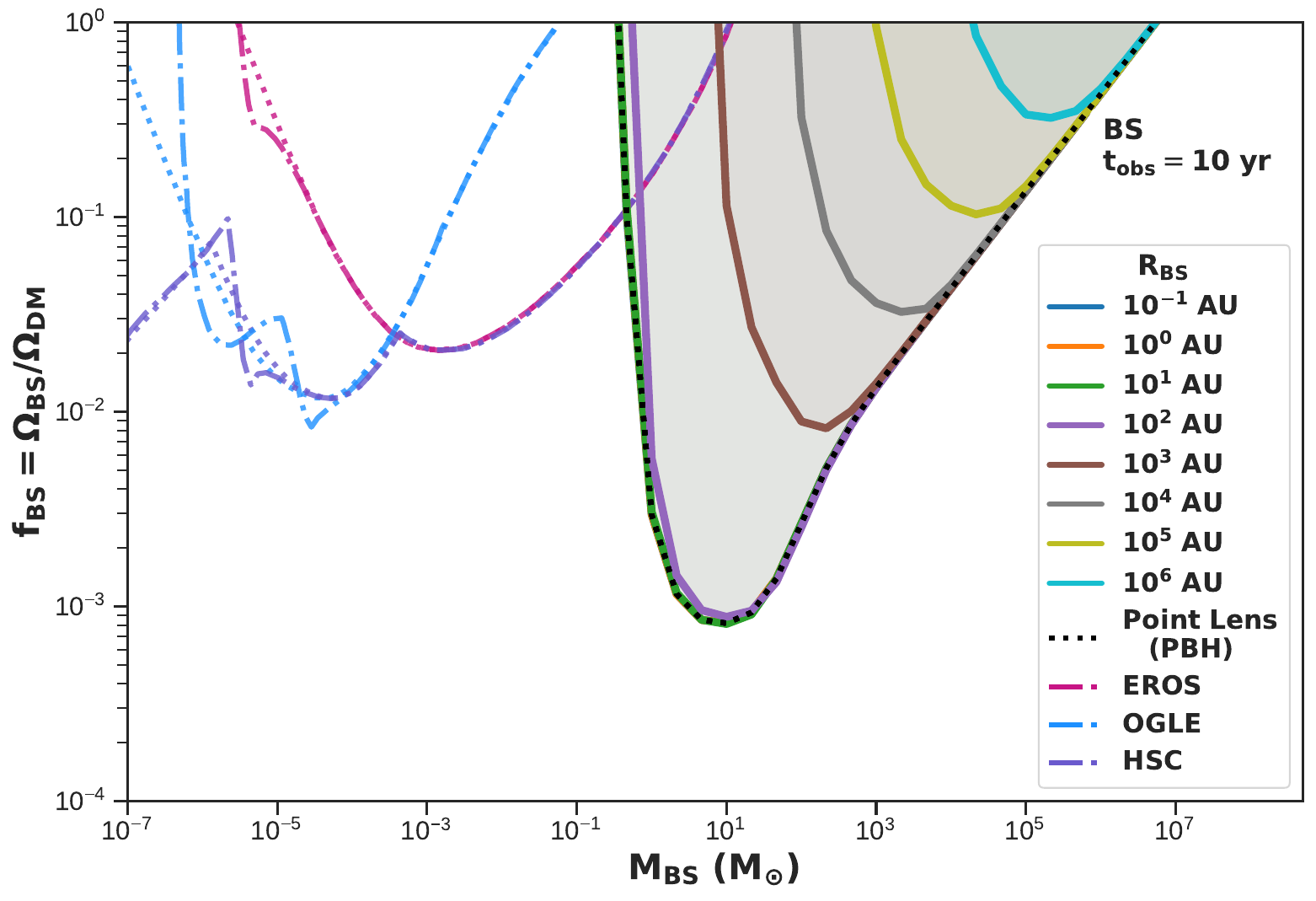}
    \caption{We present the expected 90\% confidence level exclusion regions in the $(M_{\text{\tiny L}} , f_{\text{\tiny L}})$   parameter space constrained by Gaia.  The top panel shows the constraints for the thin-wall Q-ball model, and the bottom panel shows the constraints for the Bose star model. Both panels assume a 10-year Gaia observation period ($t_{\text{\tiny obs}} = 10$ years). The solid colored curves correspond to specific lens radii, $R_{\text{\tiny L}} =\{0.1, 1, 10, 10^2, 10^3, 10^4, 10^5, 10^6\}~\text{AU}$, while the dotted black line denotes the point-lens approximation. For comparison, previous PML constraints for $R_{\text{\tiny L}}\sim 1 R_{\odot}\approx 5\times 10^{-3}~ \text{AU} $ are also shown for  EROS\,\cite{Ansari:2023cay,Croon:2020wpr} (red, dash-dot), OGLE\,\cite{Ansari:2023cay,Croon:2020wpr} (royal blue, dash-dot), and HSC-Subaru\,\cite{Ansari:2023cay,Croon:2020ouk} (red, dash-dot) surveys. The PML constraints were derived at the $95\%$ confidence level for UDS lenses and at the $90\%$ confidence level for BS lenses. Both calculations assume an isothermal profile for the Milky Way's dark matter distribution. See text for more details.}
    \label{fig:uds_exclusion}
\end{figure}

Having addressed the statistical background by imposing a cut of $\bar{t}_{\text{\tiny e}}\gtrsim 2~\text{years}$, we now define "anomalous events" as signals uniquely attributable to EADOs and not reproducible by standard astrophysical models. We assume the number of observed AML events follows Poisson statistics. Therefore, if Gaia detects $N_{\text{\tiny obs}}$ such events, the probability of this observation at $\mathcal{C}$ confidence level, given an expected rate of $N_{\text{\tiny L}}$, is given by 
\begin{equation}\label{eq:pois_cl}
\sum_{i=0}^{N_{\text{\tiny obs}}}P(i,N_{\text{\tiny L}})=1-\mathcal{C}\,.
\end{equation}
Here, $P(i,N_{\text{\tiny obs}}=)N_{\text{\tiny L}}^ie^{-N_{\text{\tiny L}}}/i!$ denotes the probability of observing $i$ events for a expected number of $N_{\text{\tiny u}}$. So if $N_{\text{\tiny obs}}$ events are observed, we can exclude models that predict a events $N_{\text{\tiny L}}$ for which the probability of observing $N_{\text{\tiny obs}}$ is less than 10\% i.e. $\sum_{i=0}^{N_{\text{\tiny obs}}}P(i,N_{\text{\tiny L}})<0.1$ with $0.9$ (or 90 \%) confidence level. This puts an upper bound on expected event rates $N_{\text{\tiny L}}$, which can be used to put the upper bound on $f_{\text{L}}$ as a function of the lens mass and its radii for both thin-wall Q-balls and boson star models.

For the present analysis, we assume that Gaia’s observations over its mission duration $t_{\text{\tiny obs}}$ are consistent with the $0$ background rate ($N_{\text{\tiny obs}}=0$), i.e., no anomalous events are detected. Under this null-detection assumption, we exclude regions of the thin-wall Q-ball (modelled as UDS) and boson star (BS) parameter space that predict more than $N_{\text{\tiny L}} > 2.3$ (using Eq.\eqref{eq:pois_cl}) events at the 90\% confidence level. 

The results of this analysis are presented in Fig.\,\ref{fig:uds_exclusion}, which shows the excluded parameter space ($f_{\text{\tiny L}}, M_{\text{\tiny L}}$) for lens radii $R_{\text{\tiny L}} =\{0.1, 1, 10, 10^2, 10^3, 10^4, 10^5, 10^6\}~\text{AU}$ in each EADO model. The top panels correspond to a thin-wall Q-ball lens, while the bottom panels correspond to a boson star lens, for Gaia's observation periods of $t_{\text{\tiny obs}} = 10$ years. The corresponding constraint for the 5-year observation periods is shown in Fig.\,\ref{fig:uds_exclusion_1} in App.\,\ref{app:sup_result}. For comparison, we show photometric microlensing (PML) surveys, including EROS\,\cite{Ansari:2023cay,Croon:2020wpr} (red, dash-dot), OGLE\,\cite{Ansari:2023cay,Croon:2020wpr} (royal blue, dash-dot), and HSC-Subaru\,\cite{Ansari:2023cay,Croon:2020ouk} (red, dash-dot), derived for radii $R \sim 1~R_{\odot}$. 

Although the PML constraints are shown for $R \sim 1~R_{\odot}$, the exclusion weakens for larger radii and is applicable in the same mass range as given for $R \sim 1~R_{\odot}$. In these studies, the constraints were based on the observation of one anomalous event for EROS, but $0$ anomalous events for OGLE and HSC Subaru. Additionally, these PML constraints were derived at the $95\%$ confidence level for UDS lenses and at the $90\%$ confidence level for BS lenses.  Both calculations assume an isothermal profile for the Milky Way's dark matter distribution.

Based on the null-detection hypothesis, Gaia is expected to place constraints on both UDS and BS lenses over a broad parameter space, spanning masses $10^{-1} - 10^{7}~\solarmass$ and radii $R_{\text{\tiny L}} \lesssim 10^6~\text{AU} \sim 5~\text{pc}$. This significantly extends the reach of previous PML studies\,\cite{Ansari:2023cay,Croon:2020ouk,Croon:2020wpr}, which were primarily sensitive to these  EADOs with masses $M_{\text{\tiny L}} \lesssim 10\,\solarmass$. Furthermore, for EADOs with masses in the $1-10~M_\odot$ range, AML can produce significantly stronger constraints than PML. This is a direct consequence of AML's sensitivity to much larger impact parameters. 

Furthermore, we find that Gaia achieves its strongest constraints for lenses with $M_{\text{\tiny L}} \simeq 10~\solarmass$ and $R_{\text{\tiny L}} \lesssim 10~\text{AU}$, where the fractional abundance can be limited to $f_{\text{\tiny L}} \lesssim \mathcal{O}(10^{-3})$. For reference, the canonical dark matter density in the solar neighbourhood is $\sim 1~\solarmass/(10^6~\text{AU})^3$. While the overall constraints on the UDS and BS models are quantitatively comparable, their observational signatures are not degenerate. As shown in Sec.\,\ref{sec:AML_UDS_BS}, the individual trajectories for the two models differ significantly during caustic crossings. 

For $R_{\text{\tiny L}} \lesssim 10~\text{AU}$, the constraints on  EADOs coincide with those of the primordial black holes (point-lens case). However, these constraints weaken significantly with increasing radius $R_{\text{\tiny L}}$, departing from the limits set on point-like PBHs. For instance, in Fig.\,\ref{fig:uds_exclusion} for $M_{\text{\tiny L}}\sim 10~\solarmass$, the constraint of $f_{\text{\tiny L}}<\mathcal{O}(10^{-3})$ on radius $R_{\text{\tiny L}}=10^2~\text{AU}$ relaxes significantly to $f_{\text{\tiny L}}<\mathcal{O}(10^{-2})$ for $R_{\text{\tiny L}}\lesssim 10^3~\text{AU}$ .  As established in Sec.\,\ref{sec:aml_eds}, this suppression occurs because, for a fixed mass, a larger EADO is more diffuse, and its dilute mass distribution produces a suppressed astrometric shift, which in turn lowers the event rate. The only exception arises for larger masses, such that specific mass–radius combinations match the point lens values.

The relative contribution of different event types depends on the lens mass. As mentioned earlier, SDLEs possess a large statistical background and are often completely excluded by imposing a cut of $\bar{t}_{\text{\tiny e}} \gtrsim 2~\text{years}$ on the event duration. The LDLEs dominate the high-mass regime, $M_{\text{\tiny L}} \sim 10^{2} - 10^{7}~\solarmass$. Intermediate-duration lensing events IDLEs provide the most significant contribution over $M_{\text{\tiny L}} \sim 10^{-1} - 10^{2}~\solarmass$, with a peak sensitivity around $M_{\text{\tiny L}} \sim \mathcal{O}(10)~\solarmass$.

Below $\sim 1~M_{\odot}$, Gaia’s sensitivity declines because both the astrometric deflections and event durations fall below the mission’s detection thresholds. At the opposite extreme, for large masses $(\sim 10^{2}~\solarmass)$, the reduced galactic number density of such lenses suppresses the lensing probability, again limiting Gaia’s sensitivity. 

In summary, our analysis demonstrates that Gaia’s astrometric microlensing capabilities can provide robust and complementary constraints on  EADOs such as thin-wall Q-balls and boson stars. By probing wide regions of the mass–radius parameter space beyond the reach of traditional photometric microlensing surveys, these studies highlight Gaia’s unique potential to test exotic objects as viable dark matter candidates. The sensitivity to different event types across distinct mass scales further underscores the richness of astrometric lensing as a probe of new physics.

\section{Summary }\label{sec:conclusion}
In this work, we present a detailed theoretical study of the astrometric microlensing signatures induced by exotic astrophysical dark objects. Due to the extended nature of these EADOs, we expect distinct AML signatures compared to point-like objects, such as primordial black holes. We focus on two representative models---thin-wall Q-balls and boson stars for EADOs, and study the AML signal generated by them. Our aim is to evaluate whether the astrometric precision of the Gaia mission enables the detection of AML events induced by EADOs and, consequently, to place meaningful constraints on their abundance within our Galaxy.

We now provide a comprehensive summary of our main results, detailing the unique AML phenomenology of EADOs and the corresponding observational prospects for the Gaia mission
\begin{itemize}
    \item In Sec.\,\ref{sec:aml_eds}, we investigate the AML signatures of the thin wall Q-ball and boson star models, uncovering a rich lensing phenomenology that departs markedly from the standard point-lens case, such as primordial black holes. We have found that for $R_{\text{\tiny L}}/R_{\text{\tiny E}}\sim 1$, EADOs can produce an astrometric shift significantly larger than that of a point lens. This enhancement occurs because, in this single-image regime, the EADO lacks the secondary, opposing image that a point lens always produces, which would otherwise reduce the net centroid shift. Additionally, since BS have more centrally concentrated mass distribution, they induce stronger deflections for light rays traversing them than thin-wall Q-balls, resulting in enhanced astrometric shifts.
    \item A primary distinctive feature of EADOs is the emergence of caustics, which exist when the lens radius lies below a critical threshold: $R_{\text{\tiny u}} < \sqrt{3/2} R_{\text{\tiny E}}$ for thin-wall Q-balls and $R_{\text{\tiny BS}}/ \lesssim 2.73 R_{\text{\tiny E}}$ for the BS (see Figs.\,\ref{fig:uds_massfunc_vs_ui} and \ref{fig:bs_massfunc_vs_ui}). Crossing these caustics produces a discontinuous jump in the image centroid, serving as a distinct feature of  EADOs. 
    \item We have shown in Fig.\,\ref{fig:uds_centshift_vs_us_2d}, these caustic features manifest as dramatic signatures in the observable AML trajectory. Most notably, the centroid can exhibit a sharp discontinuity, jumping to the side opposite the source (see bottom-right panel in Fig.\,\ref{fig:uds_centshift_vs_us_2d}). This phenomenon is a smoking gun signature of EADO lensing, and is not expected from standard astrophysical objects. Additionally, we found that for the BS lens, caustics form at larger impact parameters, implying more observations of caustic crossing events compared to thin-wall Q-balls. 
    \item The Gaia DR3 catalogue\,\cite{Gaia_2021}, which contains high-precision astrometric parameters for over 1.5 billion stars, serves as the foundational stellar map for our analysis. We utilized this catalogue in Sec.\,\ref{sec:aml_lens_prob_no} to compute the probability that a given source star will experience a detectable AML event from an EADOs.  By convolving this probability with the stellar distribution from Gaia DR3 and adopting a standard NFW dark matter halo profile, we estimated the total number of AML events expected over $5$ and $10$ years of Gaia observations, aligning with the observation times of Gaia DR4 (expected late 2026) and DR5 release (expected after 2030).
    \item Our analysis shows that Gaia can put constraints on the population of  EADOs with masses in the range $10^{-1}-10^{7}\,\solarmass$ and radii $R_{\text{\tiny L}} \lesssim 10^{6}~\text{AU}$ (see Fig.\,\ref{fig:uds_bs_Nu_vs_M}). For both the UDS and BS models, the number of events peaks around $R_{\text{\tiny L}} \lesssim 10~\text{AU}$ and $M_{\text{\tiny L}} \sim 1-10\,\solarmass$, corresponding to as many as $\sim 2500$ events for $t_{\text{\tiny obs}} = 5$ years and $ \sim 6000 $ events for $t_{\text{\tiny obs}} = 10$ years. While our analysis predicts a comparable total event rate for both the UDS and BS objects, their observational signatures are not degenerate. As established, the individual AML trajectories remain distinguishable for caustic-crossing events.
\item Finally, under the conservative assumption of no anomalous detections by Gaia, we derived projected 90\% confidence-level exclusion limits on the abundance of  EADOs in Sec.\,\ref{sec:gaia_detectibility}. Our results indicate that Gaia can place novel stringent constraints on both UDS and BS across a wide mass range, $10^{-1} - 10^{7}~\solarmass$ and radii $R_{\text{\tiny L}} \lesssim 10^6~\text{AU}$ (see Fig.\,\ref{fig:uds_exclusion}). These constraints significantly broaden the scope of previous PML studies,\, \cite{Ansari:2023cay,Croon:2020ouk,Croon:2020wpr}, which were mainly limited to probing these objects with masses $M_{\text{\tiny L}} \lesssim 10\,\solarmass$ and radii $R_{\text{\tiny L}} \lesssim 1\,~\text{AU}$. 
\item Furthermore, our results show that Gaia provides its most stringent constraints for lenses with $M_{\text{\tiny L}} \sim 1-10~\solarmass$ and $R_{\text{\tiny L}} \lesssim 10~\text{AU}$, where the fractional abundance can be constrained up to $f_{\text{\tiny L}} \lesssim \mathcal{O}(10^{-3})$. The EADO constraints are found to converge to the point-mass (PBH) limit for $R_{\text{\tiny L}}\lesssim 10~\text{AU}$. For larger radii, the constraints weaken significantly. This suppression is a direct consequence of the extended-lens regime, where the lens's mass distribution is diffused. This results in a weaker astrometric deflection and, consequently, a lower detection probability. 
\end{itemize}

This study demonstrates that astrometric microlensing provides a powerful new probe of the nature of possible EADOs, complementing traditional photometric surveys. Gaia’s unprecedented precision enables sensitivity to a wider range of impact parameters, allowing it to probe extended dark structures such as Q-balls, boson stars, and other exotic objects. The distinct AML signatures associated with different exotic candidates highlight the potential of this technique not only for detection but also for discriminating between different EADOs.

\section{Discussion and future outlook }\label{sec:discussion}
While our analysis provides meaningful constraints, it is based on several simplifying assumptions. These assumptions can be relaxed or improved, thereby strengthening the robustness and scope of future studies. For example, in our analysis, we assumed a uniform temporal sampling of stellar positions by Gaia. A more realistic approach would involve incorporating Gaia’s actual scanning law\,\cite{Gaia:2016zol}, which specifies the true time-sampling pattern across different regions of the sky. Additionally, our present approach adopts a simplified treatment of the astrometric error, $\sigma_{\text{\tiny a}}(m_{\text{\tiny G}})$, by accounting only for the component along Gaia’s scanning direction. A more refined numerical analysis could improve the estimate of $P_{\text{\tiny Star}}$ by incorporating both the along-scan and across-scan astrometric uncertainties\,\cite{Gaia:2023fqm}, which are closely tied to Gaia’s detailed scanning strategy.

In our analysis, we adopted a simplified rectilinear motion model for target stars. In reality, stellar trajectories are non-rectilinear due to intrinsic parallax and the Earth’s orbital motion. Moreover, a substantial fraction of Galactic stars reside in binary systems, introducing additional astrometric perturbations from their orbital motion. To incorporate these effects more accurately into the $P_{\text{\tiny Star}}$ calculations, a robust numerical simulation of stellar trajectories would be required. Stellar velocity information, which can be directly obtained from the Gaia catalogue\,\cite{Gaia:2023fqm}, could further aid in disentangling genuine lensing signals from these intrinsic stellar motions.
   
  The present exclusion limits are based solely on Gaia’s AML signatures. However, combining them with Gaia’s photometric data\,\cite{Gaia_2021}—even at lower precision—could enhance sensitivity and strengthen the resulting constraints.  For instance, the simultaneous observation of both a photometric and astrometric signal for the same event provides a powerful method to mitigate instrumental systematics and reduce false positives.
  
    Our study has explored AML signals caused by thin-wall Q-balls and mini-boson stars. It would be valuable to extend this analysis to other EADOs, such as axion stars\,\cite{Braaten:2015eeu}, axion miniclusters\,\cite{Kolb:1993hw}, self-interacting boson stars\,\cite{Colpi:1986ye}. This would allow for a more comprehensive search and provide broader constraints on the nature of dark matter.

  A key step is to translate the astrophysical exclusion limits on lens mass and radius into constraints on the fundamental microphysical parameters of the underlying theories. For boson stars, this involves mapping the allowed regions in mass–radius space to the boson mass, while for Q-balls it requires relating them to the scalar potential parameters. This framework can be further extended to probe the self-interaction coupling of boson stars\,\cite{Kolb:1993hw}, and, in the case of axion stars and miniclusters\,\cite{Braaten:2015eeu,Kolb:1993hw}, to express the exclusion curves in terms of the axion mass and its decay constant. In this way, Gaia’s observational bounds can be recast as direct tests of specific dark matter models, thereby bridging astrophysics and particle physics.

While the simplifying assumptions in this analysis are sufficient to understand the fundamental impact of these objects, incorporating more realistic treatments will be crucial for strengthening future constraints. These improved models will be especially vital for analyzing the full time-series astrometric solutions for Milky Way stars, which are expected to be released in Gaia DR5 and DR6 data release. Looking ahead, upcoming missions such as the Nancy Grace Roman Space Telescope\,\cite{Spergel:2015sza} will extend the reach of AML to lower-mass lenses and fainter signals, offering unprecedented opportunities to test a wide range of exotic dark matter scenarios and their fundamental parameters.

\acknowledgments
LB acknowledges support from a Senior Research Fellowship, granted by the Human Resource Development Group, Council of Scientific and Industrial Research, Government of India. HV is supported by the funding for the Roman Galactic Exoplanet Survey Project Infrastructure Team which is provided by the Nancy Grace Roman Space Telescope Project through the National Aeronautics and Space Administration grant 80NSSC24M0022, by The Ohio State University through the Thomas Jefferson Chair for Space Exploration endowment, and by the Vanderbilt Initiative in Data-intensive Astrophysics (VIDA).

\appendix
\section{Density profile of Boson star}
\label{app:BS}
In this appendix, we outline the procedure for deriving the density distribution of a mini-boson star (see \cite{Kling:2017hjm,Kling:2020xjj} for comprehensive treatments). As discussed in Sec.\,\ref{sec:eds_rev}, boson stars are self-gravitating configurations of a complex scalar field. Their dynamics can be described by the action (we set $c=\hbar=1$ for simplicity),
\begin{equation}\label{eq:app_bs_action}
    \mathcal{S}=\int d^4x \sqrt{-g}\left(\frac{\mathcal{R}}{4 \pi G_{\text{\tiny N}}}-g^{\mu\nu}\partial_\mu \Phi^* \partial_\nu \Phi - V(\Phi)\right)\;.
\end{equation}
Here, $\Phi$ is the complex scalar field $\mathcal{R}$ is the Ricci scalar, $g$ is the determinant of the spacetime metric $g_{\mu\nu}$, $G_{\text{\tiny N}}$ is Newton's gravitational constant, and $V(\abs{\Phi}^2)$ represents the potential of the scalar field.  

By varying the action in Eq.\,\eqref{eq:app_bs_action} with respect to the metric $g_{\mu\nu}$, one obtains the Einstein field equations,
\begin{equation}\label{eq:app_bs_einstein_eq}
			R_{\mu\nu} - \frac{1}{2}g_{\mu\nu}R = 8\pi G T_{\mu\nu}\;,
\end{equation} 
where $R_{\mu\nu}$ is the Ricci tensor, $R$ is the Ricci scalar, and $T_{\mu\nu}$ is the energy-momentum tensor of the scalar field, which is given by
\begin{equation}\label{eq:app_bs_energy_momentum}
			T_{\mu\nu} = \partial_\mu \Phi^* \partial_\nu \Phi + \partial_\nu \Phi^* \partial_\mu \Phi - g_{\mu\nu} \left( g^{\alpha\beta} \partial_\alpha \Phi^* \partial_\beta \Phi - V(\Phi) \right)\;.
\end{equation}
Variation of the action (see Eq.\,\eqref{eq:app_bs_action}) with respect to the complex conjugate of the scalar field $\Phi^*$ yields the Klein–Gordon equation in curved spacetime,
\begin{equation}\label{eq:app_bs_kg_eq}
			 \frac{1}{\sqrt{-g}} \partial_\mu \left( \sqrt{-g} g^{\mu\nu} \partial_\nu \right)\Phi - \frac{\partial V}{\partial \Phi^*} = 0\;,
\end{equation}
 which governs the dynamics of the scalar field in the gravitational background determined by the Einstein equations (see Eq.\,\eqref{eq:app_bs_einstein_eq}). The boson star is described by a stationary, spherically symmetric, and asymptotically flat solution of Eq.\,\eqref{eq:app_bs_einstein_eq}. In the non-relativistic and weak-field regime (Newtonian approximation), the metric in spherical polar coordinates $(r,\theta,\phi)$ is given by,
 \begin{equation}\label{eq:app_bs_metric_weak_non_rel_c}
			ds^2 = -\left(1+2\Psi_{\text{\tiny N}}(r)\right)dt^2 + dr^2 + r^2d\Omega^2\;.
		\end{equation}
where $\Psi_{\text{\tiny N}}(r)$ is the Newtonian potential and $d\Omega^2=d\theta^2+\sin^2\theta d\phi^2$. The ground-state configuration of the boson star is spherically symmetric and is assumed to take the stationary form, i.e.
		\begin{equation}\label{eq:app_bs_kg_weak_ground}
			\Phi(t,r) = \left(\frac{N}{2\omega}\right)^{1/2}\phi_{\text{\tiny BS}}(r)e^{i \omega t}\;,
		\end{equation}
where $\omega\equiv  m+e$ is the energy of the ground state, with $e~(\ll m)$  the binding energy, $N$ the number of bosonic particles in the configuration, and $\phi_{\text{\tiny BS}}(r)$ is the ground state radial wavefunction. For a mini-boson star, the potential is given by $V(\Phi)=m^2\abs{\Phi}^2$. Therefore, utilising Eqs.\,\eqref{eq:app_bs_energy_momentum}, \eqref{eq:app_bs_metric_weak_non_rel_c} and \eqref{eq:app_bs_kg_weak_ground}, the Einstein equation Eq.\eqref{eq:app_bs_einstein_eq} and the Klein-Gordon equation \eqref{eq:app_bs_kg_eq} can be simplified to obtain
\begin{equation}\label{eq:app_bs_einstein_tt_final}
			\nabla^2 \Psi_{\text{\tiny N}} = 4\pi G \rho_{\text{\tiny BS}}\;.	
\end{equation}
\begin{equation}\label{eq:app_bs_kg_weak_eq_final_sch}
			 -\frac{1}{2m}\nabla^2 \phi_{\text{\tiny BS}} + m\Psi_{\text{\tiny N}}\phi_{\text{\tiny BS}} =  e\phi_{\text{\tiny BS}}\;.
\end{equation}
where $M_{\text{\tiny BS}}\approx N m$ is the total mass of the mini-boson star, and $\rho_{\text{\tiny BS}}$ is the density profile of the boson star is then given by
\begin{equation}\label{eq:app_bsmini_bos_star_density}
			\rho_{\text{\tiny BS}}({r})=M_{\text{\tiny BS}}\phi_{\text{\tiny BS}}^2(r)\;.
		\end{equation}
Eqs.\,\eqref{eq:app_bs_einstein_tt_final}, \eqref{eq:app_bs_kg_weak_eq_final_sch}, and \eqref{eq:app_bsmini_bos_star_density} together constitute the Schrödinger–Poisson system for a mini-boson star, which governs its ground-state configuration. An analogous set of equations for a self-interacting boson star can be derived by modifying the scalar potential $V(\abs{\Phi})$\,\cite{Kling:2020xjj}.  Importantly, Eqs.\,\eqref{eq:app_bs_einstein_tt_final}, \eqref{eq:app_bs_kg_weak_eq_final_sch} and \eqref{eq:app_bsmini_bos_star_density} exhibit a scaling symmetry under the following transformation,
\begin{equation}\label{eq:app_bs__scaling_final}
r\to \Lambda r\;, \quad e\to \Lambda^{-2} e\;, \quad \phi_{\text{\tiny BS}}\to \Lambda^{-3/2} \phi_{\text{\tiny BS}}\;, \quad \Psi_{{\text{\tiny N}}}\to \Lambda^{-2} \Psi_{\text{\tiny N}}, \quad M_{\text{\tiny BS}}\to \Lambda   M_{\text{\tiny BS}}\;.
\end{equation}
where $\Lambda$ is the scale factor. 

We numerically solve Eqs.\,\eqref{eq:app_bs_einstein_tt_final}, \eqref{eq:app_bs_kg_weak_eq_final_sch}, and \eqref{eq:app_bsmini_bos_star_density} once, and subsequently employ the scaling relation in Eq.\,\eqref{eq:app_bs__scaling_final} to generate the complete family of boson star solutions. The resulting numerical density profiles for mini-boson stars, obtained from Eq.\,\eqref{eq:app_bsmini_bos_star_density}, are shown in Fig.\,\ref{fig:UDS_BS_dens_prof}.

\section{Microlensing due to an extended mass distribution}
\label{app:mlrev}
In this appendix, we present a detailed derivation of the formulae relevant for astrometric microlensing by an extended mass distribution (see \,\cite{Schneider_1992,Hog:1994xt,Walker_1995,Miyamoto_1995,Gould:1996nb,Dominik_2000,Lee:2010pj} for comprehensive treatments). The lensing configuration is illustrated in Fig.\,\ref{fig:gr_lens_fig}. We consider a continuous mass distribution of mass density $\rho(\vec{r})$, total mass $M_{\text{\tiny L}}$ and radius $R_{\text{\tiny L}}$. The deflection angle at impact parameter $\vec{\chi}$ is given by integrating the infinitesimal deflections contributed by each mass element $dM'$, and it is given by,
 \begin{equation}\label{eq:app_mlrev_def_ang_cont_0}
	\vec{\hat{\theta}}_{\text{\tiny D}}(\vec{\chi})=\frac{4 G_{\text{\tiny N}} }{c^2}\int\,dM'(\vec{\chi}',z)\,\frac{\vec{\chi}-\vec{\chi}'}{\abs{\vec{\chi}-\vec{\chi}'}^2}\;.
\end{equation}
where $\vec{\chi}'$ denotes the projection of the position vector $\vec{r}$ onto the lens plane shown in Fig.\,\ref{fig:gr_lens_fig}. 

Under the thin lens approximation (i.e. $R_{\text{\tiny L}}\ll (D_{\text{\tiny L}},D_{\text{\tiny S}})$), we may project the 3-D mass distribution of the lens onto the 2-D lens plane.
 The resulting surface mass density, in a spherical cylindrical coordinate system $(\chi,\phi,z)$, is given by
		 \begin{equation}\label{eq:app_mlrev_surf_dens}
		 	\sigma(\vec{\chi}')=\int\,dz\,\rho(\vec{\chi}',z)\;,
		 \end{equation}
where $\vec{\chi}'\equiv(\chi'\cos \phi,\chi'\sin \phi)$. In the spherical cylindrical coordiantes, $dM'(\vec{\chi}',z)=d^2{\vec{\chi}'}dz\rho(\vec{\chi}',z)$.  Therefore,  using Eqs.\eqref{eq:app_mlrev_def_ang_cont_0} and \eqref{eq:app_mlrev_surf_dens}, the deflection angle in becomes,
		 \begin{equation}\label{eq:app_mlrev_def_ang_cont_1}
		 	\vec{\hat{\theta}}_{\text{\tiny D}}(\vec{\chi}')=\frac{4 G_{\text{\tiny N}} }{c^2}\int\,d^2{\vec{\chi}'} \sigma(\vec{\chi}') \,\frac{\vec{\chi}-\vec{\chi}'}{\abs{\vec{\chi}-\vec{\chi}'}^2}\;.
		 \end{equation}
For an axially symmetric lens, since $\rho(\vec{\chi}',z)=\rho(\chi',z)$, and therefore the deflection angle in Eq.\,\eqref{eq:app_mlrev_def_ang_cont_1} simplifies to
\begin{equation}\label{eq:app_mlrev_def_ang_cont}
    \vec{\hat{\theta}}_{\text{\tiny D}}(\chi)=\frac{4 G_{\text{\tiny N}} \tilde{M}_{\text{\tiny L}}(\chi) }{c^2\chi}\hat{\chi}\;,
\end{equation}
		  where $\tilde{M}_{\text{\tiny L}}(\chi) $ is the mass enclosed inside the cylindrical tube with radius $\chi$ and center axis coinciding with the axis of symmetry, and we call it the mass function, which is given by
  \begin{equation}\label{eq:app_mlrev_mass_encl}
 \tilde{M}_{\text{\tiny L}}(\chi)=2\pi\int_{\text{\tiny 0}}^{\chi}\,d{\chi'}\,\chi'\sigma(\chi')\;.
  \end{equation}
Notice that in Eq.\,\eqref{eq:app_mlrev_def_ang_cont}, for an axially symmetric lens the deflection vector is aligned with the impact-parameter vector $\vec{\chi}$, i.e., it points radially outward from the lens centre toward the source. Consequently, the lens centre, the (unlensed) source position, and the image position are collinear when projected onto the lens plane. Furthermore, the magnitude of the deflection angle  $\vec{\hat{\theta}}_{\text{\tiny D}}(\chi)$ is directly proportional to the enclosed projected mass $\tilde{M}_{\text{\tiny L}}(\chi)$ and inversely proportional to the magnitude of the impact parameter $\chi$. 

From the geometry in Fig\,\ref{fig:gr_lens_fig}, we can obtain the relation between the angular positions of the source, image, and the deflection angle. It is given by
\begin{equation}\label{eq:app_mlrev_lens_eq_0}
    \vec{\theta}_{\text{\tiny S}}=\vec{\theta}_{\text{\tiny I}}-\frac{D_{\text{\tiny S}}-D_{\text{\tiny L}}}{D_{\text{\tiny S}}}\vec{\hat{\theta}}_{\text{\tiny D}}\;.
\end{equation}
Using Eqs.\eqref{eq:app_mlrev_def_ang_cont} and \eqref{eq:app_mlrev_lens_eq_0}, we obtained the lens equation for an axially symmetric lens,
\begin{equation}\label{eq:app_mlrev_lens_eq_2}
\theta_{\text{\tiny S}}=\theta_{\text{\tiny I}}-\frac{D_{\text{\tiny S}}-D_{\text{\tiny L}}}{D_{\text{\tiny S}}D_{\text{\tiny L}}}\frac{4 G_{\text{\tiny N}} {\tilde{M}_{\text{\tiny L}}(\theta_{\text{\tiny I}})} }{c^2\theta_{\text{\tiny I}}}\;.
\end{equation}
Here, we have used $\chi=D_{\text{\tiny L}}\theta_{\text{\tiny I}}$. Remember, since the deflection is in the direction of the impact parameter, the angular positions of the source, image, and deflection angle are collinear. This allows the vector relation in Eq. \eqref{eq:app_mlrev_lens_eq_2} to be simplified into a scalar equation.\footnote{For a mass distribution that lacks axial symmetry about the line of sight, this simplification no longer holds.}
		  
Furthermore, one may define a characteristic angular scale and a corresponding length scale for the system, referred to as the Einstein angle $ \theta_{\text{\tiny E}}$ and the Einstein radius  $R_{\text{\tiny E}}$, respectively, and are given by
 \begin{eqnarray}\label{eq:app_mlrev_eins_ang}
    \theta_{\text{\tiny E}}&=&\left[\frac{4 G_{\text{\tiny N}} M_{\text{\tiny L}} }{c^2}\frac{D_{\text{\tiny S}}-D_{\text{\tiny L}}}{D_{\text{\tiny S}}D_{\text{\tiny L}}}\right]^{1/2}\;,\\
    &\simeq&2.85 \text{ mas}\,\left(\frac{D_{\text{\tiny S}}}{D_{\text{\tiny L}}}-1\right)^{1/2}\left(\frac{D_{\text{\tiny S}}}{\text{1 }  \text{kpc}}\right)^{-1/2}\left(\frac{M_{\text{\tiny L}}}{\text{1 } \solarmass}\right)^{1/2}\nonumber \;, \\    
    R_{\text{\tiny E}}&\equiv&D_{\text{\tiny L}}\theta_{\text{\tiny E}} \simeq 2.87 \text{ AU}\,\left(1-\frac{D_{\text{\tiny L}}}{D_{\text{\tiny S}}}\right)^{1/2}\left(\frac{D_{\text{\tiny L}}}{\text{1 }  \text{kpc}}\right)^{1/2}\left(\frac{M}{\text{1 } \solarmass}\right)^{1/2}\;.
    \label{eq:aml_rev_eins_rad}
 \end{eqnarray}
Using the definition of the $ \theta_{\text{\tiny E}}$ in Eq.\,\eqref{eq:app_mlrev_eins_ang}, we may re-write the lens equation Eq.\,\eqref{eq:app_mlrev_lens_eq_2} into a more compact form as,
\begin{equation}\label{eq:app_mlrev_lens_eq_eins_ang}
    \theta_{\text{\tiny S}}=\theta_{\text{\tiny I}}-\frac{ \tilde{M}_{\text{\tiny L}}(\theta_{\text{\tiny I}}) }{{M}_{\text{\tiny L}} }\frac{\theta_{\text{\tiny E}}^2}{\theta_{\text{\tiny I}}}\;.
 \end{equation}
 For a given mass distribution and angular position of source, $\theta_{\text{\tiny S}}$, one may use above equation to determine the angular position of image(s) $\theta_{\text{\tiny I}}$. Notice that the lens equation effectively depends only on the Einstein angle $\theta_{\text{\tiny E}}$ and the 2-D enclosed mass fraction $ \tilde{M}_{\text{\tiny L}}(\theta_{\text{\tiny I}})/{M}_{\text{\tiny L}}$. Depending upon the angular source position $\theta_{\text{\tiny S}}$, the Einstein angle $\theta_{\text{\tiny E}}$ and the mass fraction $ \tilde{M}_{\text{\tiny L}}(\theta_{\text{\tiny I}})/{M}_{\text{\tiny L}}$, there may be multiple solutions to the lens equation and hence multiple images of the same source.
 
Additionally, each of these images is magnified relative to the unlensed source. The magnification is defined as the ratio of the solid angle of the image to that of the source and is given by
  \begin{equation}\label{eq:app_mlrev_axial_mag}
    \mathfrak{m}_{\text{\tiny I}}=\abs{\left(\frac{\theta_{\text{\tiny I}}}{\theta_{\text{\tiny S}}}\frac{d \theta_{\text{\tiny I}}}{d {\theta}_{\text{\tiny S}}}\right)}\;.
  \end{equation}
For certain source positions, the magnification of one or more images can formally diverge. The locus of these source positions $\theta_{\text{\tiny S,caust}}$ is called the caustics, while the corresponding locus of image positions,  $\theta_{\text{\tiny I,crit}}$, are the critical curves. In Sec.\,\ref{sec:aml_eds}, these curves mark the boundaries where pairs of lensed images are created or annihilated.

According to Liouville's theorem\,\cite{Schneider_1992}, gravitational lensing conserves the surface brightness of a source. Consequently, when the multiple images cannot be individually resolved, the change in magnification manifests as an amplification in the apparent brightness of the source. This amplification is proportional to the total magnification, which is given by
  \begin{equation}\label{eq:app_rev_axial_mag_tot}
    \mathfrak{m}_{\text{\tiny T}}=\sum_{I}\abs{\left(\frac{\theta_{\text{\tiny I}}}{\theta_{\text{\tiny S}}}\frac{d \theta_{\text{\tiny I}}}{d {\theta}_{\text{\tiny S}}}\right)}\;.
  \end{equation}
In addition to this photometric magnification, gravitational lensing also induces an astrometric shift in the centroid of the image. This centroid is formally defined as a magnification-weighted average of the individual angular image positions and is given by 
\begin{equation}\label{eq:app_rev_rev_cent_pos}
    \vec{\theta}_{\text{\tiny cent}}=\frac{\sum_{I}\vec{\theta}_{\text{\tiny I}} \mathfrak{m}_{\text{\tiny I}}}{\sum \mathfrak{m}_{\text{\tiny I}}}\;.
\end{equation}
In the absence of a lens, the centroid coincides with the true source position. However, in the presence of the lens, the centroid is displaced from the source position by an astrometric shift, defined as
\begin{equation}\label{eq:app_rev_cent_shift}
    \Delta{\vec{\theta}}_{\text{\tiny cent}}\equiv \vec{\theta}_{\text{\tiny cent}}-\vec{\theta}_{\text{\tiny S}}=\frac{\sum_{I}\vec{\theta}_{\text{\tiny I}} \mathfrak{m}_{\text{\tiny I}}}{\sum \mathfrak{m}_{\text{\tiny I}}}-\vec{\theta}_{\text{\tiny S}}\;.
\end{equation}
This astrometric shift $\Delta\vec{\theta}_{\text{\tiny cent}}$, is the primary observable in astrometric microlensing. For an axially symmetric lens, the centroid shift $\Delta\vec{\theta}_{\text{\tiny cent}}$ is always collinear with the image and source positions, due to the fact that the deflection occurs along the direction of the impact parameter. The temporal evolution of this astrometric shift, as the source and lens move relative to one another, traces a path on the sky whose shape is determined by the properties of the lens. 
 
We now model an event by assuming a point source moves on a rectilinear trajectory relative to the lens, with a constant transverse velocity  $v_{\text{\tiny S,T}}$ to the line of sight. We choose coordinates on the lens plane so that the source moves along the $x$ direction, and the minimum angular separation lies along the $y$ direction. Then the angular position of the source can be written compactly as  
\begin{equation}\label{eq:app_rev_src_traject}
		 \vec{\theta}_{\text{\tiny S}}(t) 
    = \begin{pmatrix}
        \dfrac{t - t_{0}}{t_{\text{\tiny E}}}\,\theta_{\text{\tiny E}} \\[8pt]
        \theta_{\text{\tiny S,0}}
    \end{pmatrix}\,, \,\quad \text{and ~}	{\theta}_{\text{\tiny S}}(t)=\sqrt{\theta_{\text{\tiny S,0}}^2+\left(\frac{t-t_{\text{\tiny 0}}}{t_{\text{\tiny E}}}\theta_{\text{\tiny E}}\right)^2}\;.
		  \end{equation}
where $t_{\text{\tiny E}}$ is the Einstein time (a timescale in microlensing events)
		 \begin{equation}\label{eq:app_rev_lens_ml_tE}
		 	t_{\text{\tiny E}}\equiv \frac{R_{\text{\tiny E}}}{v_{\text{\tiny S,T}}}\simeq 0.69\text{ yr}\,\left(1-\frac{D_{\text{\tiny L}}}{D_{\text{\tiny S}}}\right)^{1/2}\left(\frac{D_{\text{\tiny L}}}{\text{1 }  \text{kpc}}\right)^{1/2}\left(\frac{M}{\text{1 } \solarmass}\right)^{1/2}\left(\frac{v_{\text{\tiny S,T}}}{300 {\text{ kms$^{-1}$}}}\right)^{-1}\;,
		 \end{equation}
 and $\theta_{\text{\tiny S,0}}$ is the angular separation at the time of closest approach, $t_{0}$. 

For this rectilinear motion, the angular separation ${\theta}_{\text{\tiny S}}$ between the source and lens varies with time, modulating both the total magnification $ \mathfrak{m}_{\text{\tiny T}}(t)$ and the astrometric shift $\Delta{\vec{\theta}}_{\text{\tiny cent}}$. This produces the transient brightening characteristic of photometric microlensing (PML). Concurrently, the evolving astrometric shift vector, $\Delta{\vec{\theta}}_{\text{\tiny cent}}$, traces a characteristic trajectory on the sky, which is a key signal in astrometric microlensing (AML).

To illustrate these effects, we first consider the simple case of a point-mass lens, for which the lens equation Eq.\,\eqref{eq:app_mlrev_lens_eq_eins_ang} simplifies to 
\begin{equation}\label{eq:app_mlrev_lens_eq_eins_ang_point}
    \theta_{\text{\tiny S}}=\theta_{\text{\tiny I}}-\frac{\theta_{\text{\tiny E}}^2}{\theta_{\text{\tiny I}}}\;.
 \end{equation}
This quadratic equation has two solutions for the image positions,
\begin{equation}\label{eq:rev_mlrev_point_lens_image_sol_1}
   \theta_{\text{\tiny I(1,2)}}=\frac{\theta_{\text{\tiny S}}\pm\sqrt{\theta_{\text{\tiny S}}^2+4\theta_{\text{\tiny E}}^2}}{2}\;.
\end{equation}
The primary image I and secondary image II are at angular positions $\theta_{\text{\tiny I,1}}$ and $\theta_{\text{\tiny I,2}}$, respectively. Image I is located towards the source; however, image II is located on the opposite side of the lens. The total magnification of the two images is given by
		 \begin{equation}\label{eq:rev_mlrev_point_lens_mag}
		 	\mathfrak{m}_{\text{\tiny T}}=\frac{\theta_{\text{\tiny S}}^2+2\theta_{\text{\tiny E}}^2}{\theta_{\text{\tiny S}} \sqrt{\theta_{\text{\tiny S}}^2+4\theta_{\text{\tiny E}}^2}}\;.
		 \end{equation}
And the centroid shift is given by 
\begin{equation}\label{eq:rev_aml_centroid_change_1}
    \Delta\vec{\theta}_{\text{\tiny C}} =\frac{\vec{\theta}_{\text{\tiny S}}}{\theta_{\text{\tiny S}}^2+2\theta_{\text{\tiny E}}^2}\theta_{\text{\tiny E}}^2\;.
\end{equation}
For large source position, i.e. $\theta_{\text{\tiny S}}\gg \theta_{\text{\tiny E}}$, the magnification falls as $\mathfrak{m}_{\text{\tiny tot}}\sim1+(\theta_{\text{\tiny S}}/\theta_{\text{\tiny E}})^{-4}$. However, the centroid shift falls much more slowly as $\Delta {\theta}_{\text{\tiny C}}\sim {\theta_{\text{\tiny S}}}^{-1}$. 

We call an AML event detected when the shift of the centroid of the apparent
source position is greater than the detector's threshold $\sigma_{\text{\tiny a}}$
\begin{equation}\label{eq:rev_aml_centroid_det_crit}
     \Delta{\theta}_{\text{\tiny C}}\ge \sigma_{\text{\tiny a}}\,.
\end{equation}
This detection criteria is satisfied in for the annulur region $\theta_{\text{\tiny S-}}<\theta_{\text{\tiny S}}<\theta_{\text{\tiny S+}}$, whose boundary is determined by solving Eqs.\,\eqref{eq:rev_aml_centroid_change_1} and \eqref{eq:rev_aml_centroid_det_crit}. The location of the inner and outer radii is given by

\begin{equation}\label{eq:rev_aml_inn_out_radii}
  \theta_{\text{\tiny S,$\pm$}}=\theta_{\text{\tiny E}}\frac{\theta_{\text{\tiny E}}\pm\sqrt{\theta_{\text{\tiny E}}^2-8\sigma_{\text{\tiny a}}^2}}{2 \sigma_{\text{\tiny a}}}\;.
\end{equation}
Notice that for $\theta_{\text{\tiny E}}<\sigma_{\text{\tiny a}}(2\sqrt{2})$, $\theta_{\text{\tiny S,$\pm$}}$ are imaginary meaing  the point lens is unable to produce $\sigma_{\text{\tiny a}}$ astrometric shift. This also corresponds to the maximum shift produced by the point lens is $\theta_{\text{\tiny E}}/(2\sqrt{2})$ at the source position $\theta_{\text{\tiny S}} = \sqrt{2} \theta_{\text{\tiny E}}$.  

\subsection{Uniform density sphere (UDS)}
\label{app:mlrev_UDS}
We now derive the lens equation for a uniform-density sphere, which will be applied in the AML analysis of Sec.\,\ref{sec:aml_eds_uds}. The mass distribution of such a sphere, with radius $R_{\text{\tiny u}}$ and total mass $M_{\text{\tiny u}}$, is characterized by the density profile
		\begin{equation}\label{eq:app_uds_dens_prof_0}
			\rho_{\text{\tiny  u}}(r)=\begin{cases}
				\frac{3M_{\text{\tiny  u}}}{4\pi R_{\text{\tiny  u}}^3}\,,& r\le R_{\text{\tiny  u}}\,, \\
				0\,,& r> R_{\text{\tiny  u}}\,.
			\end{cases}
		\end{equation}
Applying Eqs.\,\eqref{eq:app_mlrev_surf_dens} and \eqref{eq:app_mlrev_mass_encl}, we obtain the 2-D projected mass function for the uniform-density sphere (UDS) lens as
		\begin{equation}\label{eq:app_uds_mass_func}
			\tilde{M}_{\text{\tiny  u}}(\theta_{\text{\tiny I}})=\begin{cases}
				M_{\text{\tiny  u}}\left[1-\left(1-\frac{\theta_{\text{\tiny I}}^2/\theta_{\text{\tiny E}}^2}{R_{\text{\tiny  u}}^2/R_{\text{\tiny E}}^2}\right)^{3/2}\right]\,,& \theta_{\text{\tiny I}}/\theta_{\text{\tiny E}} <  R_{\text{\tiny  u}}/R_{\text{\tiny E}}\,, \\
				M_{\text{\tiny  u}}\,,& \theta_{\text{\tiny I}}/\theta_{\text{\tiny E}} \ge  R_{\text{\tiny  u}}/R_{\text{\tiny E}} \,.
			\end{cases}
		\end{equation}
Here, we have used $\chi=D_{\text{\tiny L}}\theta_{\text{\tiny I}}$.  Using Eqs.\,\eqref{eq:app_mlrev_lens_eq_eins_ang} and \eqref{eq:app_uds_mass_func}, we get the lens equation 
\begin{equation}\label{eq:app_eds_uds_lens_eq_1}
    \theta_{\text{\tiny S}}=\begin{cases}
        \theta_{\text{\tiny I}}-\frac{\theta_{\text{\tiny E}}^2}{\theta_{\text{\tiny I}}}\left[1-\left(1-\frac{\theta_{\text{\tiny I}}^2/\theta_{\text{\tiny E}}^2}{R_{\text{\tiny  u}}^2/R_{\text{\tiny E}}^2}\right)^{3/2}\right]\,,& \theta_{\text{\tiny I}}/\theta_{\text{\tiny E}} <  R_{\text{\tiny  u}}/R_{\text{\tiny E}}\,, \\
        \theta_{\text{\tiny I}}-\frac{\theta_{\text{\tiny E}}^2}{\theta_{\text{\tiny I}}}\,,& \theta_{\text{\tiny I}}/\theta_{\text{\tiny E}} \ge  R_{\text{\tiny  u}}/R_{\text{\tiny E}}\,.
    \end{cases}
\end{equation}
Note that Eq.\,\eqref{eq:aml_eds_uds_lens_eq_1} is defined piecewise. As a result, while each branch of the equation may yield multiple mathematical solutions, only the solution satisfying its specific condition is physically meaningful. It is convenient to further simplify the lens equation, Eq.\,\eqref{eq:app_eds_uds_lens_eq_1}, to better understand the behavior of its solutions. The simplified lens equation is given by 
\begin{eqnarray}\label{eq:app_eds_uds_lens_eq_2}
        \theta_{\text{\tiny I}}^5+\theta_{\text{\tiny I}}^3\theta_{\text{\tiny E}}^2 \left(\frac{R_{\text{\tiny  u}}^6}{R_{\text{\tiny  E}}^6}-\frac{3R_{\text{\tiny  u}}^2}{R_{\text{\tiny  E}}^2}\right)-2 \theta_{\text{\tiny I}}^2\theta_{\text{\tiny S}}\theta_{\text{\tiny E}}^2 \frac{R_{\text{\tiny  u}}^6}{R_{\text{\tiny  E}}^6}\nonumber \quad \quad&&\\ +\theta_{\text{\tiny I}}\theta_{\text{\tiny E}}^4 \left(\frac{R_{\text{\tiny  u}}^6}{R_{\text{\tiny  E}}^6} \frac{\theta_{\text{\tiny S}}^2}{\theta_{\text{\tiny E}}^2}-\frac{2R_{\text{\tiny  u}}^6}{R_{\text{\tiny  E}}^6}+\frac{3R_{\text{\tiny  u}}^4}{R_{\text{\tiny  E}}^4}\right)+2 \theta_{\text{\tiny S}}\theta_{\text{\tiny E}}^4\frac{R_{\text{\tiny  u}}^6}{R_{\text{\tiny  E}}^6}&=&0\;, \quad \nonumber  \theta_{\text{\tiny I}}/\theta_{\text{\tiny E}} <  R_{\text{\tiny  u}}/R_{\text{\tiny E}}\,, \\
        \theta_{\text{\tiny I}}^2-\theta_{\text{\tiny S}}\theta_{\text{\tiny I}}-\theta_{\text{\tiny E}}^2&=& 0\,,\quad  \theta_{\text{\tiny I}}/\theta_{\text{\tiny E}} \ge  R_{\text{\tiny  u}}/R_{\text{\tiny E}}\,.
\end{eqnarray}
Note that the first equation is a quintic and, as such, does not admit a general analytical solution. Therefore, $\theta_{\text{\tiny I}}/\theta_{\text{\tiny E}} < R_{\text{\tiny u}}/R_{\text{\tiny E}}$, it has five possible solutions. Applying Descartes' rule of signs \cite{barnard1959higher}, there can be either two or zero negative roots and exactly one positive root.

For $\theta_{\text{\tiny I}}/\theta_{\text{\tiny E}} \ge R_{\text{\tiny u}}/R_{\text{\tiny E}}$, the second equation is quadratic and thus admits two solutions. However, the actual number of physically relevant images for the piecewise lens equation, Eq.\,\eqref{eq:app_eds_uds_lens_eq_2}, must be determined numerically, and the details are provided in Sec.\,\ref{sec:aml_eds_uds}. 

Interestingly, when the observer, source, and the lens are collinear, i.e. $\theta_{\text{\tiny S}}=0$, the lens equation admits the solution,
		\begin{equation}\label{eq:app_uds_lens_eq_sol_us_0}
			 \theta_{\text{\tiny I,(1)}}=0\;, \quad \theta_{\text{\tiny I(2,3)}}=\begin{cases}
					\pm \theta_{\text{\tiny E}}\,,&  R_{\text{\tiny  u}}<R_{\text{\tiny E}}\;, \\
					\pm \frac{\theta_{\text{\tiny E}}R_{\text{\tiny  u}}}{R_{\text{\tiny  E}}^5\sqrt{2}}\left(3R_{\text{\tiny  E}}^4-R_{\text{\tiny  u}}^4-\sqrt{(R_{\text{\tiny  u}}^2-R_{\text{\tiny  E}}^2)^3(\tilde{R}_{\text{\tiny  u}}^2+3R_{\text{\tiny  E}}^2)}\right)^{1/2}\,,& R_{\text{\tiny  E}}<R_{\text{\tiny u}}<\sqrt{\frac{3}{2}}R_{\text{\tiny  E}}\;,\\
                   \text{No secondary images}\,,& R_{\text{\tiny u}}\ge\sqrt{\frac{3}{2}}R_{\text{\tiny  E}}\;, \\
			\end{cases}
		\end{equation}
For $R_{\text{\tiny u}} < R_{\text{\tiny E}}$, one image forms at $\theta_{\text{\tiny I}} = 0$, accompanied by a ring-like image (the Einstein ring) at $\theta_{\text{\tiny I}} = \theta_{\text{\tiny E}}$. In the regime $R_{\text{\tiny E}} < R_{\text{\tiny u}} < \sqrt{3/2},R_{\text{\tiny E}}$, Image I remains at the same position, but the Einstein ring appears at a smaller angular radius. When $R_{\text{\tiny u}} \ge \sqrt{3/2},R_{\text{\tiny E}}$, the Einstein ring disappears entirely, leaving only the primary image.

Furthermore, by applying Eqs.\,\eqref{eq:app_eds_uds_lens_eq_2} and \eqref{eq:app_mlrev_axial_mag}, we identify the caustics ($\theta_{\text{\tiny S,caust}}$) and critical curves ($\theta_{\text{\tiny I,crit}}$), which correspond to angular source and image positions where the total magnification formally diverges. For the UDS lens, these features exist only when $R_{\text{\tiny u}} < \sqrt{3/2},R_{\text{\tiny E}}$, and their locations are given by
\begin{eqnarray}\label{eq:app_eds_uds_crit_cur_caust}
    \abs{\theta_{\text{\tiny I,crit}}}&=& \frac{R_{\text{\tiny  u}}\theta_{\text{\tiny E}}}{2\sqrt{2}R_{\text{\tiny  E}}^3}\left(\sqrt{R_{\text{\tiny  u}}^8-32R_{\text{\tiny  u}}^2R_{\text{\tiny  E}}^6+48R_{\text{\tiny  E}}^8}-R_{\text{\tiny  u}}^4\right)^{1/2}\;,\\
    \abs{\theta_{\text{\tiny S,caust}}}&=&\frac{\theta_{\text{\tiny E}}}{8\sqrt{2}R_{\text{\tiny  u}}R_{\text{\tiny  E}}^3\left(\sqrt{R_{\text{\tiny  u}}^8-32R_{\text{\tiny  u}}^2R_{\text{\tiny  E}}^6+48R_{\text{\tiny  E}}^8}-R_{\text{\tiny  u}}^4\right)^{1/2}}\left(4R_{\text{\tiny  u}}^6+32R_{\text{\tiny  E}}^6\right. \\
    &&\quad -4R_{\text{\tiny  u}}^2\sqrt{R_{\text{\tiny  u}}^8-32R_{\text{\tiny  u}}^2R_{\text{\tiny  E}}^6+48R_{\text{\tiny  E}}^8}-\left.\sqrt{2}\left(R_{\text{\tiny  u}}^4+8R_{\text{\tiny  E}}^4-\sqrt{R_{\text{\tiny  u}}^8-32R_{\text{\tiny  u}}^2R_{\text{\tiny  E}}^6+48R_{\text{\tiny  E}}^8}\right)^{3/2}\right)\;.\nonumber
\end{eqnarray}

The critical curve is located at the local maximum of the lens equation (see Fig.\,\ref{fig:uds_massfunc_vs_ui}). Physically, this corresponds to the angular position where the deflection due to the lens reaches an extremum, and it plays a key role in the formation of multiple images. When the source crosses the corresponding caustic, two additional images appear or disappear, leading to a sudden increase in image multiplicity. This behavior highlights the unique astrometric features of extended lenses compared to the point-lens case, where the critical curve and caustics are determined solely by the Einstein radius.

\section{Calculation of lensing probability for AML}
\label{app:lens_prob}
In this appendix, we present the derivation of the probability for a star to undergo astrometric microlensing (AML). This formulation provides the foundation for calculating the AML probabilities for UDS and BS lenses discussed in Sec.\,\ref{sec:aml_lens_prob_no_lens_prob}. For this analysis, we first consider a simplified AML configuration in which the source moves in rectilinear motion with tangential speed $v_{\text{\tiny S,T}}$\footnote{The gravitational microlensing is sensitive to the velocity tangential to the lens plane.} relative to a lens. This framework can be readily extended to cases where both the lens and source are in motion\,\cite{Verma:2022pym}. We also neglect parallax effects arising from Earth's orbital motion (see, e.g.\,\cite{Dominik_2000,Li_2012}).

The motion of the source is projected onto the lens plane, yielding an angular velocity $\mu \approx v_{\text{\tiny S,T}} / D_{\text{\tiny S}}$. The corresponding effective transverse speed in the lens plane is then $v_{\text{\tiny T}} \equiv \mu D_{\text{\tiny L}}=v_{\text{\tiny S,T}} D_{\text{\tiny L}}/ D_{\text{\tiny S}}$. This parameter is crucial for determining the characteristic duration of lensing events detailed in Sec.\,\ref{sec:aml_lens_prob_no} and, consequently, for estimating their occurrence probability.

As discussed in Sec.\,\ref{sec:aml_lens_prob_no}, Gaia can effectively detect AML events when the displacement of a star’s trajectory exceeds the instrument’s astrometric resolution, i.e.
\begin{equation}\label{eq:app_AML_lens_class_lens_crit}
   \abs{\Delta \theta_{\text{\tiny cent}}}>\sigma_{\text{\tiny a} }(m_{\text{\tiny G} })\,.
\end{equation}
Examining the behaviour of centroid shift of extended lens in Figs.\,\ref{fig:uds_massfunc_vs_ui} and \ref{fig:bs_massfunc_vs_ui}, we find that the detection criterion in Eq.\,\ref{eq:app_AML_lens_class_lens_crit} is satisfied within an annular region defined by $\theta_{\text{\tiny S-}} < \theta_{\text{\tiny S}} < \theta_{\text{\tiny S+}}$ around the lens. The boundaries $\theta_{\text{\tiny S+}}$ and $\theta_{\text{\tiny S-}}$ correspond to the source positions where $\lvert \Delta \theta_{\text{\tiny cent}}(\theta_{\text{\tiny S+}}) \rvert = \lvert \Delta \theta_{\text{\tiny cent}}(\theta_{\text{\tiny S-}}) \rvert = \sigma_{\text{\tiny a}}(m_{\text{\tiny G}})$. 

The time $t_{\text{\tiny e}}$ that a source spends inside this annular region along its trajectory sets the characteristic timescale of the AML event. The average event duration can then be defined by averaging $t_{\text{\tiny e}}$ over all trajectories crossing the annulus\,\cite{Verma:2022pym}. Accordingly, the mean event duration is $\bar{t}_{\text{\tiny e}} = \bar{l}D_{\text{\tiny S}}/v_{\text{\tiny S,T}}$, where $\bar{l}$ is the average chord length within the annular region. Hence, the average AML event timescale is given by
\begin{equation}\label{eq:app_AML_lens_t_e}
   \bar{t}_{\text{\tiny e}}=\frac{\pi\left(\theta_{\text{\tiny S+}}^2-\theta_{\text{\tiny S-}}^2\right) D_{\text{\tiny S}}}{2\theta_{\text{\tiny S+}}v_{\text{\tiny S,T}}}\;.
\end{equation}
For most astrometric shifts, we see that $\theta_{\text{\tiny S-}}\ll\theta_{\text{\tiny S+}}$, hence, we can conveniently take $\bar{t}_{\text{\tiny e}}\sim\pi \theta_{\text{\tiny S+}}D_{\text{\tiny S}}/(v_{\text{\tiny S,T}})$.  As described in Sec.\,\ref{sec:aml_lens_prob_no}, based on event duration $t_{\text{\tiny e}}$,  total observation time $t_{\text{\tiny obs}}$, and observation cadence $t_{\text{\tiny s}}$, we can classify astrometric microlensing events into three categories: short-duration lensing events (SDLEs), intermediate-duration lensing events (IDLEs), and long-duration lensing events (LDLEs). We now proceed to calculate the lensing probability for each event type. Initially, this calculation is carried out under the simplifying assumption of a uniform velocity distribution for the lenses. We will subsequently refine the analysis to incorporate a Maxwell–Boltzmann velocity distribution.

It is important to note that our kinematic framework, which may, for simplicity, assume a stationary lens, is fully general. All lensing observables (such as $t_e$ and the probability cross-section) depend only on the relative transverse angular velocity, $\mu = |\vec{v}_{\text{\tiny S,T}}/D_{\text{\tiny S}} - \vec{v}_{\text{\tiny L,T}}/D_{\text{\tiny L}}|$. Therefore, our results are independent of the specific reference frame and apply universally, whether it is the source, the lens, or both that are in motion.

We will first calculate conditional probability $p_{\text{\tiny AML}}^{\text{\tiny c}}$ for a given star, characterized by its apparent magnitude ($m_{\text{\tiny G}},D_{\text{\tiny S}}, \alpha, \delta$) to undergo AML due to a lens situated at a distance $D_\text{\tiny L}$ along the line of sight, within a patch of the sky subtending a solid angle $\Delta \Omega$. We consider this patch to be sufficiently large so that it encompasses the full trajectory of the source star during the observation period. As the source moves across this patch over the total observation time $t_{\text{\tiny obs}}$, it traverses an angular distance $v_{\text{\tiny S,T}} t_{\text{\tiny obs}}/D_{\text{\tiny S}}$ on the sky.

Now, suppose a lens lies along the line of sight within this patch such that the source star’s trajectory passes within the angular impact parameter $\theta_{\text{\tiny S+}}$ with respect to the lens.  In this case, a segment of the trajectory will experience an astrometric shift exceeding Gaia’s threshold resolution $\sigma_{\text{\tiny a} }(m_{\text{\tiny G} })$, thereby producing a detectable AML event. The actual event duration depends on the specific trajectory of the star, particularly its minimum impact parameter relative to the lens. For practical purposes, however, we consider an average event duration $\bar{t}_{\text{\tiny e}}$ (see Eq.\,\eqref{eq:app_AML_lens_t_e}), which represents a typical AML event time averaged over a distribution of impact parameters.

We can now compute the conditional probability $p_{\text{\tiny AML}}^{\text{\tiny c}}$ for IDLEs ($t_{\text{\tiny s}} < \bar{t}_{\text{\tiny e}} < t_{\text{\tiny obs}}$). In this case, only a portion of the source star’s trajectory exhibits a significant deviation from straight-line motion. The effective area on the sky where the trajectory is appreciably deflected can be approximated as an imaginary rectangle with angular width $2\theta_{\text{\tiny S+}}$ as it moves on the sky. The length of this rectangle corresponds to the angular distance traversed by the source over the total observation time $v_{\text{\tiny S,T}} t_{\text{\tiny obs}}/D_{\text{\tiny S}}$. Consequently, the area of this rectangle is $2\theta_{\text{\tiny S+}}\times v_{\text{\tiny S,T}} t_{\text{\tiny obs}}/D_{\text{\tiny S}} $. Therefore, the conditional probability $p_{\text{\tiny AML}}^{\text{\tiny c}}$ is the ratio of this rectangle to the total patch under study,
\begin{equation}\label{eq:app_AML_prob_IDLE}
 p_{\text{\tiny IDLE}}^{\text{\tiny c}}=\frac{2\theta_{\text{\tiny S+}}v_{\text{\tiny S,T}} t_{\text{\tiny obs}}}{D_{\text{\tiny S}}\Delta \Omega}\;.
\end{equation}
For SDLEs ($ \bar{t}_{\text{\tiny e}}  < t_{\text{\tiny s}}$), Gaia will typically miss most events because the event duration is shorter than the observation cadence. However, there remains a chance that a star is observed coincidentally during an SDLE, producing a measurable astrometric deviation from its rectilinear trajectory. This happens when the sampling happens coincidentally during the events. In such cases, the conditional probability $p_{\text{\tiny AML}}^{\text{\tiny c}}$ is suppressed by the conditional factor $\bar{t}_{\text{\tiny e}}/t_{\text{\tiny S}}$. which accounts for the likelihood of the event coinciding with a sampling instance. Therefore, for SDLEs , the conditional probability $p_{\text{\tiny AML}}^{\text{\tiny c}}$ is given by
\begin{equation}\label{eq:app_AML_prob_SDLE}
 p_{\text{\tiny SDLE}}^{\text{\tiny c}}=\frac{2\theta_{\text{\tiny S+}}v_{\text{\tiny S,T}} t_{\text{\tiny obs}}}{D_{\text{\tiny S}}\Delta \Omega}\frac{\bar{t}_{\text{\tiny e}}}{t_{\text{\tiny S}}}\;.
\end{equation}
For LDLEs ($\bar{t}_{\text{\tiny e}} > t_{\text{\tiny obs}}$), the event duration is larger than Gaia's observation time. As a result, only a portion of the star’s lensed trajectory is sampled during the observation period. In this scenario, there is generally no rectilinear reference segment of the trajectory because the star is undergoing astrometric microlensing throughout the entire observation. A practical detection criterion can then be defined by requiring that the difference in astrometric shift between the start and end of the mission exceeds Gaia’s astrometric resolution, $\sigma_{\text{\tiny a} }(m_{\text{\tiny G} })$. This criterion effectively sets the width of the imaginary rectangle discussed earlier a $2\sqrt{\frac{v_{\text{\tiny S,T}}t_{\text{\tiny obs}}}{D_{\text{\tiny S}}\sigma_{\text{\tiny a} }(m_{\text{\tiny G} })} }\theta_{\text{\tiny E}}$\,\cite{Dominik_2000}. Consequently,  the conditional probability $p_{\text{\tiny AML}}^{\text{\tiny c}}$ for LDLEs is given by

\begin{equation}\label{eq:app_AML_prob_LDLE}
 p_{\text{\tiny SDLE}}^{\text{\tiny c}}=2\sqrt{\frac{v_{\text{\tiny S,T}}t_{\text{\tiny obs}}}{D_{\text{\tiny S}} \sigma_{\text{\tiny a} }(m_{\text{\tiny G} })} } \theta_{\text{\tiny E}}\times\frac{v_{\text{\tiny S,T}} t_{\text{\tiny obs}}}{D_{\text{\tiny S}}\Delta \Omega}\;.
\end{equation}
Combining Eqs.\,\eqref{eq:app_AML_prob_IDLE}, \eqref{eq:app_AML_prob_SDLE}, and \eqref{eq:app_AML_prob_LDLE}, the conditional probability $p_{\text{\tiny AML}}^{\text{\tiny c}}$ of AML of a given star due to a lens situated at a distance $D_\text{\tiny L}$ is given by

\begin{equation}\label{eq:app_AML_lens_cond_prob}
p_{\text{\tiny AML}}^{\text{\tiny c}} =
\begin{cases}
  \displaystyle
  \frac{2 \theta_{\text{\tiny S+}}}{  \Delta \Omega}
  \,   \frac{v_{\text{\tiny S,T}}t_{\text{\tiny obs}}}{D_{\text{\tiny S}}} \, 
  \, \frac{\bar{t}_{\text{\tiny e}}}{t_{\text{\tiny s}}}
  & ; \, \bar{t}_{\text{\tiny e}} < t_{\text{\tiny s}}, \\[1.2em]

  \displaystyle
  \frac{2 \theta_{\text{\tiny S+}}}{  \Delta \Omega}
  \,   \frac{v_{\text{\tiny S,T}}t_{\text{\tiny obs}}}{D_{\text{\tiny S}}}
  & ; \, t_{\text{\tiny s}} < \bar{t}_{\text{\tiny e}} < t_{\text{\tiny obs}}, \\[1.2em]

  \displaystyle
 \frac{2 \theta_{\text{\tiny E}}}{ \Delta \Omega \sqrt{\sigma_{\text{\tiny a} }(m_{\text{\tiny G} })} }
  \,   \left(\frac{v_{\text{\tiny S,T}}t_{\text{\tiny obs}}}{D_{\text{\tiny S}}}\right)^{3/2}
  
  & ; \, t_{\text{\tiny obs}} < \bar{t}_{\text{\tiny e}} .
\end{cases}
\end{equation}
Eq.\,\eqref{eq:app_AML_lens_cond_prob} allows us to systematically calculate the probability for any specific star, taking into account both the geometry of the lensing system and the temporal sampling of observations.
Since all lensing observables depend only on the relative transverse angular velocity, $\mu = |\vec{v}_{\text{\tiny S,T}}/D_{\text{\tiny S}} - \vec{v}_{\text{\tiny L,T}}/D_{\text{\tiny L}}|$, the preceding equations can be generalized simply by replacing the source's transverse angular velocity (${v}_{\text{\tiny S,T}}/D_{\text{\tiny S}}$) with $\mu$, which gives 

\begin{equation}\label{eq:app_AML_lens_cond_prob_1}
p_{\text{\tiny AML}}^{\text{\tiny c}} =
\begin{cases}
  \displaystyle
 \frac{2 \theta_{\text{\tiny S+}}}{  \Delta \Omega}
  \,   \mu t_{\text{\tiny obs}} \, 
  \, \frac{\bar{t}_{\text{\tiny e}}}{t_{\text{\tiny s}}}
  & ; \, \bar{t}_{\text{\tiny e}} < t_{\text{\tiny s}}, \\[1.2em]

  \displaystyle
 \frac{2 \theta_{\text{\tiny S+}}}{  \Delta \Omega}
  \,   \mu  t_{\text{\tiny obs}}
  & ; \, t_{\text{\tiny s}} < \bar{t}_{\text{\tiny e}} < t_{\text{\tiny obs}}, \\[1.2em]

  \displaystyle
  \frac{2 \theta_{\text{\tiny E}}}{ \Delta \Omega \sqrt{\sigma_{\text{\tiny a} }(m_{\text{\tiny G} })} }
  \,   \left(\mu  t_{\text{\tiny obs}}\right)^{3/2}
  
  & ; \, t_{\text{\tiny obs}} < \bar{t}_{\text{\tiny e}} .
\end{cases}
\end{equation}
Therefore, the average event duration in Eq.\,\eqref{eq:app_AML_lens_t_e}, in this generalized case can be written as 
\begin{equation}\label{eq:app_AML_lens_t_e_1}
   \bar{t}_{\text{\tiny e}}=\frac{\pi\left(\theta_{\text{\tiny S+}}^2-\theta_{\text{\tiny S-}}^2\right) }{2\theta_{\text{\tiny S+}}\mu}\;.
\end{equation}

So far, the conditional probability has been calculated under the simplifying assumption that dark lenses have the same transverse velocity with respect to the source stars or vice versa. In reality, dark matter lenses possess their own intrinsic velocity distribution. To account for this, we incorporate a simplistic galactic dark matter velocity distribution, which we model using a Maxwell-Boltzmann profile (see \cite{Griest_1991} for similar treatment), expressed as
\begin{equation}\label{eq:app_max_dist}
			f(v)d^3\vec{v}=\frac{1}{(\pi v_{\text{\tiny c}}^2 )^{3/2}}e^{-v^2/v_{\text{\tiny c}}^2}d^3\vec{v}
\end{equation}
		 where $v_{\text{\tiny c}}=\sqrt{2}~v_{\text{\tiny disp}}\approx 220~\text{Km~s}^{-1} $, and   $v_{\text{\tiny disp}}$ denotes the velocity dispersion of dark matter in the Solar neighborhood. Also, $v\equiv \abs{\vec{v}}$ is the magnitude of the lens velocity. To account for the intrinsic motion of dark lenses, the conditional probability for a star to undergo astrometric microlensing is obtained by averaging $p_{\text{\tiny AML}}^{\text{\tiny c}}$ in Eq.\,\eqref{eq:app_AML_lens_cond_prob} over the Maxwell-Boltzmann velocity distribution given in Eq.\,\eqref{eq:app_max_dist} and is given by
\begin{equation}\label{eq:app_avfg_lens_conf_prob}
   \expval{p_{\text{\tiny AML}}^{\text{\tiny c}}}=\int d^3\vec{v}_{\text{\tiny L}} ~f(v_{\text{\tiny L}}) ~p_{\text{\tiny AML}}^{\text{\tiny c}}\;.
\end{equation}
For simplicity, we assume the sources are stationary in further calculation. Now, $p_{\text{\tiny AML}}^{\text{\tiny c}}$ depends upon the magnitude of the tangential velocity $v_{\text{\tiny L,T}}$ along the lens plane (see Eq.\,\eqref{eq:app_AML_lens_cond_prob_1}). It is therefore convenient to perform the velocity integral in cylindrical coordinates $(v_{\text{\tiny L,T}},\phi,v_{\text{\tiny L,z}})$. Here, $v_{\text{\tiny L,z}}$ is the component along the line of sight, $v_{\text{\tiny L,T}}$ is the tangential speed, and $\phi$ is the azimuthal angle in the lens plane. In these coordinates, the phase space volume element becomes $ d^3\vec{v}_{\text{\tiny L}}=v_{\text{\tiny L,T}}dv_{\text{\tiny L,T}}dv_{\text{\tiny L,z}}d\phi$. Integrating Eq.\,\eqref{eq:app_avfg_lens_conf_prob} over $v_{\text{\tiny L,z}}$ and $\phi$, we obtain
\begin{equation}\label{eq:app_max_dist_0}
    \expval{p_{\text{\tiny AML}}^{\text{\tiny c}}}=\frac{2}{v_{\text{\tiny c}}^2}\int dv_{\text{\tiny L,T}} ~v_{\text{\tiny L,T}}e^{-v_{\text{\tiny L,T}}^2/v_{\text{\tiny c}}^2} ~p_{\text{\tiny AML}}^{\text{\tiny c}}\;.
\end{equation}

Taking $v_{\text{\tiny S,T}}=0$ in Eq.\,\eqref{eq:app_AML_lens_t_e_1} we can find the relation between average time duration with the tangential velocity i.e
\begin{equation}\label{eq:app_max_v_t_rel}
    v_{\text{\tiny L,T}}=\frac{\pi D_{\text{\tiny L}}(\theta_{\text{\tiny S+}}^2-\theta_{\text{\tiny S-}}^2)}{2\theta_{\text{\tiny S+}}\bar{t}_{\text{\tiny e}}}\;.
\end{equation}
It is important to note that $p_{\text{\tiny AML}}^{\text{\tiny c}}$, as given in Eq.\,\eqref{eq:app_AML_lens_cond_prob_1}, depends on the event duration and, consequently, the event type (SDLE, IDLE, or LDLE). Consequently, the integral over the tangential velocity (i.e. Eq.\,\eqref{eq:app_max_dist_0}) must be performed piecewise, as the different event types make distinct contributions to the final averaged probability, $\expval{p_{\text{\tiny AML}}^{\text{\tiny c}}}$. To make this convenient, we change the variable of integration from $v_{\text{\tiny L,T}}$ to $\bar{t}_{\text{\tiny e}}$ using Eq.\,\eqref{eq:app_max_v_t_rel}. This allows us to explicitly integrate the distinct contributions from each event type.
Performing this substitution in Eqs.\,\eqref{eq:app_max_dist_0} and \eqref{eq:app_AML_lens_cond_prob_1} and integrating over the appropriate range for each event type yields,

\begin{align}\label{eq:app_AML_lens_cond_prob_bolt}
    \expval{p_{\text{\tiny AML}}^{\text{\tiny c}}}&= \frac{\pi (\theta_{\text{\tiny S+}}^2-\theta_{\text{\tiny S-}}^2)  
     t_{\text{\tiny obs}}}{\Delta \Omega \, t_{\text{\tiny s}}}  
     e^{-t_{\text{\tiny c}}^2/t_{\text{\tiny s}}^2}&&\text{\textbf{(SDLE)}} \nonumber\\
    &\quad +\, \frac{\pi  (\theta_{\text{\tiny S+}}^2-\theta_{\text{\tiny S-}}^2)  
     t_{\text{\tiny obs}}}{ \Delta \Omega \, t_{\text{\tiny c}}}  
     \big( g(t_{\text{\tiny obs}})-g(t_{\text{s}}) \big)&& \text{\textbf{(IDLE)}} \nonumber\\
    &\quad +\, \sqrt{\frac{\pi^3 \theta_{\text{\tiny E}}^2 (\theta_{\text{\tiny S+}}^2-\theta_{\text{\tiny S-}}^2)^{3}  
     t_{\text{\tiny obs}}^3}{2 \sigma_{\text{\tiny a} }(m_{\text{\tiny G} }) \, \Delta \Omega^2 \, \theta_{\text{\tiny S+}}^{3}t_{\text{\tiny c}}^3}}
     \Big( \Gamma(\tfrac{7}{4}) - \Gamma\left(\frac{7}{4}, \frac{t_{\text{\tiny c}}^2}{t_{\text{\tiny obs}}}\right) \Big) \quad && \text{\textbf{(LDLE)}}\;.
\end{align}

Here, the function $g(t)$ is defined as $g(t)=t_{\text{\tiny c}}\exp\left(-t_{\text{\tiny c}}^2/t^2\right)/t-\sqrt{\pi}\text{Erf}\left(t_{\text{\tiny c}}/t\right)/{2}$ and $t_{\text{\tiny c}}=\pi D_{\text{\tiny L}}(\theta_{\text{\tiny S+}}^2-\theta_{\text{\tiny S-}}^2)/(2\theta_{\text{\tiny S+}}v_{\text{\tiny c}})$. Alternatively, we may perform the integration in Eq.\,\eqref{eq:app_max_v_t_rel} directly over the variable $v_{\text{\tiny L,T}}$, by converting the piecewise conditional probability function from the time domain to the velocity domain. Following this procedure yields the same result as given in Eq.\,\eqref{eq:app_AML_lens_cond_prob_bolt}.

As detailed in Sec.\,\ref{sec:gaia_detectibility}, AML events with durations $\bar{t}_{\text{\tiny e}}<2~\text{years}$ are significantly contaminated by statistical backgrounds arising from random astrometric noise. To obtain robust constraints, we therefore apply a conservative cut on the event duration, $\bar{t}_{\text{\tiny e,cut}}=2~\text{years}$. As previously discussed, this criterion excludes all SDLEs and a subset of IDLEs for which $\bar{t}_{\text{\tiny e}}<2~\text{years}$. Since the total observation time $t_{\text{\tiny obs}}>2~\text{years}$, the class of LDLEs remains unaffected. We incorporate this cut into Eq.\,\eqref{eq:app_max_dist_0} by setting the minimum event duration to $\bar{t}_{\text{\tiny e,cut}}=2~\text{years}$, and following the same procedure as before, we obtain  

\begin{align}\label{eq:app_AML_lens_cond_prob_bolt_cut}
    \expval{p_{\text{\tiny AML}}^{\text{\tiny c}}}&= 
    \frac{\pi  (\theta_{\text{\tiny S+}}^2-\theta_{\text{\tiny S-}}^2)  
     t_{\text{\tiny obs}}}{ \Delta \Omega \, t_{\text{\tiny c}}}  
     \big( g(t_{\text{\tiny obs}})-g(\bar{t}_{\text{\tiny e,cut}}) \big)&&\text{\textbf{(IDLE)}}\nonumber\\
    &\quad +\, \sqrt{\frac{\pi^3 \theta_{\text{\tiny E}}^2 (\theta_{\text{\tiny S+}}^2-\theta_{\text{\tiny S-}}^2)^{3}  
     t_{\text{\tiny obs}}^3}{2 \sigma_{\text{\tiny a} }(m_{\text{\tiny G} }) \, \Delta \Omega^2 \, \theta_{\text{\tiny S+}}^{3}t_{\text{\tiny c}}^3}}
     \Big( \Gamma(\tfrac{7}{4}) - \Gamma\left(\frac{7}{4}, \frac{t_{\text{\tiny c}}^2}{t_{\text{\tiny obs}}}\right) \Big) \quad && \text{\textbf{(LDLE)}}\;.
\end{align}

The total probability $P_{\text{\tiny Star}}$ for a source star to undergo an astrometric microlensing event is obtained by integrating the conditional probability $\expval{p_{\text{\tiny AML}}^{\text{\tiny c}}}$ along the entire line of sight to the source. For a lens located within a distance interval $(D_{\text{\tiny L}}, D_{\text{\tiny L}}+\Delta D_{\text{\tiny L}})$, the contribution to the total probability is weighted by the number of lenses present within the angular patch $\Delta \Omega$. Summing these contributions over all lens distances from the observer $0$ to the sources $D_{\text{\tiny S}}$ then gives the overall probability,
\begin{equation}\label{eq:app_lens_tot_prob}
		P_{\text{\tiny Star }} =  \int_{0}^{D_{\text{\tiny S}}} dD_{\text{\tiny L}} D_{\text{\tiny L}}^2  \Delta \Omega \frac{f_{\text{\tiny DM}}\rho_{\text{\tiny DM}}(D_{\text{\tiny L}},\alpha,\delta)}{M_{\text{\tiny L }}} \expval{p_{\text{\tiny AML}}^{\text{\tiny c}}}\;.
\end{equation}
We consider that the dark matter (DM) density within the Milky Way halo follows the standard spherically symmetric Navarro–Frenk–White (NFW) profile\,\cite{Navarro:1995iw}, which is expressed as
\begin{equation}\label{eq:app_AML_lens_NFW}
		\rho_{\text{\tiny DM}} = \frac{\rho_0}{\frac{r}{r_s}\left(1+ \frac{r}{r_s}\right)^2}\;,
\end{equation}
where,  $r\equiv\sqrt{R_{\text{\tiny e}}^2+D_{\text{\tiny l}}^2-2D_{\text{\tiny L}}R_{\text{\tiny e}}\cos\alpha\cos\delta }$, represents the distance of the lens from the galactic center, and $R_{\text{\tiny e}}=8.5$ kpc, is the distance from the earth to the galactic center. The NFW profile parameters are $\rho_0 =1.06\times10^7 ~M_{\odot}/\textrm{kpc}^3 $ and a scale radius  $r_s = 12.5 \textrm{ kpc}$ \,\cite{Sofue:2011kw}. 

\section{Supplementary plots}
\label{app:sup_result}

 		\begin{figure}[H]
 			\begin{center}
 				\includegraphics[scale=0.45]{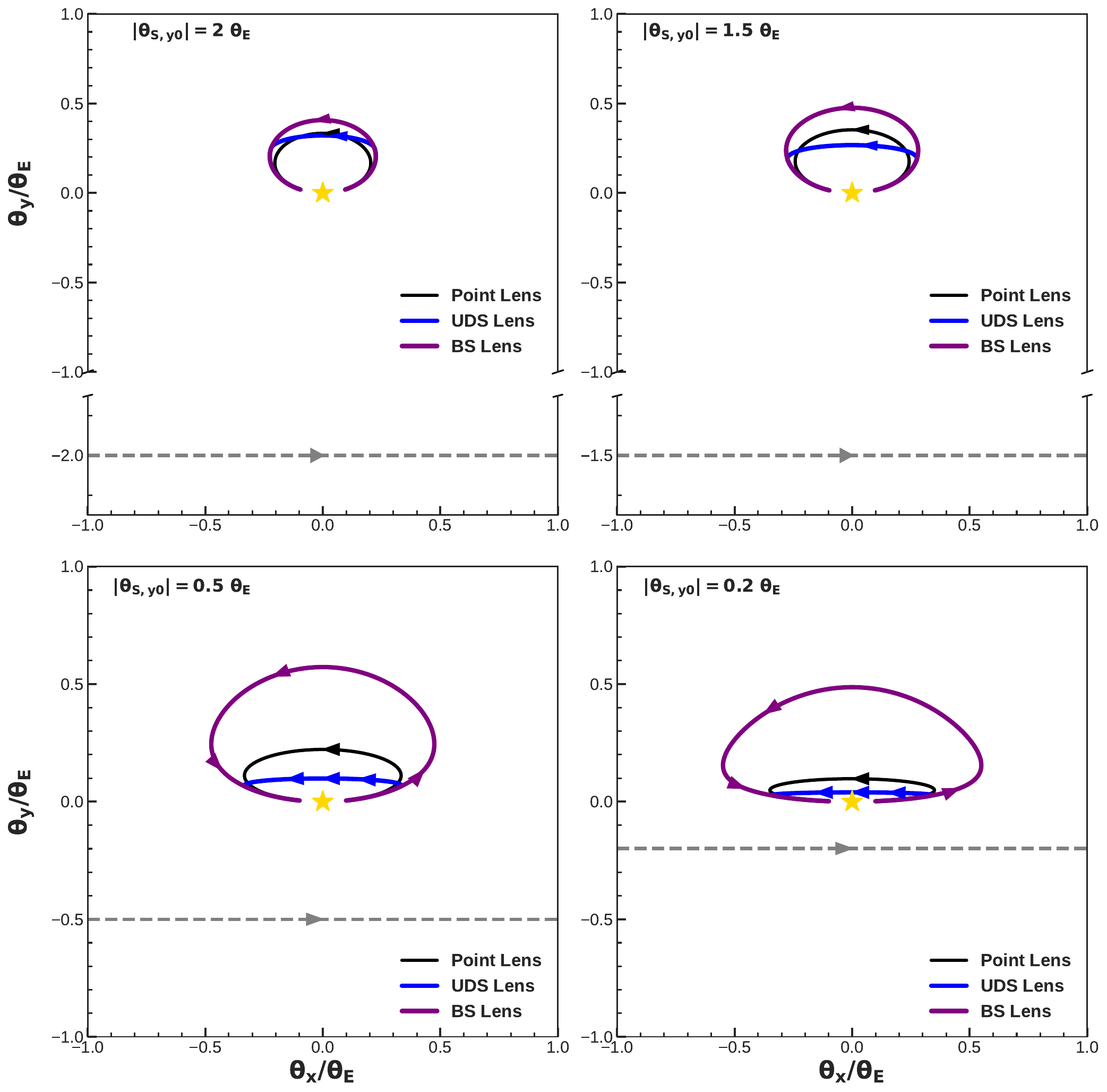}
 			\end{center}
 			\caption{Same as Fig.\,\ref{fig:uds_centshift_vs_us_2d}, but for dilute lens with $R_{\text{\tiny L}}/R_{\text{\tiny E}}=3$. We display a comparison of AML signals due to UDS (blue) and BS (purple) lenses. As before, the blue and purple curves trace the centroid-shift trajectories of the source induced by UDS and BS lenses, respectively, for lenses moving along the $x$-axis. Four cases are shown with minimum impact parameters $\theta_{\text{\tiny S,y0}}/\theta_{\text{\tiny E}} = 2$ (\textit{top left}), $1.5$ (\textit{top right}), $0.5$ (\textit{bottom left}), and $0.2$ (\textit{bottom right}). The gray line marks the lens trajectory, while black curves indicate the corresponding centroid shifts for a point lens.  The source star is located at the origin. All axes indicate angular positions on the sky, normalized by $\theta_{\text{\tiny E}}$. Plot displays that the AML signal remains quantitatively distinct for the BS and UDS models, even in the dilute-lens regime where no caustic crossings occur.} 
 			\label{fig:uds_centshift_vs_us_2d_1}
 		\end{figure}

\begin{figure}[H]
    \centering
    \includegraphics[scale=0.375]{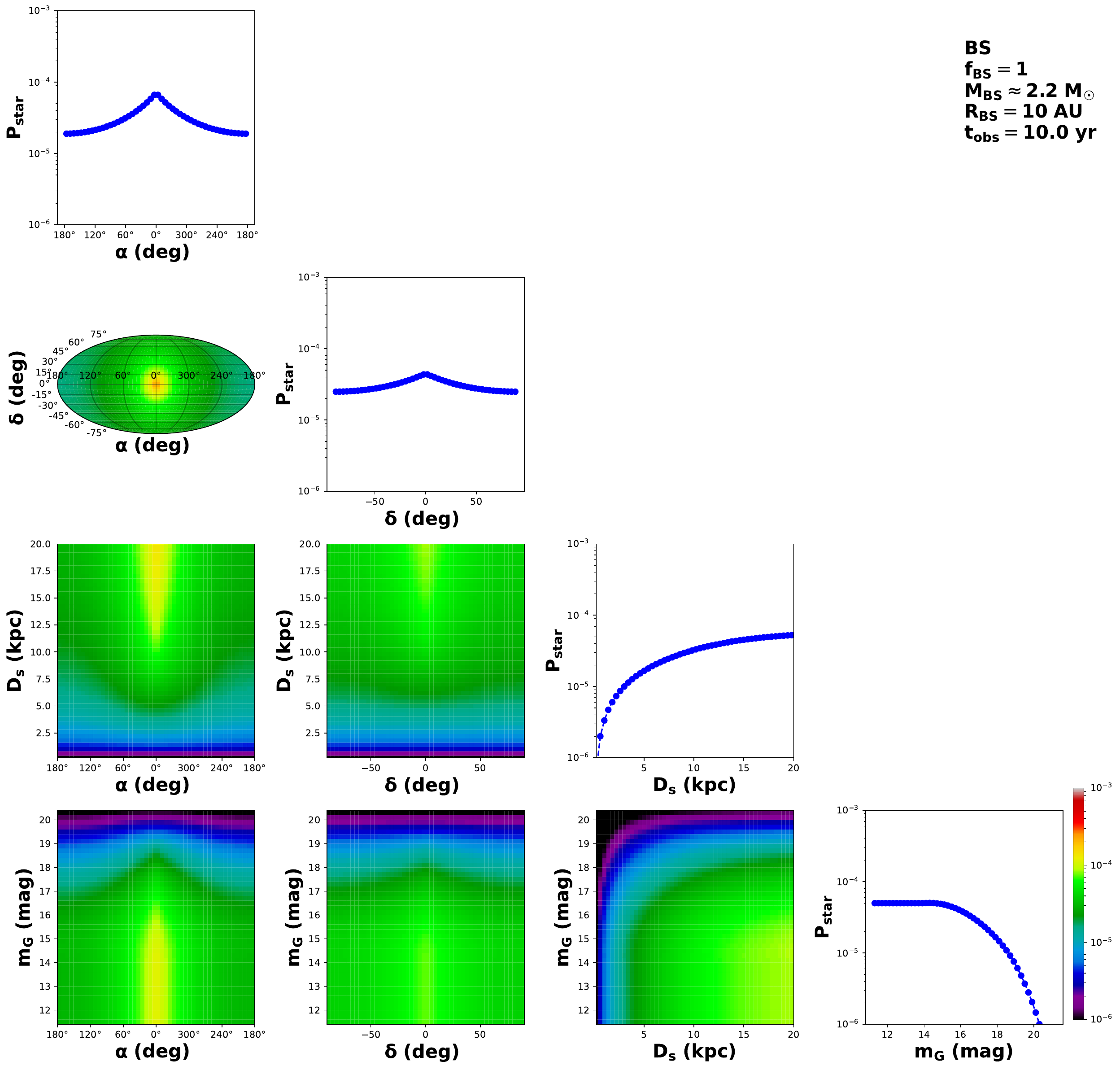}
    \caption{Same as the Fig.\,\ref{fig:uds_prob_dist_plot}, but for a boson star (BS) lens with parameters $(f_{\text{\tiny BS}},M_{\text{\tiny BS}},R_{\text{\tiny BS}})=(1,2.2~\solarmass,10~\text{AU})$ and an observational time of $t_\textrm{obs}=10~\text{yr}$. The 2D histogram plots (located at off-diagonal position) display $P_{\text{\tiny Star}}$ as a function of two source star parameters chosen from $(D_s,\alpha,\delta,m_G)$, while averaging uniformly over the remaining two parameters. The source parameters are obtained on the grid defined in Tab.\,\ref{tab:gaia_data_binned}.}
    \label{fig:bs_prob_dist_plot}
\end{figure}

\begin{figure}[H]
    \centering
    \includegraphics[scale=0.375]{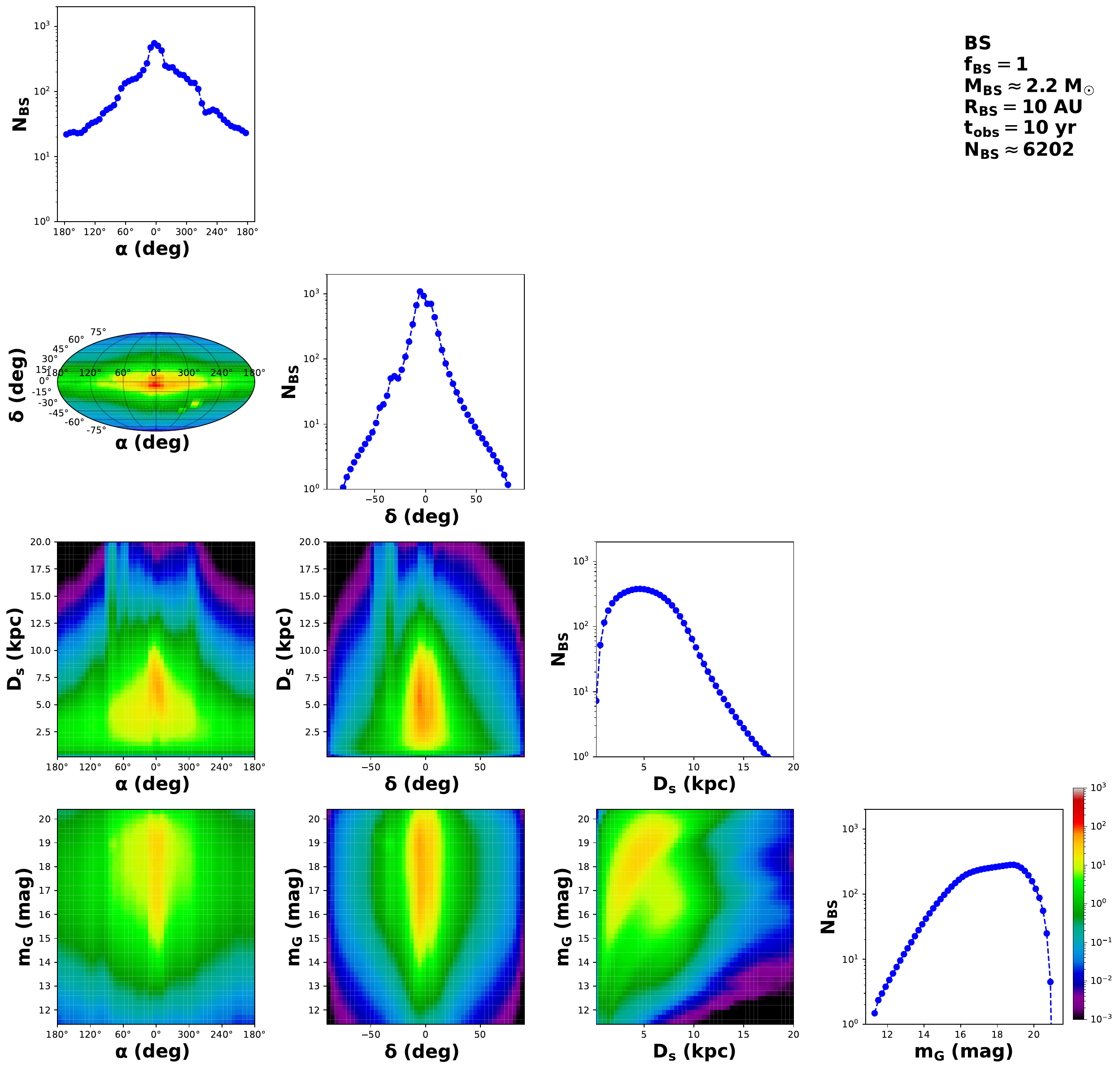}
    \caption{Same as the Fig.\,\ref{fig:uds_ANL_dist_plot}, but for a boson star (BS) lens with parameters $(f_{\text{\tiny BS}},M_{\text{\tiny BS}},R_{\text{\tiny BS}})=(1,2.2~\solarmass,10~\text{AU})$ and an observational time of $t_\textrm{obs}=5~\text{yr}$. Again, each off-diagonal 2D histogram shows $N_{\text{\tiny u}}$ as a function of two source star parameters from $(D_s,\alpha,\delta,m_G)$, having been summed over the remaining two parameters. And the diagonal 1D histogram plots present $N_{\text{\tiny u}}$ as a function of a single parameter, summed over the other three. Similar to the UDS lens, we expect approximately $6200$ AML events caused by the BS lens to be observed by Gaia over its 10-year mission. Further details are provided in the text.}
    \label{fig:bs_ANL_dist_plot}
\end{figure}

\begin{figure}[H]
    \centering
    \includegraphics[scale=0.28]{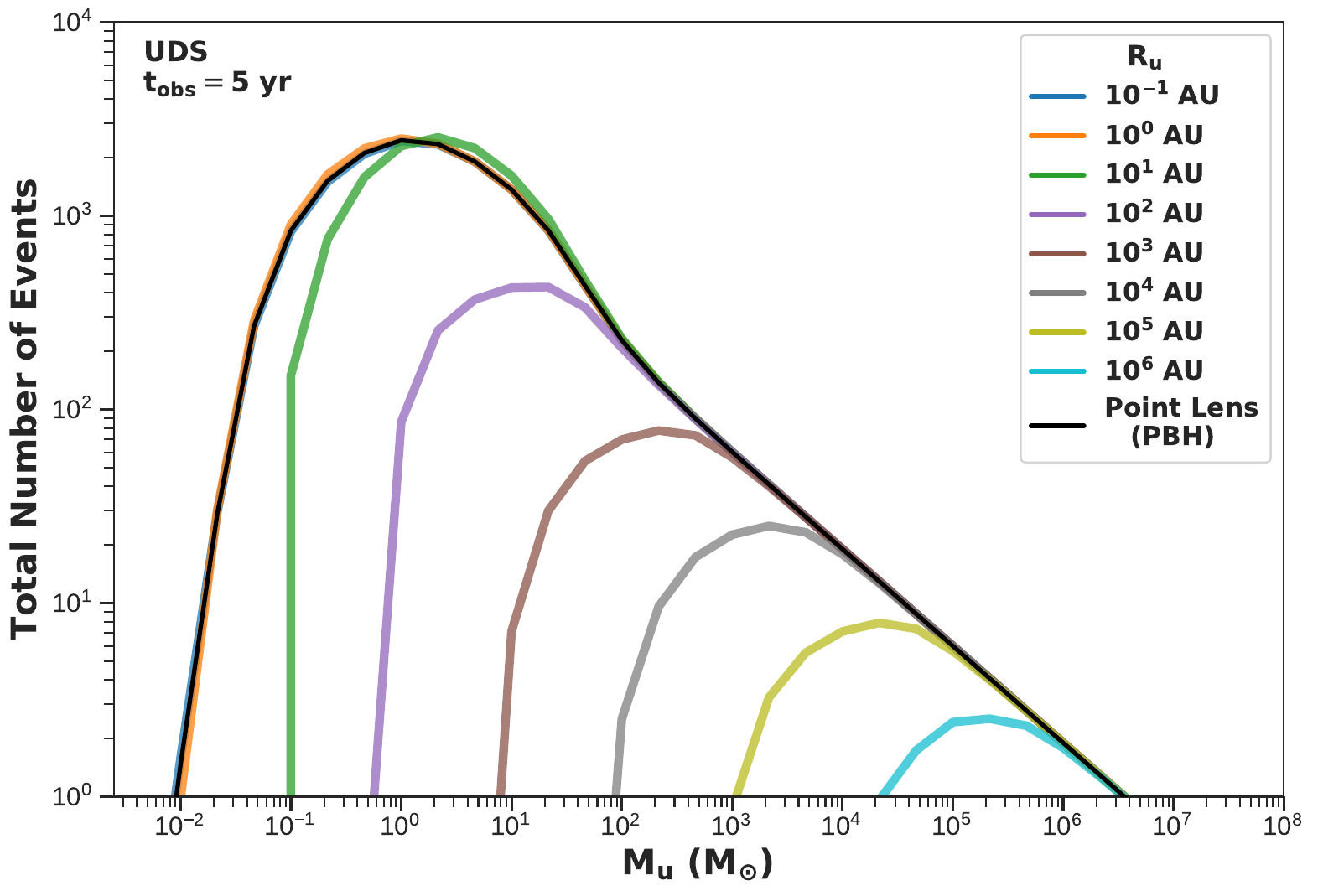}
    \includegraphics[scale=0.28]{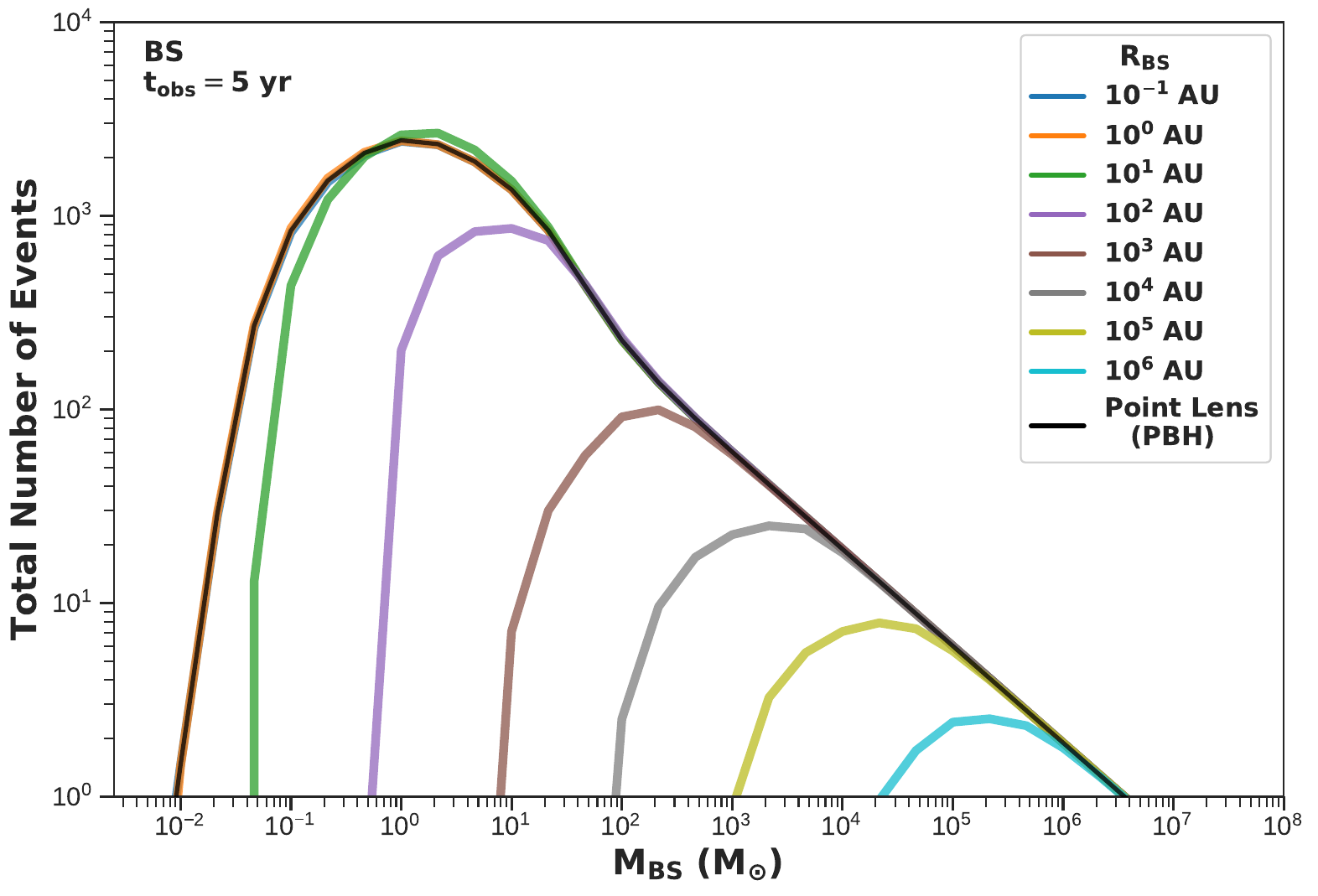}
    \includegraphics[scale=0.28]{uniform_sphere_Nl_vs_M_tobs_10_draft_v2.pdf}
    \includegraphics[scale=0.28]{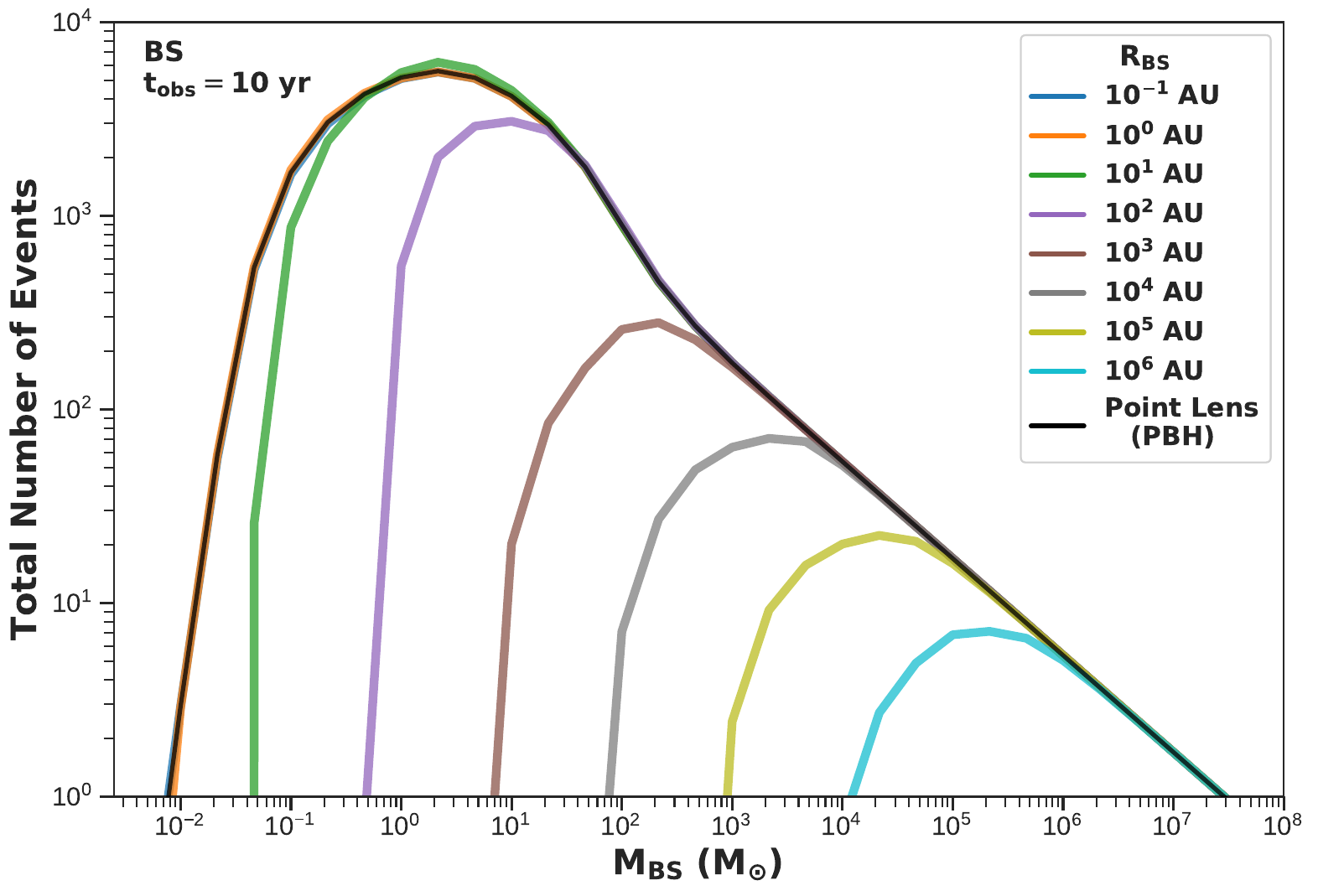}
    \caption{Same as the top panel of Fig.\,\ref{fig:uds_bs_Nu_vs_M}, the 2x2 grid compares the UDS model (top row) with the BS model (bottom row) for two observation periods: $t_{\text{\tiny obs}} = 5$ years (left column) and $t_{\text{\tiny obs}} = 10$ years (right column). All calculations assume $f_{\text{\tiny L}}=1$. Within each panel, the colored curves represent different lens radii, $R_{\text{\tiny L}} =(0.1,1,10,10^2,10^3,10^4,10^5,10^6)$, while the black line shows the point-lens approximation. These figures highlight how finite-size effects, which depend on both lens mass and radius, lead to deviations from the point-lens predictions.}
    \label{fig:uds_bs_Nu_vs_M_1}
\end{figure}

\begin{figure}[H]
    \centering
    \includegraphics[scale=0.45]{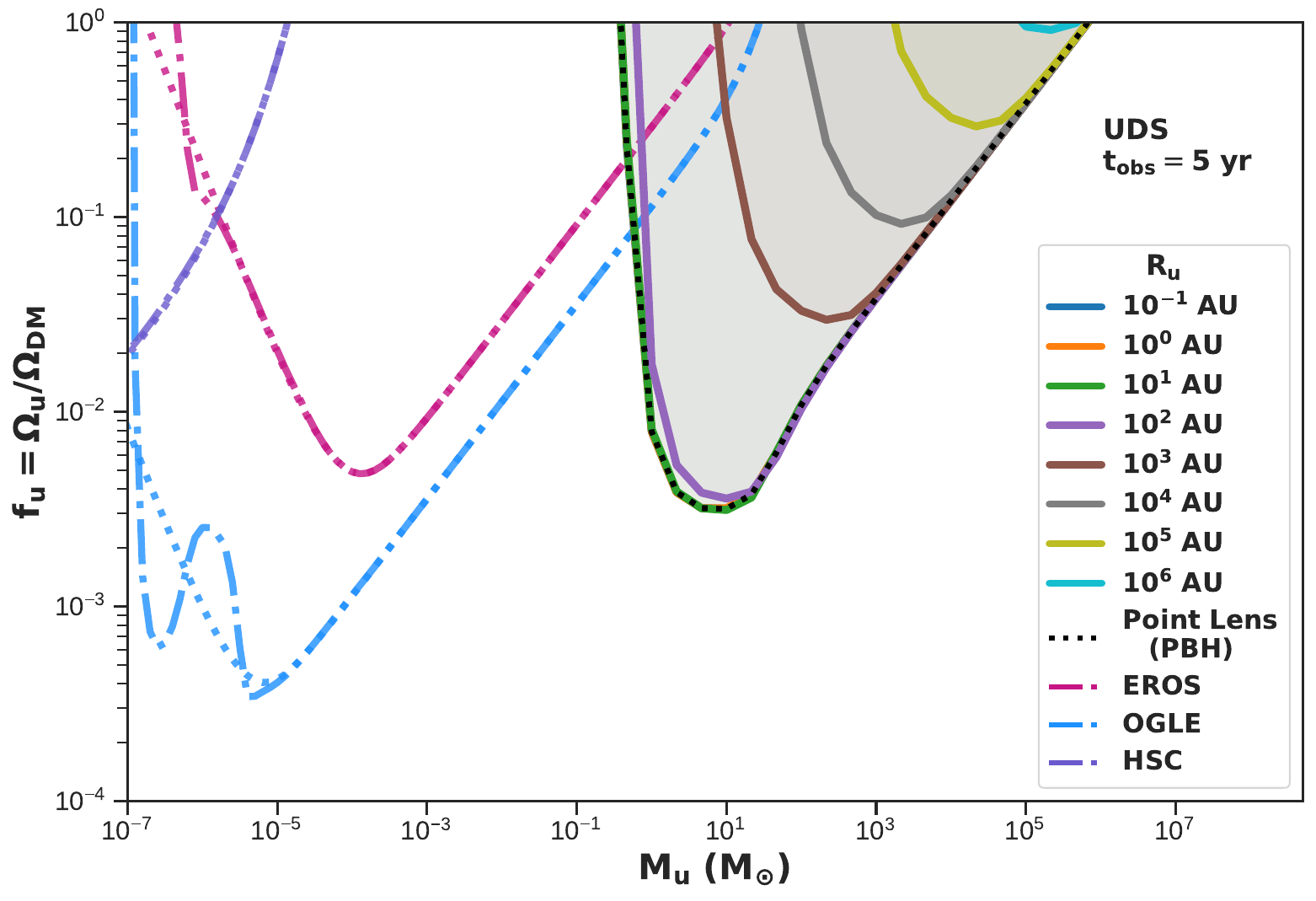}
    \includegraphics[scale=0.45]{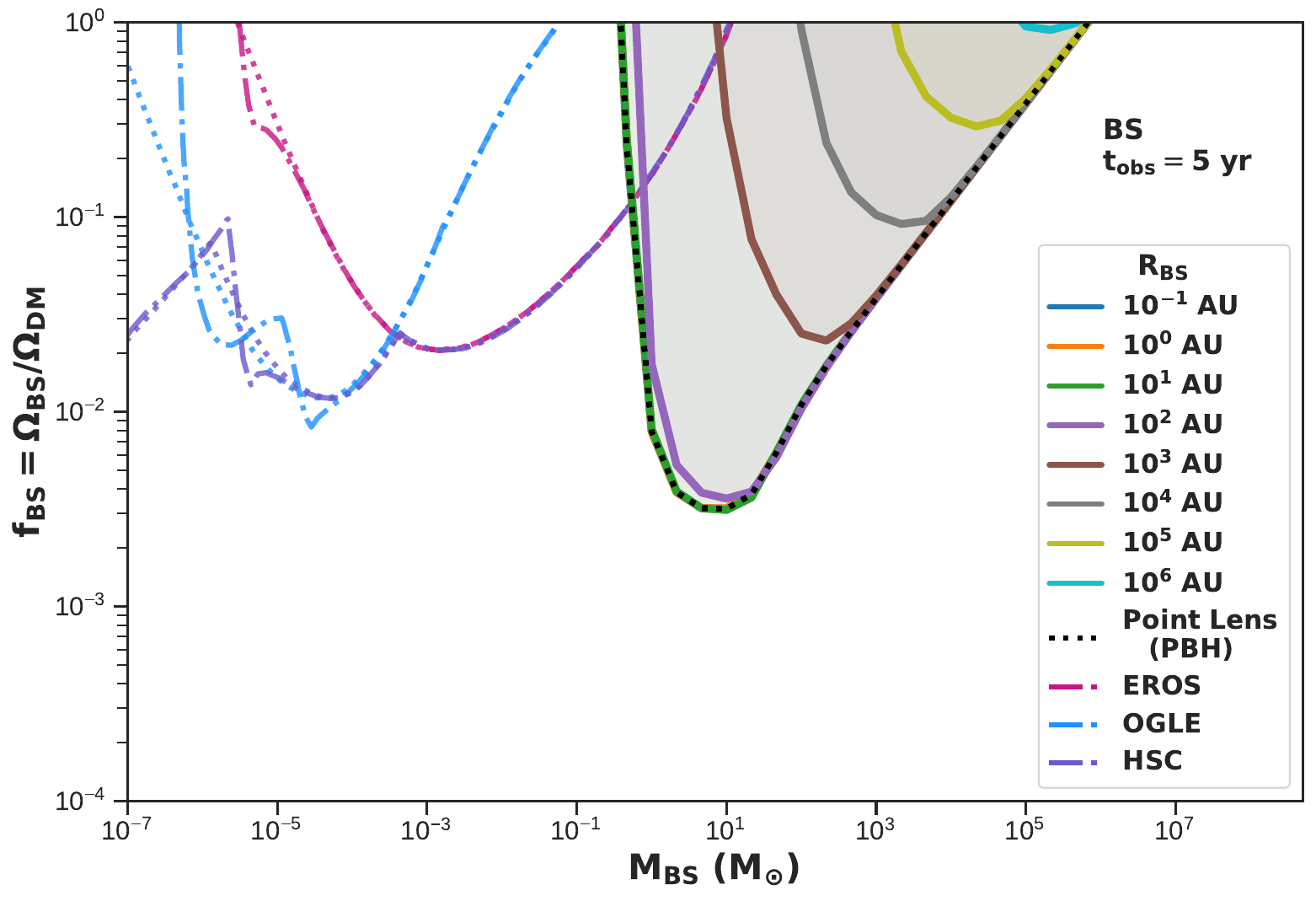}
    \caption{Same as Fig.\,\ref{fig:uds_exclusion}, but for  observation time of $t_{\text{\tiny obs}} = 5$ years. The top panel displays the results for the UDS lens model, while the bottom panel displays the results for the BS model. Each colored curve represents a lens radius, $R_{\text{\tiny L}} =(0.1, 1, 10, 10^2, 10^3, 10^4, 10^5, 10^6)~\text{AU}$, while the solid black line denotes the point-lens approximation. For comparison, existing PML constraints are also shown from EROS EROS\,\cite{Ansari:2023cay,Croon:2020wpr} (red, dash-dot), OGLE\,\cite{Ansari:2023cay,Croon:2020wpr} (royal blue, dash-dot), and HSC-Subaru\,\cite{Ansari:2023cay,Croon:2020ouk} (red, dash-dot) surveys.  We get peak constraint $f_{\text{\tiny L}}\lesssim \mathcal{O}(10^{-2})$ for EADOs in mass range $M_{\text{\tiny L}}\sim1-10~\solarmass$ and radii $R_{\text{\tiny L}}\lesssim 10~$ AU. }
    \label{fig:uds_exclusion_1}
\end{figure}

\newpage 
\bibliographystyle{JHEP.bst}
\bibliography{ECOs_Astrometric_Lensing} 
\end{document}